%% file: 00_main.tex
\documentclass[12pt]{article}

\usepackage[titletoc]{appendix}
\usepackage[margin={1.0in}]{geometry}
\usepackage{rotating, lscape}
\usepackage{graphicx, color, subfig}
\usepackage{threeparttable, booktabs, longtable, threeparttablex, tabularx}

\usepackage{pdflscape} 
\usepackage{makecell}

\usepackage{natbib}

\usepackage{authblk}
\usepackage{mathtools}
\usepackage{dsfont}
\usepackage{comment}

\usepackage{lmodern} 
\usepackage[sc]{mathpazo}
\usepackage{courier}
\usepackage[utf8]{inputenc}
\usepackage[T1]{fontenc}

\ExplSyntaxOn
\cs_new:Npn \expandableinput #1
  { \use:c { @@input } { \file_full_name:n {#1} } }
\ExplSyntaxOff

\usepackage{setspace}
\usepackage{amssymb,amsmath,mathrsfs,amsthm, bbm}
\usepackage{microtype}
\usepackage{float}
\usepackage{multirow}
\usepackage{pdflscape}

\usepackage[hidelinks,colorlinks=true,linkcolor=blue,citecolor=black,urlcolor=blue]{hyperref} 

\usepackage[nameinlink]{cleveref} 

\hypersetup{pdfstartview={Fit}} 

\RequirePackage[font=small,format=plain,labelfont=bf,textfont=it]{caption}

\begin{document}

\title{\vspace{-0.1cm} \LARGE \bf Failing Banks}

\author{\vspace{0.25cm} Sergio Correia, Stephan Luck, and Emil Verner\textsuperscript{*}}

\date{\vspace{0.2cm} June 5, 2025}

\pagenumbering{gobble}
\maketitle

\begin{abstract}


Why do banks fail? We create a panel covering most commercial banks from 1863 through 2024 to study the history of failing banks in the United States. Failing banks are characterized by rising asset losses, deteriorating solvency, and an increasing reliance on expensive noncore funding. These commonalities imply that bank failures are highly predictable using simple accounting metrics from publicly available financial statements. Failures with runs were common before deposit insurance, but these failures are strongly related to weak fundamentals, casting doubt on the importance of non-fundamental runs. Furthermore, low recovery rates on failed banks’ assets suggest that most failed banks were fundamentally insolvent, barring strong assumptions about the value destruction of receiverships. Altogether, our evidence suggests that the primary cause of bank failures and banking crises is almost always and everywhere a deterioration of bank fundamentals.



\vspace{0.5cm}


\end{abstract}

\let\oldthefootnote\thefootnote
\renewcommand{\thefootnote}{\fnsymbol{footnote}}
\footnotetext[1]{Correia: Federal Reserve Bank of Richmond, \href{mailto:sergio.correia@rich.frb.org}{sergio.correia@rich.frb.org}; Luck: Federal Reserve Bank of New York, \href{mailto:stephan.luck@ny.frb.org}{stephan.luck@ny.frb.org};
Verner: MIT Sloan School of Management and NBER, \href{mailto:everner@mit.edu}{everner@mit.edu}.
We thank our discussants Jorge Abad, Matt Baron, Asaf Bernstein, Elena Carletti, Mark Carlson, Tim Eisert, Daniel Ferreira, Matt Jaremski, Joseph Mason, Andrew Metrick,  Manju Puri, Eva Schliephake, Skander van den Heuvel, and Kaspar Zimmermann for their invaluable comments. 
We also thank Viral Acharya, Rosalind Bennett, Michael Bordo, Harry Cooperman, Natalie Cox, Darrell Duffie, Thomas Eisenbach, Huberto Ennis, Robin Greenwood, Kinda Hachem, Ing-Haw Cheng, Sam Hanson, Barney Hartman-Glaser, Zhigou He, Ben Hebert, Bev Hirtle, \`{O}scar Jord\`{a}, Nobu Kiyotaki, James McAndrews, Stephen Morris, Thomas Philippon, Matt Plosser, Andrei Shleifer, Hyun Shin, Jon Steinsson, Adi Sunderam, Eugene White, and participants at various conferences and seminars 
for useful comments. We also would like to especially thank Natalie Girshman and Francis Mahoney for help with archival work and Rosalind Bennett for sharing the FDIC's Failure Transaction Database. We thank Gabriel Leonard and Tiffany Fermin for excellent research assistance. This project has received support from the MIT Research Support Committee's Ferry Fund. The opinions expressed in this paper do not necessarily reflect those of the Federal Reserve Bank of New York or the Federal Reserve Bank of Richmond.}


\let\thefootnote\oldthefootnote

\doublespacing

\clearpage
\pagenumbering{arabic}


\input{01_introduction.tex}

\input{02_conceptual.tex}

\input{03_data.tex}

\input{05_prediciting.tex}

\input{06_runs.tex}

\input{07_recovery_rates.tex}

\input{09_conclusion.tex}

\singlespacing
\bibliographystyle{chicago}
\bibliography{literature}

\begin{appendices}
\clearpage
   	
\appendix

\setcounter{footnote}{0}
\setcounter{equation}{0} \renewcommand{\theequation}{A.\arabic{equation}}
\pagenumbering{arabic} \setcounter{page}{1}
\renewcommand{\thepage}{A.\arabic{page}} 
\clearpage
\renewcommand\thefigure{\thesection.\arabic{figure}}
\renewcommand\thetable{\thesection.\arabic{table}}
\setcounter{figure}{0}
\setcounter{table}{0}

\begin{center}
\LARGE { \singlespacing \bf Failing Banks \\
\textit{Online Appendix}}
 \\

\author{\vspace{1.0cm} \large Sergio Correia, Stephan Luck, and Emil Verner\textsuperscript{*} \\ \vspace{0.7cm}}

\normalsize
\date{ \today } %
\end{center}
\let\oldthefootnote\thefootnote
\renewcommand{\thefootnote}{\fnsymbol{footnote}}
\footnotetext[1]{Correia: Federal Reserve Bank of Richmond, \href{mailto:sergio.correia@rich.frb.org}{sergio.correia@rich.frb.org}; Luck: Federal Reserve Bank of New York, \href{mailto:stephan.luck@ny.frb.org}{stephan.luck@ny.frb.org};
Verner: MIT Sloan and NBER, \href{mailto:everner@mit.edu}{everner@mit.edu}.}

\let\thefootnote\oldthefootnote

\begin{itemize}
    \item Appendix A: Evolution of the U.S. Banking System and Bank Failures
    \item Appendix B: Additional Tables and Figures
    \item Appendix C: Data Appendix
\end{itemize}

\input{99_historical_context.tex}
\setcounter{figure}{0}
\setcounter{table}{0}

\input{99_appendix_figures.tex}

\input{99_appendix_tables.tex}
\setcounter{figure}{0}
\setcounter{table}{0}
\setcounter{figure}{0}
\setcounter{table}{0}

\input{99_appendix_data.tex}

\clearpage

\setcounter{figure}{0}
\setcounter{table}{0}

\end{appendices}
\end{document}

%% file: 01_introduction.tex

\clearpage
\setcounter{page}{1}

\section{Introduction}

Bank failures are a recurrent feature of banking. In the United States, 19\% of all national banks in existence from 1863 to 1934 and 13\% of all commercial banks in existence from 1935 to 2023 failed at some point during these periods. Bank failures often lead to real economic disruptions \citep{Bernanke1983}, and there is a strong association between systemic banking crises featuring widespread bank failures and severe macroeconomic downturns \citep{Reinhart2009}.


What causes bank failures? Theory offers two main explanations. Under the \textit{bank runs view}, bank failures are the consequence of runs in which depositors collectively withdraw from otherwise solvent \citep{Diamond1983} or troubled but solvent banks \citep{Goldstein2005}. Bank runs are cited as an important cause of bank failures in prominent accounts of the Great Depression \citep{FriedmanSchwartz}, the 2008 Global Financial Crisis \citep{Bernanke2018}, and the bank failures in spring 2023. Under an alternative \textit{solvency view}, bank failures are caused by poor fundamentals, such as realized credit risk, interest rate risk, or fraud, that trigger insolvency \citep[e.g.,][]{Temin1976,Wicker1996,Calomiris1997,AdmatiHellwig2014,Gennaioli2018}. While these two views are not mutually exclusive, the \textit{bank runs view} emphasizes the runnable nature of bank liabilities as a key element to understanding bank failures, whereas, in the \textit{solvency view}, failure is caused by losses, irrespective of whether a bank run occurs or not.

This longstanding debate raises several important questions. Which type of failures are empirically most relevant? Are bank failures primarily a result of bank runs or are they more commonly caused by insolvency? When runs do occur, do they commonly cause the failure of otherwise solvent banks, or do they primarily trigger the failure of insolvent banks?

Understanding the potential determinants of bank failures empirically, however, is challenging. Government interventions such as deposit insurance and lending of last resort reduce the scope for bank runs to cause bank failures in modern times \citep{MetrickSchmelzing2021}.  A common argument for these interventions is precisely to prevent failures caused by runs, especially on otherwise solvent banks. Thus, observed bank failures in modern times may be biased towards failures involving poor fundamentals.

To overcome this challenge, we study the history of failing banks in the United States from 1863 to 2024. We construct a new database with balance sheet information for most banks in the U.S. since the Civil War. Our data consist of a historical sample that covers all national banks from 1863 to 1941 and a modern sample that covers all commercial banks from 1959 to 2024. Altogether, our data contain balance sheets for around 37,000 distinct banks, of which more than 5,000 fail. This long sample thus covers failures both before and after the founding of the Federal Reserve System  and the introduction of deposit insurance from the Federal Deposit Insurance Corporation (FDIC). This dataset, therefore, allows us to study bank failures during historical episodes in which bank runs could plausibly have been a common cause of bank failures.

We present three sets of findings that inform the determinants of bank failures throughout the history of the U.S. banking system. First, we document that bank failures are strongly related to weak bank fundamentals. As a result, bank failures are highly predictable throughout the sample. Second, we show that large deposit outflows, indicative of bank runs, were common in pre-FDIC bank failures. Nevertheless, failures with bank runs are strongly connected to weak bank fundamentals. Third, we argue that low recovery rates in pre-FDIC failures imply that most banks that failed with a run were fundamentally insolvent, unless one assumes large bank value destruction from failure. Overall, our findings suggest that the \textit{solvency view} goes a long way to understanding the primary cause of most bank failures. 

We begin by documenting that failing banks are characterized by poor and deteriorating fundamentals. First, failing banks see a rise in non-performing loans and gradually deteriorating solvency several years before failure. Second, in the run-up to failure, banks increasingly rely on expensive and risk-sensitive noncore funding, such as time deposits and wholesale funding. Furthermore, failing banks undergo a boom-bust pattern in assets during the decade before failure. Asset losses thus often follow a period of rapid loan growth.

These facts imply that bank failures are highly predictable based on weak fundamentals captured by accounting metrics from publicly available financial statements. The future probability of bank failure rises significantly in measures of insolvency risk and reliance on noncore funding. For example, a bank in the top 5\textsuperscript{th} percentile of both insolvency risk and noncore funding reliance has a probability of failure over the next three years of over 25\% in both the historical and modern sample. This amounts to a 10- to 25-fold increase in the probability of failure relative to the average bank, a large differential. 

We formally quantify the extent of predictability by estimating simple regression models in which we predict whether a bank will fail based on proxies of bank fundamentals and macroeconomic conditions. We assess predictability based on the area under the receiver operating characteristic curve (AUC), a common measure of performance for binary classifiers. In the historical, pre-FDIC sample, the AUC for predicting failure within the next year is 86\%, indicating a substantial degree of predictability. In the modern sample, after the introduction of deposit insurance, the predictability of bank failures is even higher, with an AUC between 90-95\%. In both the historical and modern samples, the predictability of failures is typically nearly as high in pseudo-out-of-sample as in in-sample forecasting exercises. 

Next, we show that bank runs were common in pre-FDIC bank failures. We compute the deposit outflow immediately before failure as the growth in deposits between the last pre-failure financial statement and failure. In the pre-FDIC sample, deposits in failing banks decline on average by 13\% immediately before failure. In contrast, deposits fall by only 2\% up to post-FDIC failures. Therefore, the deposit insurance regime features significantly fewer failures involving bank runs. Nevertheless, while large deposit outflows were common in pre-FDIC failures, we find that failures with bank runs are as predictable as failures without runs. Thus, failures with runs are similar to other failures in that they tend to occur in banks with weak fundamentals.

Weak bank fundamentals not only predict individual bank failures. They also forecast waves of bank failures during systemic banking crises. We aggregate the out-of-sample forecasts of individual bank failure risk to predict the aggregate bank failure rate. The $R^2$ of a regression of the actual bank failure rate on the predicted aggregate failure rate is 40\%. Thus, spikes in bank failures during systemic banking crises are, to a large extent, accounted for by deteriorating fundamentals.

In the final part of the paper, we examine recovery rates in pre-FDIC failures. We present a simple framework comparing a bank's recovery rate to its leverage to gauge whether a failed bank was fundamentally insolvent. Assessing the degree of insolvency is informative about whether a run could plausibly have caused the failure of a fundamentally solvent bank. However, it requires making assumptions about the wedge between the value of the assets in the bank and the value of the assets for the next best user. We first establish that recovery rates were low in pre-FDIC failures, averaging 52\% of the book value of assets. We document that low recovery rates reflect, to a significant extent, unrealized losses on assets. Based on our framework, we then show that the majority of failed banks appear to be fundamentally insolvent. Moreover, the share of bank failures that involved a run on a bank that was \textit{not} fundamentally insolvent---failures for which runs are a plausible cause of failure---is likely to be modest. For instance, under the extreme assumption that there is no value destruction from failure at all, runs on weak but solvent banks can account for less than 8\% of pre-FDIC failures. Under an equally extreme assumption that failure destroys 20\% of bank value, this share rises to 22\%.

Taken together, our evidence suggests that the primary cause of bank failures is almost always and everywhere a deterioration of bank solvency. The erosion of a bank's capitalization ultimately results in either a run or a supervisory decision to close a bank, with runs being more common in the historical data. Importantly, both depositors and supervisors seem slow to react to information about bank fundamentals, thus making bank failures predictable. The predictability of bank failures, in turn, suggests that non-fundamental, self-fulfilling runs on otherwise healthy banks, as in \cite{Diamond1983}, are an uncommon cause of bank failures, as such runs should strike randomly \citep[see, e.g.,][]{Gorton1988,Greenwood2023}.

Our finding that failures with runs typically occur in banks with weak fundamentals is consistent with the predictions of theories of fundamental-based panic runs \citep{Goldstein2005,Morris2003,Rochet2004}.  However, our findings on low recovery rates also suggest that the majority of pre-FDIC failures with runs involved fundamentally insolvent banks. Thus, we argue that runs were more commonly an important trigger of failure for already insolvent banks than a primary cause of failure of potentially weak but solvent banks.

The high predictability of bank failures and the finding that runs typically close insolvent banks suggests that runs happen later than standard theoretical benchmarks would predict. In models of panic runs with rational and forward-looking depositors exposed to large losses from failure, bank failures cannot be highly predictable, as attentive depositors would act on this information and withdraw their funds, reducing the predictability in the first place. Thus, the fact that these banks have not failed yet and we can observe high predicted failure probabilities suggests that depositors are often slow to react to an increased risk of bank failure. This fact, in turn, points to a role for behavioral frictions such as neglect of downside risk \citep[e.g.,][]{GSV2012} and sleepy or inattentive depositors \citep[see, e.g.][]{HANSON2015449,Jiang2023}.

Finally, our interpretation that solvency rather than the runnable nature of bank liabilities is key to understanding banking failures is supported by classifications of causes of bank failures provided by contemporary bank examiners from the Office of the Comptroller of the Currency (OCC). Notwithstanding the common occurrence of large deposit outflows in the run-up to failure, most pre-FDIC bank failures were classified by the OCC as being caused by losses, fraud, or external economic shocks. Despite popular narratives about bank runs playing a key role in the historical U.S.\ banking system, runs and liquidity issues account for less than 2\% of failures classified by the OCC.

\paragraph{Related literature.} 
Our paper relates to two strands of literature on bank failures and financial crises. 

First, we relate to micro-level studies of bank failures, runs, and banking crises, such as empirical studies of the Great Depression \citep[e.g.,][]{Calomiris1997,Calomiris2003a,Mitchener2019}, the 2008 Global Financial Crisis \citep[e.g.,][]{Gorton2012,Krishnamurthy2014,Schmidt2016}, the recent banking stress in March 2023 \citep[e.g.,][]{Jiang2023,Metrick2023,CiprianiEtAl2024}, and other episodes featuring bank runs \citep{Iyer2012,Frydman2015,Iyer2016,Artavanis2022}.\footnote{Several of these studies focus on explaining banking failures during specific episodes in the U.S. \cite{Calomiris2003a} find that fundamentals explain bank failures in the Great Depression, rather than panic-driven depositor flight. Using state-level data \cite{Alston1994} find that failures in the 1920s were highest in states that saw the largest growth in agricultural acreage during WWI, and most failing banks were small and rural. Studies using recent Call Report data find that highly levered banks, banks with low earnings, low liquidity, and risky asset portfolios are more likely to fail \citep[e.g.,][]{Cole1995,Cole1998,Wheelock2000,BergerBouwman2013}.} The novelty of our approach is to bring together evidence from 160 years of micro-level data that spans a range of institutional and regulatory regimes. Studying the close-to-complete history of the banking system in the United States allows us to generalize the insight that weak fundamentals are typically a necessary condition for bank failures across various institutional settings, both during financial crises and during quiet periods. The richness of the data further allows us to provide robust facts about the predictability of bank failures, deposit outflows before failure, and asset recovery rates in failure. Contrasting these facts with testable predictions of models of bank failures and runs, we argue that insolvency is the most common primary cause of failure, while runs most commonly trigger the failure of fundamentally insolvent banks. Moreover, while existing micro-level studies usually condition on a crisis, our long sample demonstrates that failures and banking crises are predictable out-of-sample.

Second, our paper is related to studies of financial crises using aggregate data. Within this literature, our paper relates most closely to studies on the nature of banking crises and the sources of bank failures and panics. \cite{Gorton1988} and \cite{Calomiris1991} study banking panics in the National Banking Era and find that panics generally followed bad macroeconomic news but were not important for bank failures. \cite{Baron2021} argue that panic runs are not necessary for banking crises, and panics are preceded by bank equity declines, reflecting the realization of bank losses. Our paper provides complementary evidence by using granular bank-level data. This allows us to show that deteriorating fundamentals are necessary for both individual and widespread bank failures, including failures with runs. \cite{Jorda2020} find that higher banking system capitalization is not associated with a lower chance of banking crises but does predict stronger recovery from crises. Our bank-level findings indicate that higher bank capitalization predicts a lower probability of failure and aggregate crises.

\paragraph{Roadmap.} The paper proceeds as follows. \Cref{sec:conceptual} provides a conceptual framework to guide our empirical analysis. \Cref{sec:data} describes the data. \Cref{sec:facts} provides basic facts about bank fundamentals and bank failures. \Cref{sec:predicting_failures} presents evidence on the predictability of bank failures. \Cref{sec:predicting_runs} studies deposit outflows in failing banks and the predictability of bank failures with runs. \Cref{sec:waves} shows that bank-level fundamentals predict the major waves of bank failures in the U.S. \Cref{sec:recovery} presents evidence on recovery rates and the share of fundamentally insolvent banks, and \Cref{sec:conclusion} concludes.

%% file: 02_conceptual.tex
\section{Conceptual Framework}

\label{sec:conceptual}

\subsection{Theory}
 We organize our empirical analysis around two views of why banks fail: the \textit{solvency view} and the \textit{bank runs view}. 

The \textit{solvency view} posits that banks fail because they become insolvent. This occurs when the expected value of assets is too low to pay off all debt claims. Insolvency can occur due to asset losses from realized credit risk, which in turn may be driven by unexpected bad shocks to bank assets or due to excessive risk taking, perhaps driven by deeper governance issues. While solvency risk is most commonly associated with credit risk, banks can also become insolvent due to interest rate risk, which can reduce asset values and also increase funding costs. Importantly, a distinct feature of the solvency view is that the runnable nature of bank liabilities is not an important factor in causing bank failures. For example, \cite{MorrisShin2016} define solvency risk as the probability of failure in a counterfactual where withdrawals are not possible.

The \textit{bank runs view}, in contrast, argues that the runnable nature of bank liabilities is an important element in explaining bank failures. In this view, bank runs play an important role in driving banks to insolvency and triggering failure. In the seminal model of \cite{Diamond1983}, banks finance illiquid assets with demandable deposits. While this benefits depositors by creating liquidity, coordination failure among depositors can lead to a self-fulfilling panic run on an otherwise solvent bank. The run leads the bank to liquidate assets at a loss, making the bank insolvent. In this model, the original cause of failure comes from the funding side and the behavior of depositors.\footnote{Other theories where runs and failures start from depositor behavior include \cite{Bryant1980}, \cite{Allen2000}, and \cite{Peck2003}.} Because a run is one of two equilibria, this framework does not predict when runs will occur, but it raises the possibility that runs can cause failure randomly and should thus be unpredictable.

The \cite{Diamond1983} framework is purposely stylized and abstracts from fundamental risk, such as asset losses. Indeed, existing time series and cross-bank studies find that banking panics and runs on individual banks generally follow bad news about fundamentals.\footnote{For aggregate time-series evidence, see, for example, \cite{Gorton1988}, \cite{Calomiris1991}, \cite{Wicker1996}, and \cite{Baron2021}. For bank-level evidence focused on specific crisis episodes, see \cite{Calomiris1997}, \cite{Calomiris2003a}, and \cite{Blickle2022}. } \cite{Goldstein2005} build on this idea and introduce shocks to bank assets in a model of bank runs. In this model, a panic run occurs when bank fundamentals are weak \citep[see also][]{MorrisShin2000,Morris2003,Rochet2004}. Formally, bank fundamentals $\theta$ are stochastic, and each depositor observes a slightly noisy signal of $\theta$.
When fundamentals are strong ($\theta > \overline{\theta}$) there is no risk of a run. If fundamentals are sufficiently weak ($\theta \leq \underline{\theta}$), the bank is insolvent and all depositors have an incentive to withdraw, irrespective of others' actions, resulting in failure through a ``fundamental run.'' Yet, moderately weak fundamentals below a threshold $\theta^*$ (i.e., $\underline{\theta} < \theta < \theta^* < \overline{\theta}$) will trigger a ``panic-based run'' in the threshold equilibrium of \cite{Goldstein2005}, leading to bank failure. In this case, absent a run, the bank's asset realization would have been sufficient to pay all creditors. Thus, runs can amplify the effects of weak fundamentals.

Theories of bank runs underpinning the \textit{bank runs view} usually have three central ingredients. First, depositors are rational and forward-looking. Hence, depositors immediately react to observable signals about sufficiently poor fundamentals \citep[e.g.,][]{Allen1998}.  Second, there are externalities among depositors. The expectation of a run by other depositors increases the incentive of a given depositor to withdraw. Third, the run leads to value destruction. In most frameworks, value destruction comes from fire selling illiquid assets \citep{Diamond1983}. Value destruction can also occur through a decline in the franchise value of deposits \citep{DSSW2023,Jiang2023,Amador2024}. 

Models of fundamental-based panic runs, such as \cite{Goldstein2005} and \cite{MorrisShin2000}, illustrate that theories of bank runs and failures can incorporate elements from both views. Moreover, dynamic interactions between solvency and funding can amplify distress: declining solvency may trigger deposit outflows and rising funding costs, which in turn further weaken solvency. This spiral can generate a ``slow run'' that erodes a bank's value \citep[e.g.,][]{LorenzoniWerning2019,LlambiasOrdonez2024}.\footnote{\cite{LlambiasOrdonez2024} model depositors who are partially inattentive. Adverse shocks to bank assets lead to gradual withdrawals. If the bank fundamentals fall below a certain threshold, then attentive depositors run. The model captures potential slow runs and the idea that failure can be preceded by persistently bad fundamentals because of depositor inattention, potentially generating predictability of runs and failures. For models of dynamic runs, see also \cite{He2012} and \cite{He2016}. \cite{LorenzoniWerning2019} provide a model of slow runs for sovereign debt crises.}

Moreover, even if the root cause of bank failure is insolvency ($\theta< \underline{\theta}$), a run can be the mechanism that triggers failure \citep[e.g.,][]{Diamond2001} and governs the efficiency of the bankruptcy. A fundamentally insolvent bank may, absent market discipline or supervisory action, be able to operate for some time until it runs out of cash. Thus, under the solvency view, even if the runnable nature of bank liabilities is not the root cause of bank failure, it can be important in determining the exact timing and mechanics of failure.

\subsection{Testable Implications}

Our empirical analysis brings forward novel evidence that sheds light on the nature and causes of bank failures. Given that the solvency and bank runs views are not mutually exclusive, we note that it is difficult to completely separate the two explanations. Nevertheless, our goal is to establish facts and assess which elements of the two views are consistent with the observed patterns.

First, we study the \textit{predictability of bank failures}. Strong predictability of failures based on weak bank fundamentals implies that fundamentals are an important factor in failures. Predictability is thus consistent with both the solvency view and the version of the bank runs view where panic-based runs are driven by weak bank fundamentals. However, predictability is less consistent with models where non-fundamental runs bring down otherwise healthy banks, as in \cite{Diamond1983}.

Further, while models of fundamental-based panic runs predict that failure is related to weak fundamentals, evidence of strong predictability also cuts against the assumption of forward-looking rational depositors. In models of fundamental-based panic runs such as \cite{Goldstein2005}, rational and attentive depositors run immediately when a sufficiently low signal $\theta$ is realized. If bank failures could be easily anticipated based on public data, then rational depositors would act on this information and run, thus reducing predictability by triggering failure soon after the first signs of distress. Therefore, predictability of failures with runs suggests that frictions, such as inattention, slow down runs. However, we emphasize that predictability does not reject the possibility that runs can close weak banks that would have remained solvent absent a run. 

Second, we study \textit{deposit outflows} in failing banks. For a bank run to represent the cause of a bank failure, deposits must actually flow out of the bank before failure. In standard theories of bank runs, deposit outflows erode solvency by forcing banks to either liquidate their otherwise valuable assets or replace deposit funding with more expensive wholesale funding \citep{Diamond1983,Allen2000,Goldstein2005}. Hence, if a bank fails with only a minimal decline in deposits, a bank run is unlikely to be the cause of failure.


Third, we study \textit{asset recovery rates} in failure. Under the solvency view, failing banks are insolvent, even absent the run. Fundamentally insolvent banks should thus have low asset quality and recovery rates in failure. In the bank runs view, the bank's asset realization would have been sufficient to pay off creditors absent the run. Therefore, asset quality in an otherwise solvent bank that fails because of the run should be less troubled than that of a fundamentally insolvent bank. An important challenge for translating this prediction into an empirical test is that the failure itself can destroy bank value due to a wedge between the value of the assets in the bank and the value for the next best user. For example, the value of bank assets may be linked to the human capital of bank managers. 
While we cannot observe the value of a failed bank in the counterfactual where it did not fail, in \Cref{sec:recovery} we present a simple framework that allows us to make inferences about the share of failed banks that were insolvent absent a run, conditional on assumptions about the potential value destruction from failure.

%% file: 03_data.tex
%
%


\section{Data}
\label{sec:data}

\paragraph{Data for historical sample (1863-1941).} We use two main data sources on bank balance sheets. Data on national bank balance sheets from 1863 through 1941 are from the Office of the Comptroller of the Currency's (OCC) Annual Report to Congress. The data provide annual bank-level information on broad balance sheet line items such as total assets, loans, deposits, and equity. In many years, the OCC also reports more detailed items that allow us to measure non-performing loans and various forms of non-deposit wholesale funding. However, the OCC did not require banks to report income statements. \Cref{fig:nb_balance_sheet_1900} and \Cref{fig:nb_balance_sheet_1933} in \Cref{app:data} provide examples of the original source. 

Data on all national banks from 1867 until 1904 are digitized and provided by \citet{Carlson2022}. For this project, we further digitize bank balance sheets from 1905 through 1941.\footnote{We also digitize additional data from 1863-1867. Moreover, note that while we collect data on national banks through 1941, most of our analysis is focused on data from before 1934 and thus before the FDIC became operational.} In both cases, balance sheets are digitized using optical character recognition (OCR), applying the methods discussed in \citet{Correia2022}. We hand-check the OCR output, with particular attention to cases where accounting identities fail to hold. 
Moreover, we compile a list of all significant bank events and their dates---including chartering, liquidations, and receiverships---from 1863 to 1935 using data manually collected by \citet{vanBelkum1968}, augmented by \citet{KeyBankData}, and which we validate by using information from the 1941 ``Alphabetical List of Banks'' \citep{AlphabeticalList}, as well as the corresponding OCC Annual Reports.


We define a national bank as failed when the OCC appoints a receiver. This definition of failure includes banks that eventually exit receivership, restore solvency, and continue operating, as well as banks that exit receivership and wind down their operations in an orderly voluntary liquidation that imposes no losses to creditors. However, this definition excludes temporary bank closures that did not involve a receiver at some point. Moreover, we exclude instances where banks briefly suspend convertibility of their debt into cash and then reopen, as was common during banking panics of the National Banking Era \citep[see, e.g.,][]{Jalil2015}. This implies that we also exclude banks that averted receivership due to cooperation through, for example, bank clearinghouses. We emphasize this distinction, since the drivers of bank runs resolved by temporary suspension of convertibility may differ from those that lead to bank failures. 

The OCC Annual Report also provides detailed information on the post-mortem developments of failed banks. These data provide information on the nominal amount of assets and deposits when a bank's business was suspended and a receiver was appointed. This allows us to estimate the outflow of deposits between the last call report and the day of failure. 
Furthermore, the OCC reports the funds ultimately collected by the receiver throughout the receivership proceedings. This allows us to calculate the recovery rate on assets and deposits. Finally, the OCC classified bank failures by the cause of failure for most failures between 1863 and 1937, except failures in 1932 and 1933. Further details on these data are provided in \Cref{sec:data_receiverships}.


For the period before the founding of the FDIC, we rely entirely on data on national banks. The main reason for focusing on national banks is the availability of consistent records provided by the OCC on both balance sheets and bank failures. However, it is important to highlight that the US banking system featured several other types of financial institutions that were chartered under state rather than federal law. National banks coexisted alongside state banks, trusts, and private banks, with the relative importance of each type of institution varying over time. For example, national banks' market share of the entire banking market ranged from around 80\% in the 1870s to around 45\% in the 1930s (see \Cref{fig:descriptives_nb_shares}). \Cref{sec:historical_context} provides further details on the historical evolution of the US banking system.


\paragraph{Data for modern sample (1959-2024).} For the modern, contemporary banking system we use the Federal Financial Institutions Examination Council (FFIEC) Consolidated Reports of Condition and Income (the modern day ``call report''). These data provide quarterly information on balance sheets  and income statements on a consolidated basis for all commercial banks operating in the United States and regulated by the Federal Reserve System, the FDIC, and the OCC. Note that most existing research based on the call report uses the data starting from 1976 onwards (forms FFIEC 031, FFIEC 041, FFIEC 051). We extend our sample further back to 1959 (forms FFIEC 010 and 011), see \Cref{sec:data_callreports} for more details. These data are digitally available at the Federal Reserve from 1959 through 2024. We also merge additional information on bank charters, such as bank founding dates, using the National Information Center (NIC) tables.


We complement the call report data with the FDIC's list of failing banks. This list documents all failures of FDIC member banks from 1934 through 2024. We define a bank failure as a bank closure that involves either a purchase of the failing bank with an assumption of some or all of its deposits or a liquidating receivership.\footnote{Note that the FDIC's definition of failure is slightly broader. The FDIC defines a bank failure as the closing of a bank by regulators or an instance of open bank assistance. In the former case, the FDIC acts as receiver of the failed bank. In the latter, the FDIC provides financial assistance to prevent failure under a systemic risk exception; the bank would likely have failed without assistance. While we drop the latter, we note that all findings are robust to broadening the failure definition to include open bank assistance.} We obtain the failure dates from the list of failing banks. Further, we obtain deposits and total assets at the time of resolution for failures from the FDIC's Failure Transaction Database. These data allow us to calculate deposit and asset growth immediately before failure for all failures since 1993.\footnote{For failures prior to 1993, the Failure Transaction Database reports total liabilities at failure but not total deposits.} 


The financial statements we use are annual until 1941. After 1959, balance sheets were reported biannually, and then quarterly starting in 1976. Unless otherwise stated, we use annual data for our analysis to ensure comparability across different eras.   When predicting bank failures, we also drop \textit{de novo} banks since the determinants of failure for these banks can be different. We define de novo banks  as banks younger than three years.





\begin{figure}[h!]
\caption{\textbf{Failing Banks, 1863-2024}}
\label{fig:bank_failures_across_time}
\centering

\includegraphics[width=0.9\textwidth]{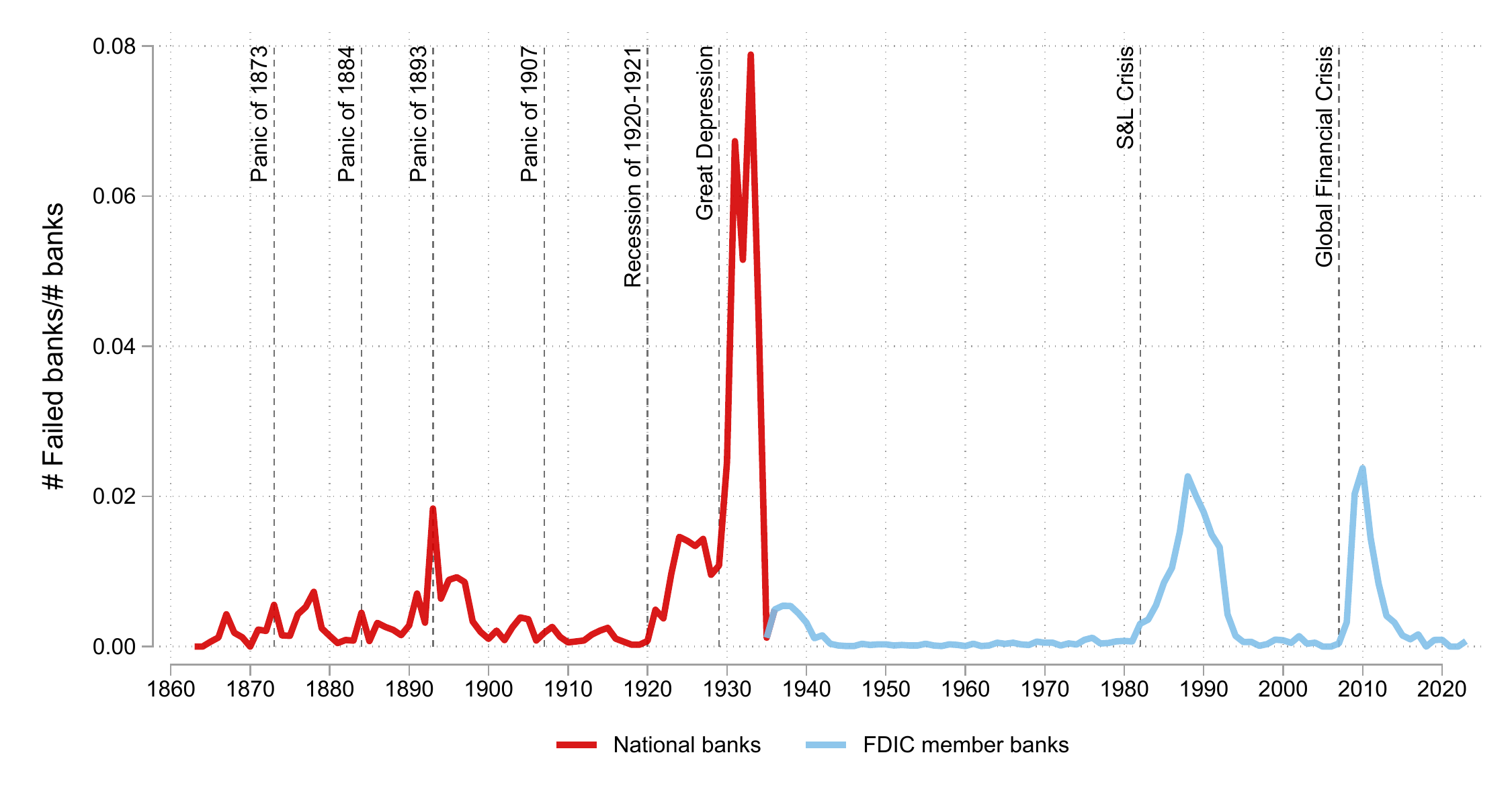}

\begin{minipage}{\textwidth}
\footnotesize
Notes: This figure plots the ratio of bank failures to the total number of banks. Vertical lines indicate selected major banking crises and economic downturns. The red line plots the failure rate for national banks, defined as national banks placed into receivership. \Cref{fig:failures_and_suspensions} in the Appendix shows suspension rates for both national banks and state-chartered institutions. The blue line plots the bank failure rate as classified by the FDIC. We restrict our sample of FDIC member banks to National Member Banks, State Member Banks, and State Nonmember Banks. The sample excludes Savings Associations, Savings Banks, and Savings and Loan Associations (Thrifts).
\end{minipage}

\end{figure}

Altogether, our sample consists of 37,361 unique banks, 14,152 for the historical sample and 23,209 for the modern one.\footnote{Note that we assign different bank identifiers in the OCC data and the Call Report data, thus treating
potentially the same bank as different entities in the historical and modern samples. Mechanically, this increases the total number of unique entities. About 3,700 national banks are counted twice as they operated in both periods.} Of these banks, 5,120 fail at some point throughout the sample, with 2,887 failures before 1935 and 2,233 failures after 1959. \Cref{fig:bank_failures_across_time} plots the rate of bank failures over time. The figure highlights that our sample includes the major financial crises in the history of the U.S., including the Great Depression and the 2008 Global Financial Crisis, as well as many quiet periods when bank failure rates were low. Moreover, our sample covers the period after the founding of the Federal Reserve in 1913 and the founding of the FDIC in 1933, as well as the period before either institution was operative. Hence, it spans an extensive period before the advent of a public lender of last resort, deposit insurance, or other forms of government interventions common in modern banking systems, such as restrictions on leverage or implicit and explicit government guarantees.

\paragraph{Other data.} We obtain consumer price index data from Global Financial Data to deflate variables that we compare across time. Further, we use aggregate outcomes such as real GDP from \citet{JordaSchularickTaylor2017} and \cite{BarroUrsua} and banking crisis dates from \cite{Baron2021}.

%% file: 05_prediciting.tex
\section{Basic Facts: Fundamentals and Bank Failures}

\label{sec:facts}

This section presents basic facts showing that weak fundamentals are strong predictors of bank failures across the entire 1863–2024 sample. 

\subsection{Measures of Bank Fundamentals}

We define bank fundamentals as observable characteristics that proxy a bank's financial health and survival prospects. We consider three bank fundamentals: insolvency risk, asset growth, and noncore funding. Insolvency risk directly proxies a bank's distance to default. Asset growth captures the risk that rapid loan growth can lead banks to overextend themselves and incur future credit losses \citep{Baron2017,Fahlenbrach2018,MullerVerner2023,Meiselman2023}.

While measures of insolvency risk and asset growth directly capture situations where bank losses reduce capitalization, we also use a funding-side measure capturing noncore funding. This measure proxies for increased reliance on costly funding, such as time deposits or non-deposit wholesale funding.\footnote{Rates on noncore deposits such as wholesale deposits, time deposits, brokered deposits, and non-deposit wholesale funding are more sensitive to bank risk, interest rates, and market conditions \citep[see, e.g.,][]{PuriJF,DSS2017}.} The noncore funding measures can both contribute to and be a signal of deteriorating solvency. First, bank equity is in part valuable because banks have access to relatively cheap deposit finance \citep[see, e.g.][]{Egan2021}. Bank equity value is reduced if a bank has to replace cheap demand deposits with non-core funding. Second, banks may use noncore funding to finance rapid and risky asset growth, presaging future losses \citep{ShinNoncore2013}. Third, once a bank starts to realize losses, it may become reliant on noncore funding to fund these losses, so noncore funding is often a signal of losses \citep[see, e.g.,][]{White1983,Calomiris1997, Calomiris2018}. Fourth, the reliance on noncore funding can be the consequence of ongoing liquidity pressures (for instance, due to a slow run) and make the bank more fragile to future funding shocks due to the increased funding fragility---both factors that erode bank equity value \citep{chen2024liquidity}.

Before proceeding, we emphasize that the measures of fundamentals are endogenous and interrelated. In the context of assessing the predictability of bank failures, we do not take a stand on the causal relation between these measures and bank failures. For instance, a bank may appear to be in bad financial condition according to its solvency metric because it is relying on expensive forms of funding. At the same time, a bank may be only able to raise expensive forms of secured wholesale funding because it is approaching insolvency. Thus, these measures should be viewed as capturing observable signals in bank financial statements that indicate a banking business is more likely to be unproductive or potentially unviable.

The exact variables we use to measure insolvency risk and reliance on noncore funding differ across samples due to differences in data availability. For the pre-1934 sample, we measure insolvency risk by surplus profits over total equity. This measure is a proxy for bank income and capitalization.\footnote{In the OCC call reports for national banks, bank equity capital is divided into initial paid-in capital, undivided profits, and the surplus fund. Paid-in capital was fixed after the founding of a bank. Undivided profits reflect accumulated net earnings that have not yet been allocated to be paid out as dividends or transferred to the surplus fund. The surplus fund is retained earnings set aside from profits to increase bank capital. ``Surplus profit'' is the sum of undivided profits and the surplus fund. Thus, surplus profit is the portion of equity that varies with retained earnings and realized losses.} For the same period, we measure noncore funding by total assets net of total deposits (including unsecured interbank deposits), equity, and national bank notes. This captures expensive, non-deposit wholesale funding.\footnote{We verify that noncore funding effectively comprises the line items ``Bills Payable'' and ``Rediscounts'' in years in which the latter are reported. These line items represent forms of short-term, expensive, and secured wholesale funding---including borrowings from the Federal Reserve after 1914 \citep[see, e.g.,][]{carlson2025young}. Note that from 1905 through 1920,  unsecured interbank funding and other liabilities are reported in the same line item. Hence, our noncore measure may under-report the use of secured wholesale funding for these years.} For 1959-2024, insolvency risk is measured by net income to assets, and noncore funding is measured by the sum of time deposits and wholesale funding relative to total assets.\footnote{For the historical sample, we do not include time deposits in noncore funding, as this item is only reported separately during 1915-1928.}

\subsection{Insolvency Risk and Bank Failures}

\begin{figure}[h!]
\centering
\caption{\textbf{Insolvency, Noncore Funding, and Future Probability of Bank Failure} \label{fig:cond_prob} }
\subfloat[Insolvency, 1863-1934]{\includegraphics[width=0.49\textwidth]{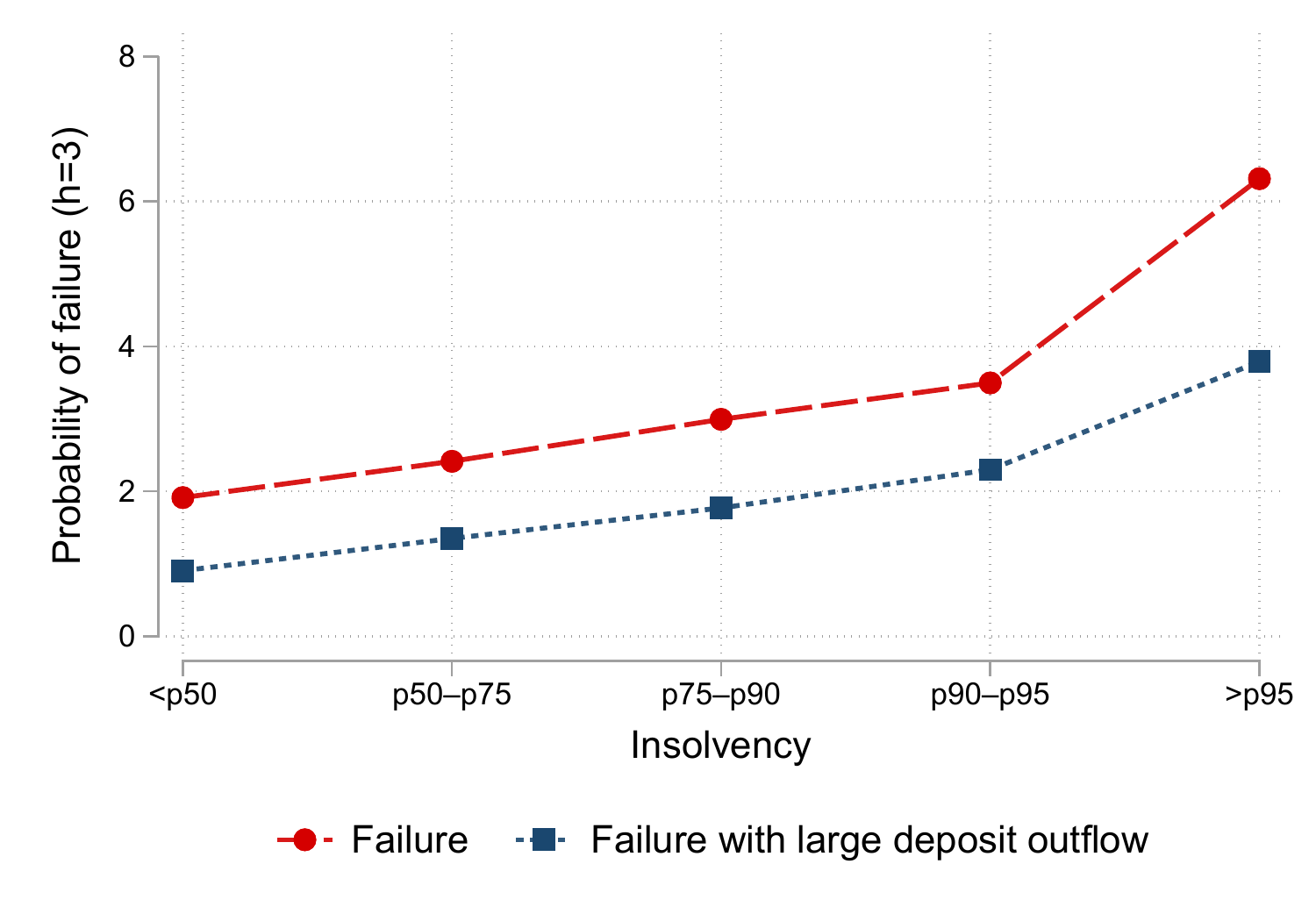}}
\hfill
\subfloat[Insolvency, 1959-2024]{\includegraphics[width=0.49\textwidth]{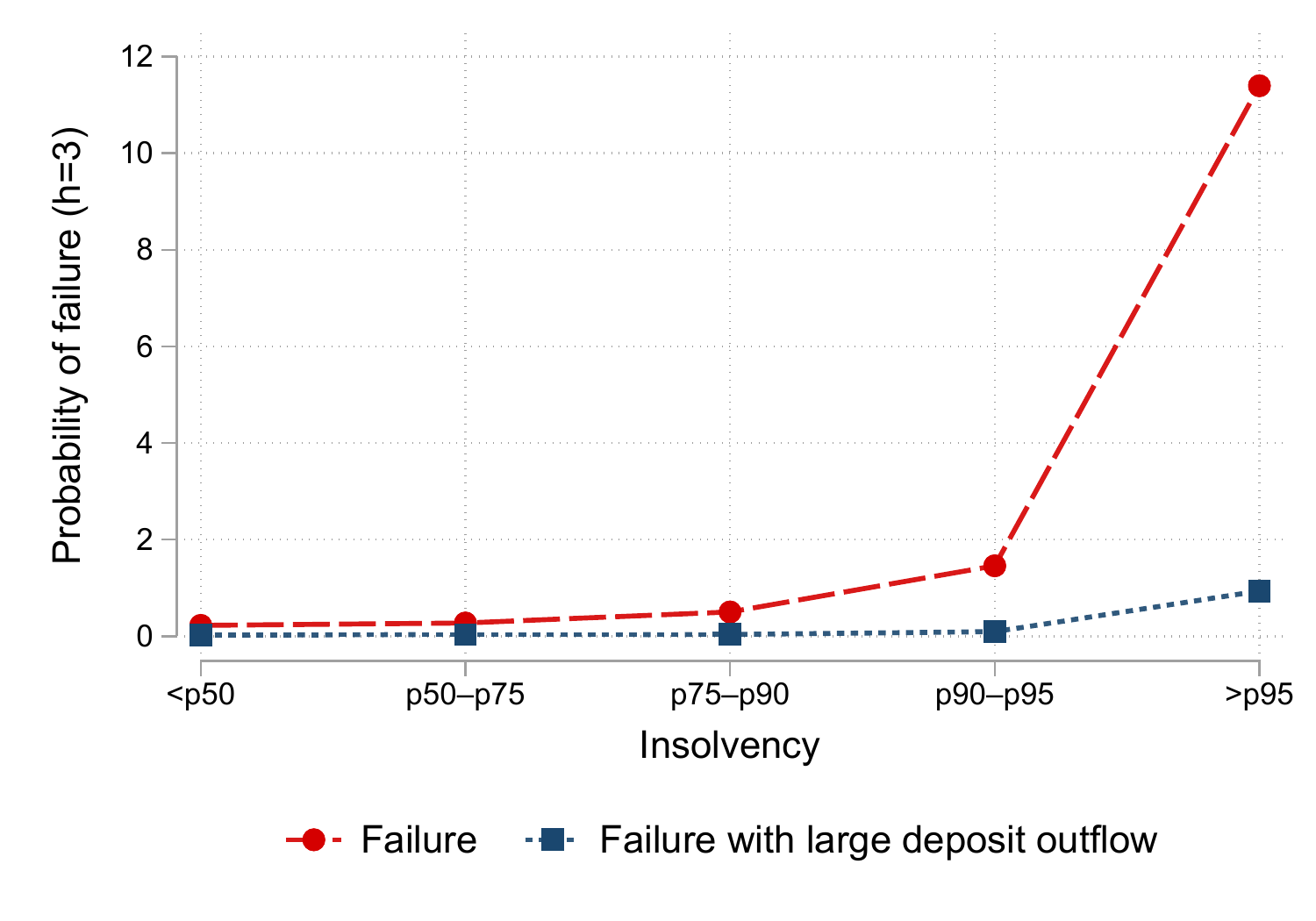}}

\subfloat[Noncore Funding, 1863-1934]{\includegraphics[width=0.49\textwidth]{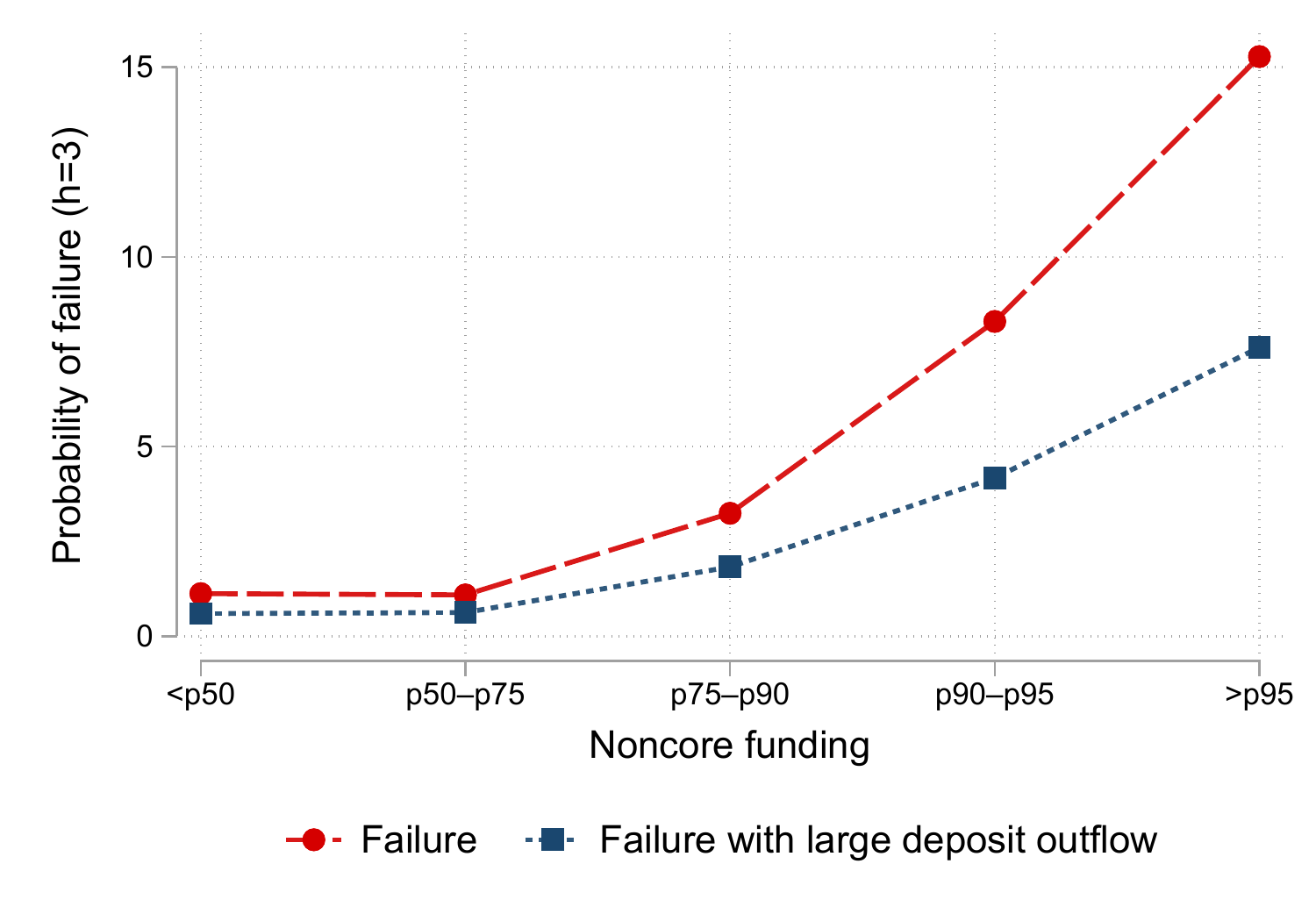}}
\hfill
\subfloat[Noncore Funding, 1959-2024 {\color{red}} ]{\includegraphics[width=0.49\textwidth]{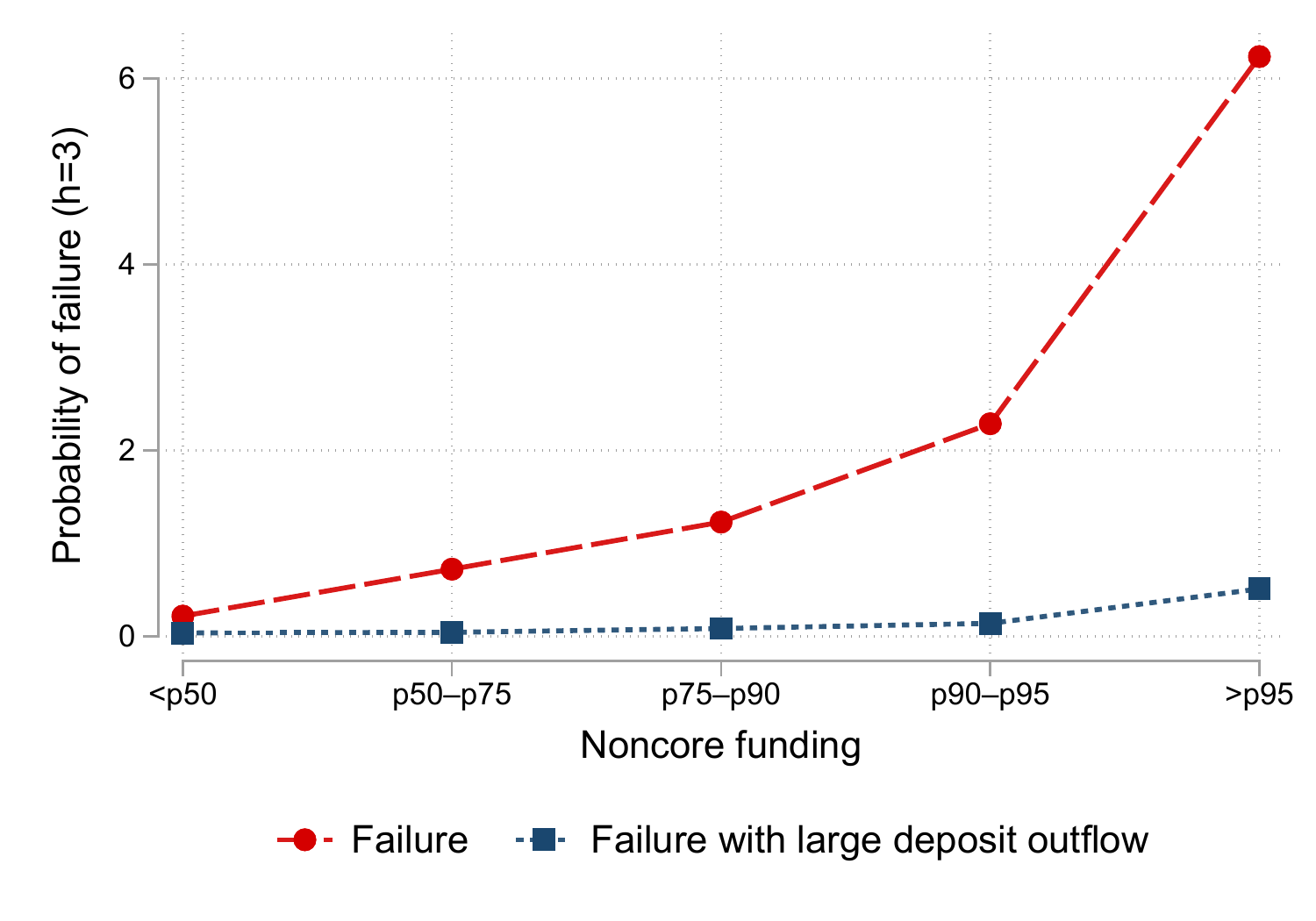}}

\begin{minipage}{\textwidth}
\footnotesize 
Notes: This figure plots the probability of bank failure within the next three years against the distribution of proxies for insolvency and noncore funding in year $t$. For the pre-FDIC era (1863-1934), insolvency is measured by surplus profit relative to equity.  This measure is a proxy for bank income and capitalization. Noncore funding is measured by total assets net of total deposits (including unsecured interbank deposits), equity, and National Bank notes, all scaled by assets. For the Modern Era (1959-2024), solvency is measured by net income-to-assets, and noncore funding is measured by time deposits and wholesale funding (``Other Borrowed Money'') to total assets. Failures with large deposit outflows are defined as those where total deposits fall by more than 7.5\% between the last call report and failure. In panels (a) and (c), failures with large deposit outflows are based on the 1880-1934 sample, as the OCC only reports deposits at the time of failure starting in 1880. In panels (b) and (d),  failures with large deposit outflows are based on the 1993-2024 sample, as the FDIC only reports deposits at the time of failure starting in 1993.
\end{minipage}
\end{figure}

\Cref{fig:cond_prob} plots the probability of failure within three years conditional on a bank's fundamentals in year $t$. Panels (a) and (b) use insolvency risk as the measure of fundamentals for the historical pre-FDIC era (1863-1934) and the modern era (1959-2024). The probability of failure within the next three years rises with exposure to insolvency risk. The relation is generally nonlinear, with the risk of failure rising rapidly in the right tail. Moving from below the 50\textsuperscript{th} percentile to above the 95\textsuperscript{th} percentile in the measure of insolvency implies an increase in the probability of failure of 4pp in the historical sample (panel a) and 10pp in the modern sample (panel b).

\begin{figure}[htpb]
\caption{\textbf{Losses and Solvency Dynamics in Failing Banks: 1863-2024} \label{fig:losses}}
\centering

\subfloat[1863-1934 \label{fig:losses_pre} ]{
\includegraphics[width=0.65\textwidth]{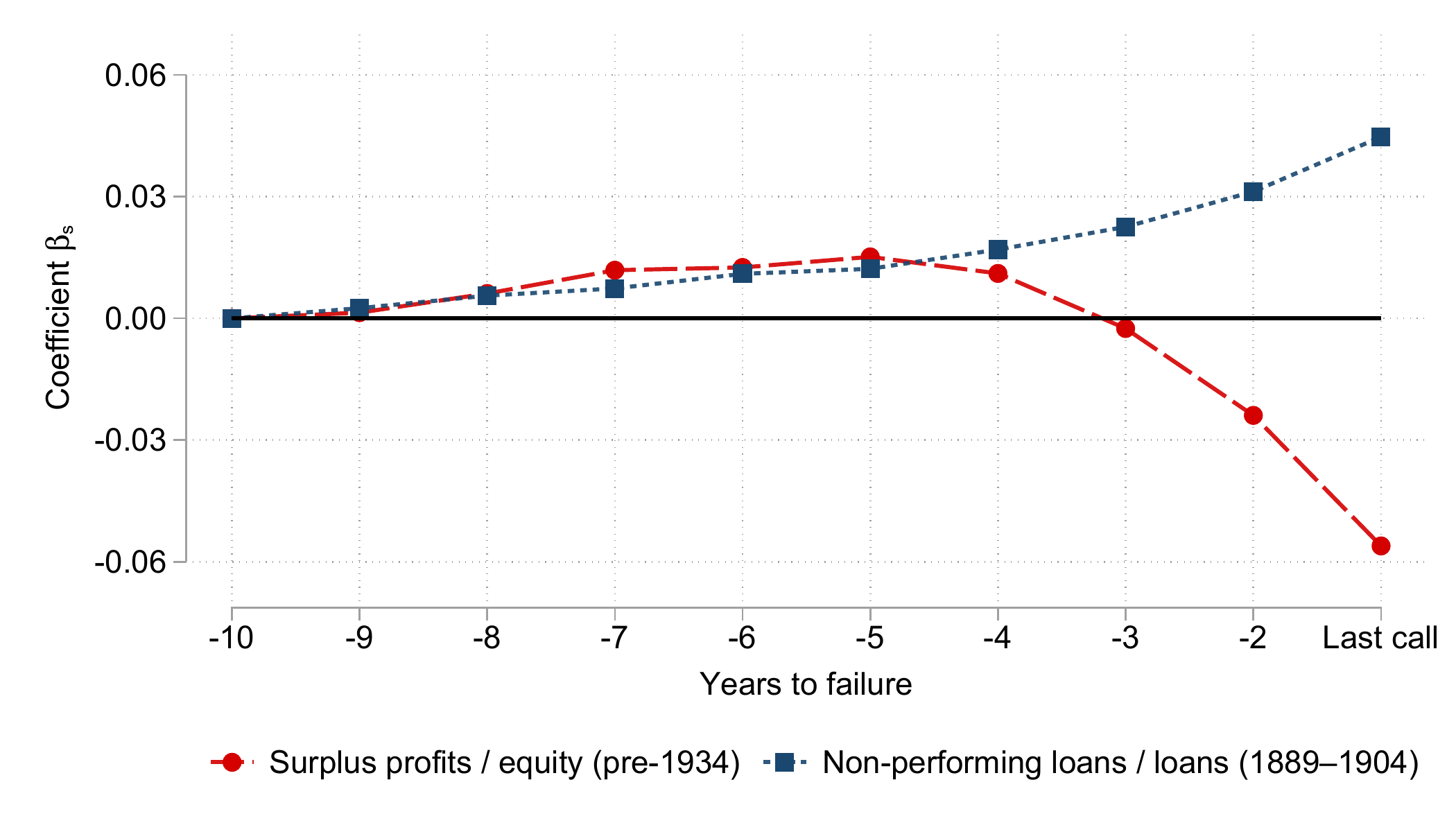}}

\subfloat[1959-2024 \label{fig:losses_post}]{
\includegraphics[width=0.65\textwidth]{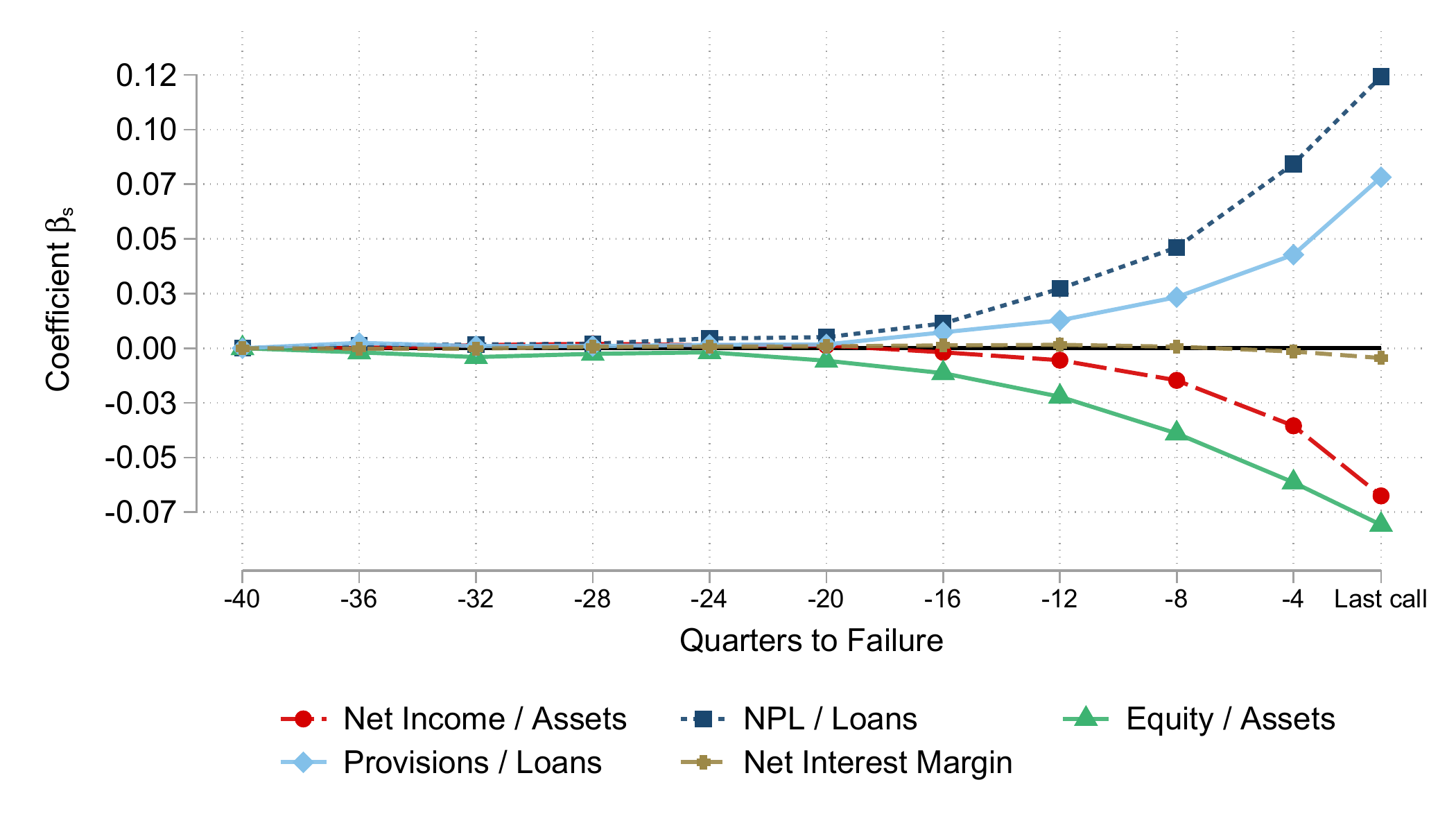}}

\begin{minipage}{\textwidth}
\footnotesize
Notes: The figure presents the sequence of coefficients from estimating \Cref{eq:dynamic},
where the dependent variable is the ratio indicated in the figure legend. The specification includes a set of bank fixed effects. The sample is restricted to failing banks and the ten years before they fail. The figure reports the dynamics in failing banks relative to ten years before failure (the omitted period). In panel (a), the sample is restricted to banks that failed from 1863 through 1934. In panel (b), the sample is restricted to banks that failed from 1959 through 2024.  In panel (a), surplus profit to equity is the sum of the surplus fund and undivided profits relative to total equity capital (paid-in capital, undivided profits, and surplus fund). Non-performing loans is proxied by the line item ``Other real estate owned (OREO)'' and is available for the 1889-1904 subsample. In panel (b), the net interest margin (NIM) is defined as total interest income net of interest expenses normalized by total assets.   
\end{minipage}
\end{figure}

Insolvency risk is strongly related to bank failures in the cross section. How does insolvency risk evolve \textit{within} failing banks? To understand the dynamics of losses and insolvency risk in failing banks before their failure, we estimate variants of the following specification:
\begin{align}
y_{b,t} = \alpha_b + \sum_{j=-10}^{0} \beta_j \times \mathbf{1}_{j=t} +\epsilon_{b,t}, \label{eq:dynamic} 
\end{align} 
where $y_{b,t}$ is a bank-level outcome, $j$ measures the number of years to failure, and $\alpha_b$ is a bank fixed effect. We restrict the sample to failing banks that are within 10 years of failure. We set the benchmark period to be $j=-10$, so all estimates are relative to ten years before failure.  The sequence of coefficients $\{ \beta_j\}$ captures the dynamics of variable $y_{b,t}$ in the ten years before failure. 

\Cref{fig:losses} presents the dynamics in indicators of loan losses and solvency in failing banks from estimation of Equation \eqref{eq:dynamic}. Panel (a) presents the results for the 1863-1934 sample. Historically, provisioning rules were less strict, so equity was not immediately impacted when loans became non-performing. Nevertheless, failing banks see a gradual deterioration in surplus profits relative to equity, indicating negative profitability and declining capitalization. Moreover, failing banks see a gradual 5 percentage point rise in ``Other Real Estate Owned,'' a proxy for non-performing loans, indicating that the decline in capitalization is at least partly driven by realized credit risk.\footnote{``Other Real Estate Owned'' (OREO) typically refers to real estate property assets that a bank holds but that are not part of its business. Often, these assets are acquired due to foreclosure proceedings as seized collateral. Below, in \Cref{tab:pred_recovery_rate}, we document that OREO as a share of loans in failing banks immediately before failure is significantly negatively correlated with asset recovery rates in failure. This measure is available for the 1889-1904 subsample.}

Panel (b)  of \Cref{fig:losses} plots the dynamics of solvency in failing banks for the 1959-2024 sample. In the five years before failure, failing banks see a 10-percentage-point rise in NPLs. This rise in NPLs translates into rising loan loss provisions, which results in a decline in realized net income. Thus, credit risk also plays an important role in the erosion of bank profitability in the modern sample. As a result, the equity-to-assets ratio declines considerably in the run-up to failure, falling by 8 percentage points. Altogether, throughout the sample from 1863-2024, failing banks see gradually rising losses and deteriorating solvency before failure.

\subsection{Noncore Funding and Bank Failures}

Panels (c) and (d) in \Cref{fig:cond_prob} plot the relation between noncore funding in year $t$ and the probability of failure over the next three years. Reliance on noncore funding is strongly related to the future likelihood of failure. Noncore funding is an especially strong signal of failure in the historical sample. For example, moving from below the 50\textsuperscript{th} percentile to above the 95\textsuperscript{th} percentile in noncore funding is associated with 
an increase in the probability of failure of 14pp in the historical sample and 5.5pp in the modern era.

\begin{figure}
\caption{\textbf{Funding Dynamics in Failing Banks: 1863-2024}}
\label{fig:funding}
\centering

\subfloat[1863-1934 \label{fig:funding_a}]{\includegraphics[width=0.6\textwidth]
{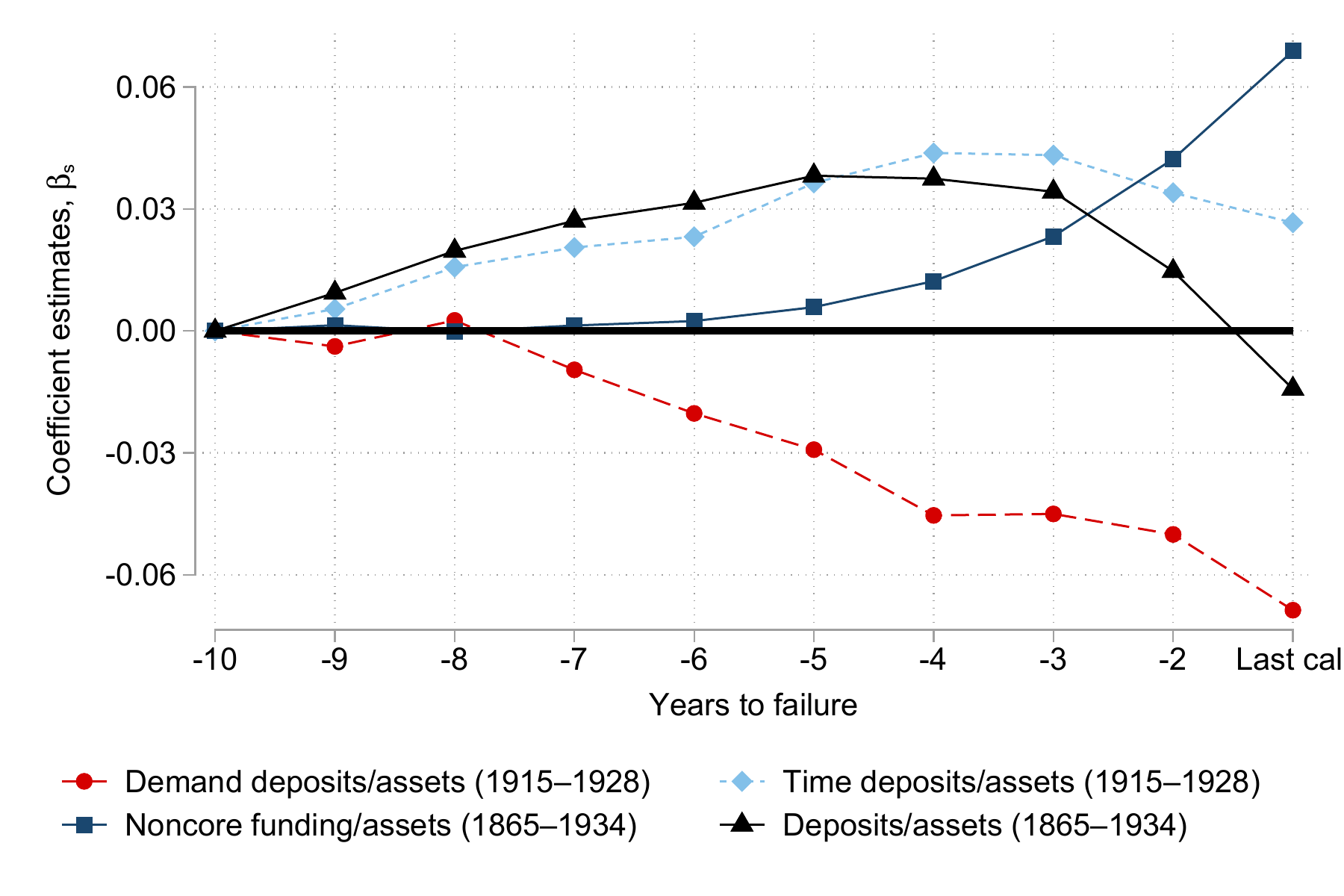}}

\subfloat[1959-2024  \label{fig:funding_c}]{\includegraphics[width=0.6\textwidth]{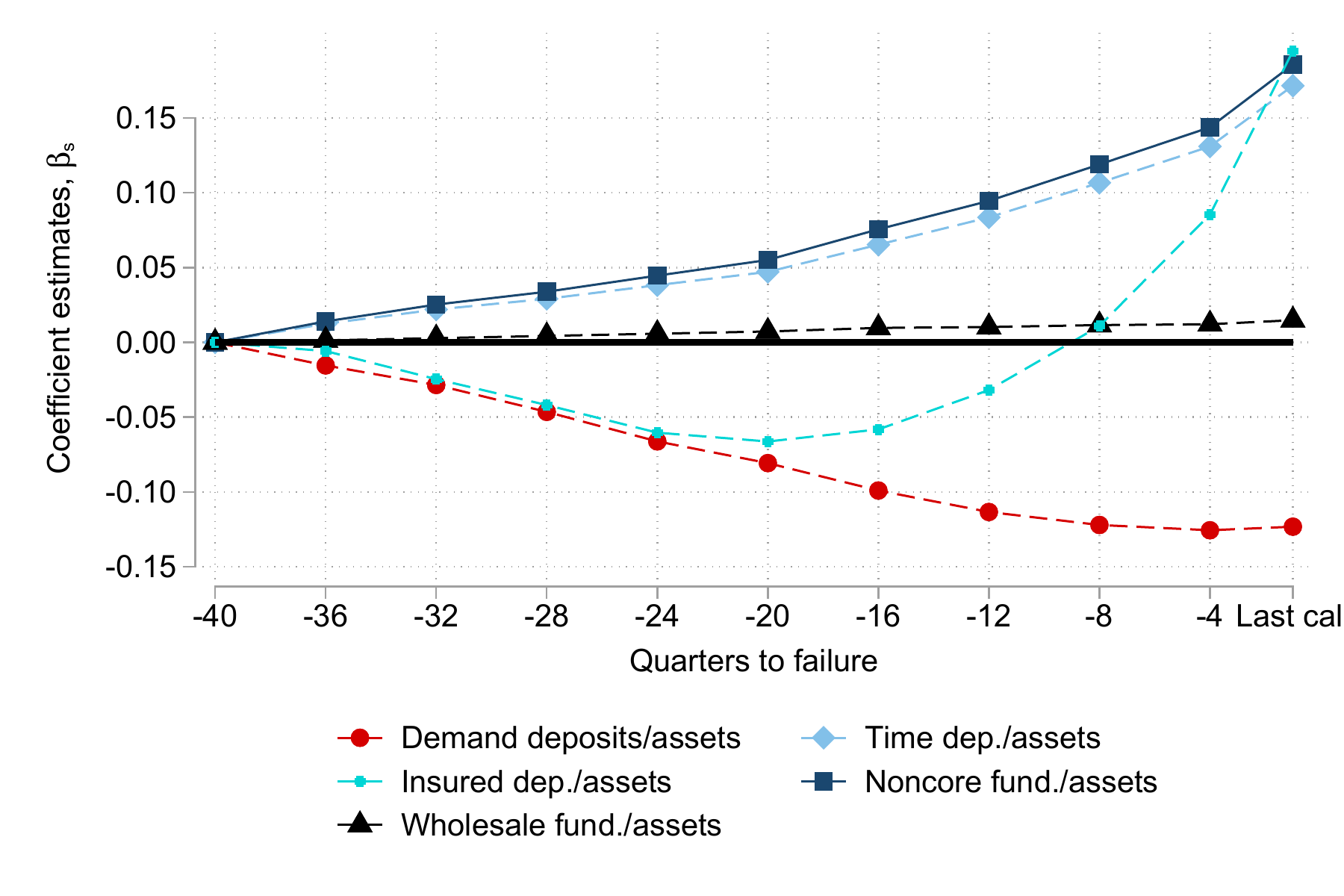}}

\begin{minipage}{\textwidth}
\footnotesize 
Notes: This figure shows the sequence of coefficients from estimating \Cref{eq:dynamic} for various funding ratios. The sample is restricted to failing banks and the ten years before they fail. In panel (a), the sample is restricted to banks that failed from 1863 through 1934.   In panel (b), the sample is restricted to banks that failed from 1959 through 2024. Further, in panel (b) we estimate the model using quarterly data but only plot the end of calendar year coefficients.  In panel (a), noncore funding is measured by total assets net of the sum total deposits, equity, and National Bank Notes, all scaled by assets. Noncore funding effectively proxies for the ``Bills Payable'' and ``Rediscounts'' line items. Total deposits are available for the entire 1863-1934 sample. Time and demand deposits are reported separately for the 1915-1928 subsample. In panel (b), noncore funding is defined as the sum of time deposits and wholesale funding. Wholesale funding is the amount reported in the call report line item ``Other Borrowed Money'' which pools various sources of bank wholesale funding, such as advances from Federal Home Loan Banks (FHLBs), other types of wholesale borrowings in the private market, and credit extended by the Federal Reserve. 

\end{minipage}
\end{figure}

To understand the evolution of funding in failing banks, \Cref{fig:funding} presents estimates of \eqref{eq:dynamic} for various funding ratios. Panel (a) presents results for the historical 1863-1934 sample. Failing banks see a gradual rise in noncore funding relative to assets. For the subsample where we observe demand and time deposits separately (1915-1928), we see there is also an increase in time deposits, while demand deposits decline as a share of assets. Demand deposits, unlike time deposits, tend to be held by less price-sensitive retail investors and are a cheaper source of financing. Thus, failing banks increasingly rely on expensive forms of funding. In the two years before failure, as failing banks start to see rising losses, deposit funding as a share of total assets starts to decline and is replaced nearly one-for-one by more expensive noncore funding, likely reducing bank profitability.  This finding is consistent with previous work showing that banks that experienced difficulties were often forced to rely on this more expensive type of funding \citep[see, e.g.,][]{White1983,Calomiris1997,Calomiris2018}.

Panel (b) in \Cref{fig:funding} presents funding dynamics in failing banks for the modern sample. Similar to the historical sample, failing banks in the modern sample increasingly rely on noncore funding, which rises by 18\% of assets in the decade before failure. Time deposits account for the largest increase. Wholesale funding also rises, though the increase is small relative to assets. 
In contrast, demand deposits decline as a share of assets in the decade before failure. As a result, failing banks see a gradual rise in interest expenses before failure (see \Cref{fig:II_IE_NIM}).

While there is an increased reliance on noncore funding, in the modern sample, insured deposits simultaneously flow \textit{into} failing banks in the five years before failure. This suggests that insured depositors do not discipline failing banks, potentially delaying failure.\footnote{These patterns are consistent with \cite{PuriJF}, who find that failing banks increasingly substitute toward expensive deposit funding but also see an inflow of insured deposits before failure. The use of noncore funding to finance rapid growth and subsequent losses is consistent with \cite{ShinNoncore2013}. Rapid growth financed by brokered deposits before failure is also a feature emphasized in previous research surveyed by \cite{FDIC2011}.} As we discuss further below, deposit dynamics differ substantially in failing banks before and after the introduction of federal deposit insurance. Despite this important difference, an important commonality is that failing banks increasingly rely on noncore funding throughout the sample.

\subsection{Interaction of Insolvency and Noncore Funding} 

Are banks more likely to fail when they have both weak solvency and are reliant on noncore funding? A bank that has weak solvency proxies \textit{and} relies on costlier and more risk-sensitive financing may see a hastier demise, as creditors raise the cost of financing or withdraw financing more quickly as losses mount. Moreover, as discussed above, noncore funding can proxy for exposure to insolvency risk, so the combination of the two measures could provide a stronger signal of impending bank failure.

\begin{figure}[ht!]
\centering
\caption{\textbf{Interaction of Insolvency and Noncore Funding for Predicting Future Bank Failures } \label{fig:interaction}}

\subfloat[1863-1934]{\includegraphics[width=0.49\textwidth]{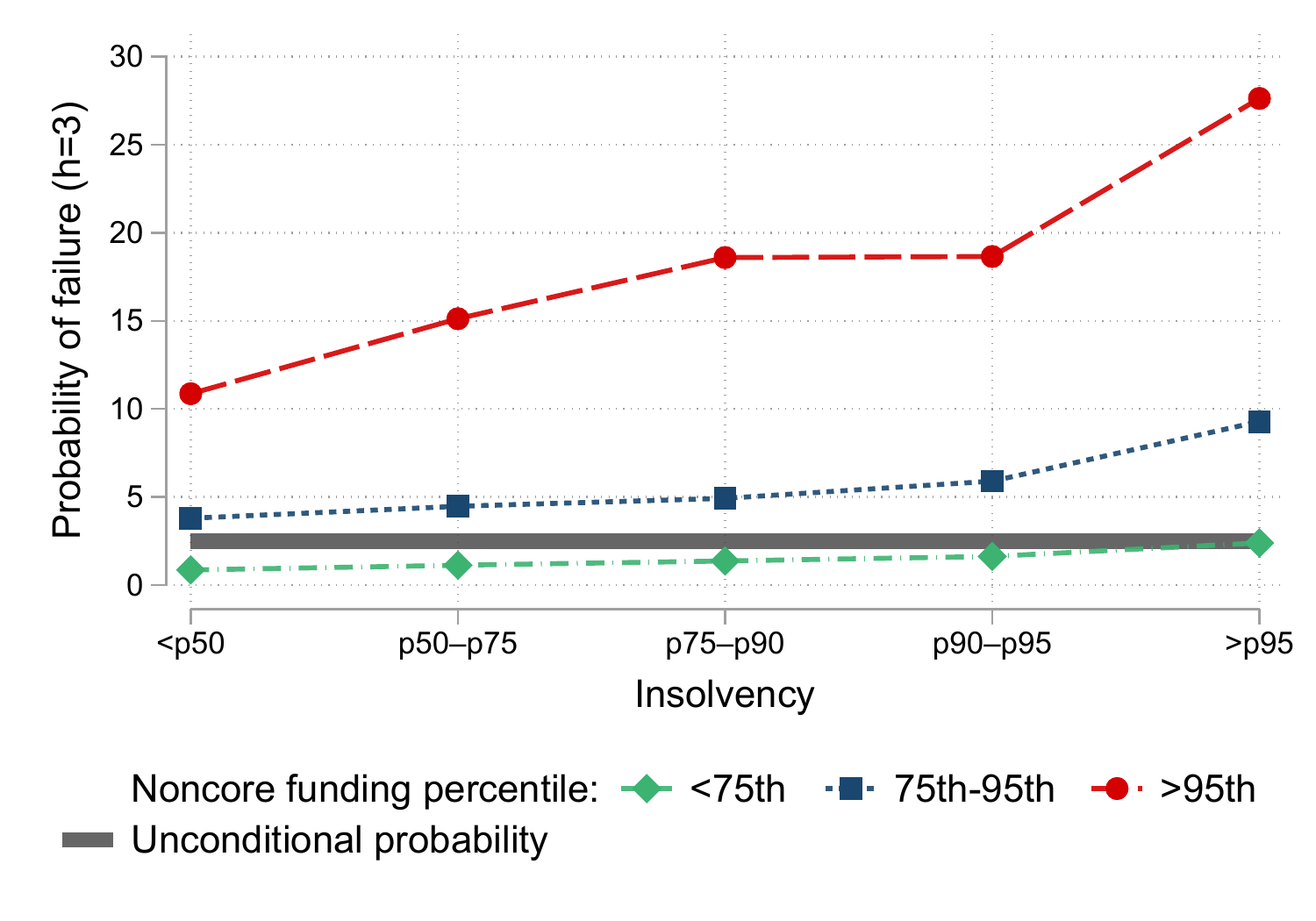}} 
\subfloat[1959-2024]{\includegraphics[width=0.49\textwidth]{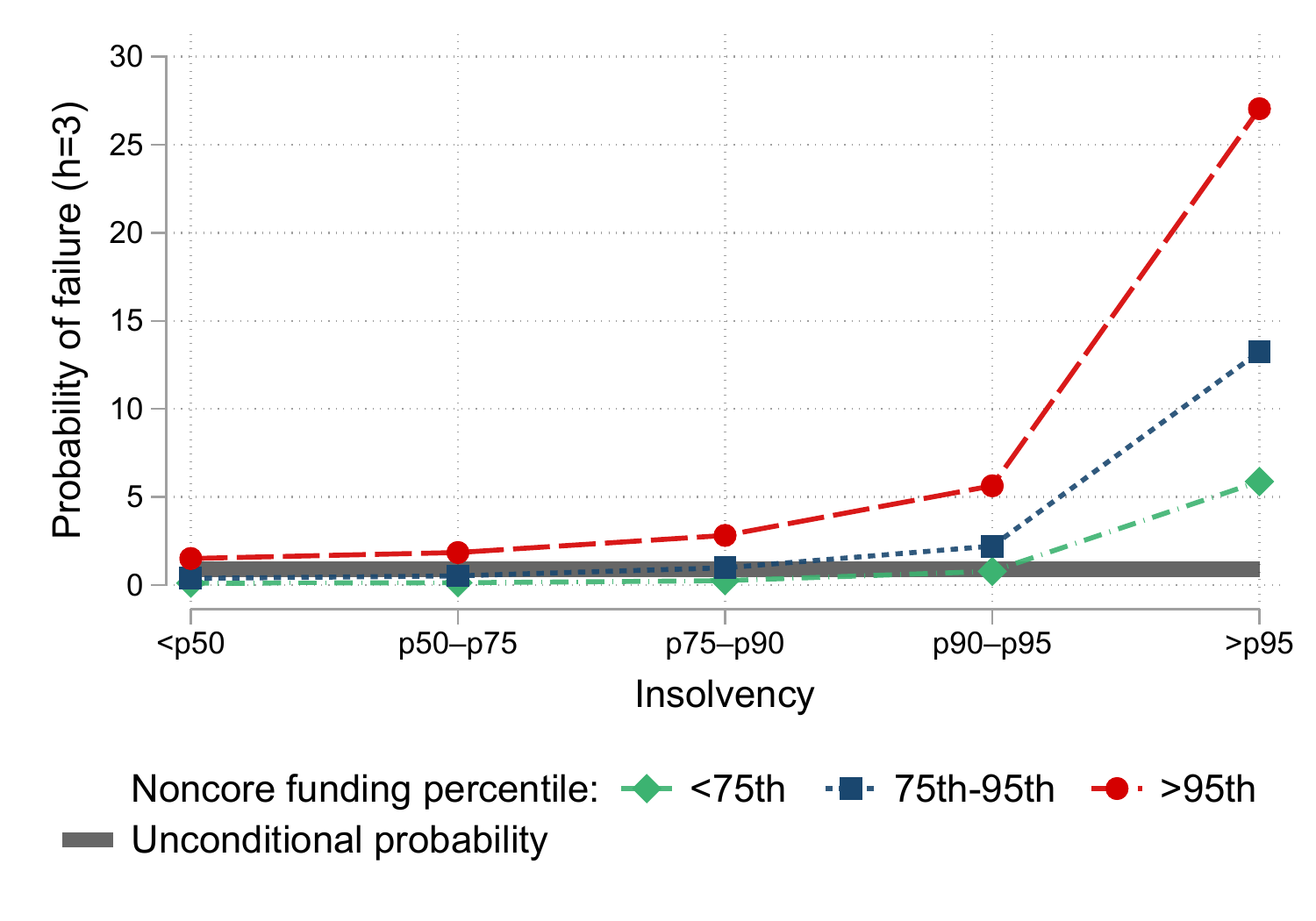}}

\begin{minipage}{\textwidth}
\footnotesize
Notes: This figure plots the probability of bank failure within the next three years against the joint distribution of proxies for insolvency and noncore funding in year $t$. For the pre-FDIC period (1863-1934), insolvency is measured by surplus profits over equity. Noncore funding is measured by total assets net of total deposits, equity, and National Bank Notes, all scaled by assets. For the Modern Era (1959-2024), insolvency is measured by net income-to-assets, and noncore funding is measured by the sum of time deposits and wholesale funding (``Other Borrowed Money'') to total assets.
\end{minipage}
\end{figure}

\Cref{fig:interaction} depicts the probability of bank failure over the next three years across the distribution of insolvency by whether funding vulnerability is below the 75\textsuperscript{th} percentile, between the 75\textsuperscript{th} and 95\textsuperscript{th}, or above the 95\textsuperscript{th} percentile. Fundamentals are again measured in year $t$. The figure confirms that banks with both high insolvency risk \textit{and} high funding vulnerability are the most likely to fail. The probability of failure for a bank that is in the top 5\textsuperscript{th} percentile of both insolvency and high funding vulnerability is 27\% in the historical sample and 27\% in the modern sample. These are large numbers, considering that the unconditional probability of failure over three years is 2.5\% in the historical sample and 1\% in the modern sample. Therefore, a bank with high insolvency risk and high funding vulnerability has a 10-20 times larger probability of failure than a randomly drawn bank. Overall, this illustrates that fundamental measures of insolvency and expensive noncore funding strongly predict future failure.

\subsection{Asset Boom and Bust and Bank Failures}

Why do banks experience gradually rising losses that eventually leads to heightened risk of failure? One reason is that rapid loan growth, potentially driven by overoptimistic expectations about default risk, is systematically associated with future credit losses \citep{Bordalo2018,Greenwood2023}.

\begin{figure}[htpb]
\caption{\textbf{Asset Dynamics in Failing Banks: 1863-2024}}
\label{fig:assets_failing_banks}
\centering

  \includegraphics[width=0.55\textwidth]{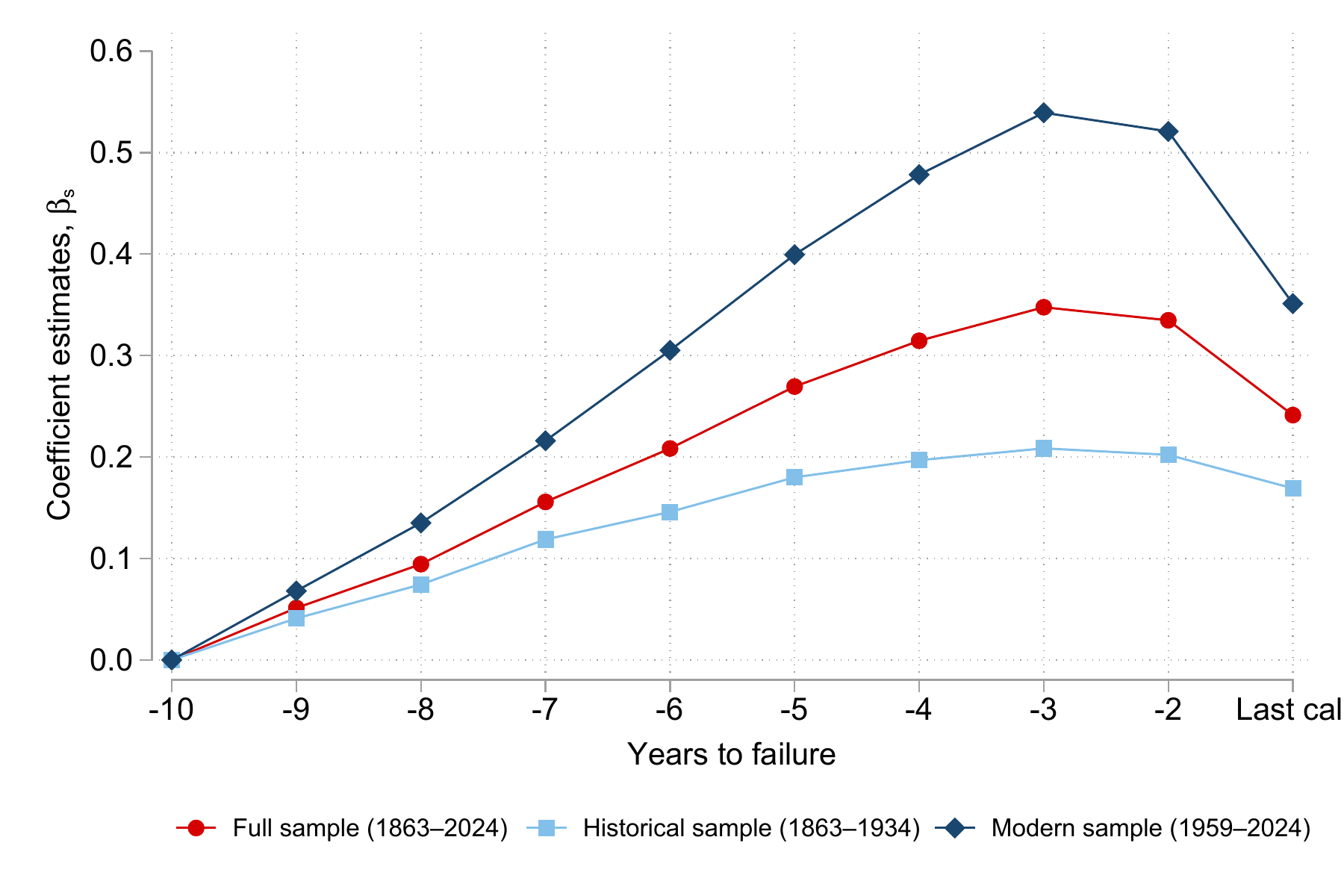}

   \begin{minipage}{\textwidth}
\footnotesize
Notes: This figure reports the sequence of coefficients from estimating \Cref{eq:dynamic} with log total assets (deflated by the CPI) as the dependent variable. The regression includes a set of bank fixed effects. The sample is restricted to failing banks and to the ten years before they fail. The sub-samples indicated in the figure legend are selected based on the years in which a bank failed.
\end{minipage}
\end{figure}

\Cref{fig:assets_failing_banks} presents the dynamics of asset growth in failing banks from estimating \Cref{eq:dynamic} with the log of real total assets as the dependent variable. The figure reveals that total assets in failing banks follows a boom-and-bust pattern in the decade before failure.\footnote{Appendix \Cref{fig:nonmonotonic} shows that the relation between asset growth and future failure is non-monotonic across quintiles of the asset growth distribution. Banks with rapid asset growth over the past three years ($t-3$ to $t$) indeed have a heightened risk of failure over the next three years. Interestingly, banks in the lowest quintile of asset growth are also more likely to fail, likely because low growth signals a failing business.} In the full sample, assets expand by 34\% in real terms from ten years to three years before failure and then contract over the last two years before failure. The boom and bust pattern is present in both the pre-FDIC sample (1863-1934) and the modern sample (1959-2024), though it is significantly more pronounced in the modern period. The boom in assets is driven mainly by loans, rather than liquid assets (see \Cref{fig:illiquid_liquid}). In the modern sample where we can observe loan types, we find that the strongest growth is in real estate lending, followed by C\&I lending (see \Cref{fig:loans}).

\par

\section{The Predictability of Bank Failures}
\label{sec:predicting_failures}

Failing banks experience deteriorating solvency, increasing reliance on noncore funding, and an asset boom-bust in the decade before failure. These patterns are robust across different institutional settings and extend to the pre-FDIC period. This section shows that these systematic patterns imply substantial predictability of bank failures. 

As discussed in \Cref{sec:conceptual}, understanding the extent of predictability of bank failures is informative for two reasons. First,  the degree of predictability of bank failures is informative about the importance of weak fundamentals in understanding bank failures. Second, the extent of predictability of bank failures in the historical sample can further help understand whether depositors are rational and forward-looking, acting quickly on signals of bank weakness, or whether frictions, such as inattention, slow down depositor reactions.

\subsection{Empirical Specification}

We conduct a formal prediction exercise to quantify the extent to which fundamentals can predict future failures, both in- and out-of-sample. Motivated by the basic facts in the previous section, we estimate simple predictive regression models of the following form:
    \begin{align}
    \text{Failure}_{b,t+1\to t+h} = \alpha &+ \beta_1 \times \text{Insolvency}_{bt} \label{eq:pred} \\ &+ \beta_2 \times \text{Noncore Funding}_{bt} \nonumber \\ &+ \beta_3 \times \text{Insolvency}_{bt} \times \text{Noncore Funding}_{bt}  \nonumber \\ &+ \beta_4 \times \text{Asset Growth}_{bt} \nonumber  \\ &+ \beta_5 \times \text{Aggregate Conditions}_t + \epsilon_{b,t+1\to t+h},  \nonumber 
    \end{align}
where $\text{Failure}_{b,t+1\to t+h}$ is an indicator variable that equals one if bank $b$ fails within $h$ years of the call report measured at time $t$. We assess predictability using both linear probability models and logit models to test the robustness of our results to the choice of functional form. 

To predict bank failures, we use the same measures of insolvency and noncore funding as in \Cref{sec:facts}. We also consider the interaction between the insolvency and noncore funding measures, given the evidence in \Cref{fig:interaction}. $\text{Asset Growth}_{bt}$ is a set of variables that capture bank-specific growth. We use five quintiles of change in log bank assets from year $t-3$ to $t$. This allows us for a non-linear relation between past growth and failure. Furthermore, for $\text{Aggregate Conditions}_t$, we include aggregate real GDP growth and inflation over the same three-year period. These latter two measures are available in the same form throughout the entire 1863-2024 sample. Finally, we control for bank age, given existing evidence that younger banks are more likely to fail \citep{White1984,Wheelock2000}. Note that we do not include bank or time fixed effects in the prediction; we only use real-time observables.

To quantify the power of these observables for predicting bank failure, we construct the receiver operating characteristic curve (ROC), a standard tool used to evaluate binary classification ability. The ROC curve traces out the true positive rate against the false positive rate as we vary the classification threshold. We then calculate the area under the ROC curve (AUC). An uninformative predictor has an AUC of 0.5, while an informative predictor has an AUC of greater than 0.5. The AUC metric is commonly used in the literature on predicting financial crises.\footnote{For reference, the in-sample AUC for predicting financial crises in aggregate data based on credit and asset price growth is typically in the range 0.65-0.75 \citep[e.g.,][]{Schularick2012,DrehmannJuselius2014,Baron2021,Greenwood2022,MullerVerner2023}. Similarly, \citet{Iyer2024} find an AUC of 0.73 when predicting local recessions with bank funding conditions.} Furthermore, we test both in-sample and pseudo-out-of-sample classification performance. The pseudo-out-of-sample AUC is constructed by estimating \Cref{eq:pred} iteratively on an expanding sample and predicting the probability of failure for each bank in $t+1\to t+h$ using only data up to year $t$. For pseudo-out-of-sample prediction exercises, we use the first 10 years of data as a training sample.

\subsection{Main Predictability Results}

\Cref{tab:AUC} presents the in-sample and out-of-sample AUC statistics based on estimates of variants of \Cref{eq:pred} for the historical pre-FDIC sample (1863-1934) and the modern sample (1959-2024). We present results for predicting failure at the 1, 3, and 5-year horizons. The corresponding regression estimates for each sample period are reported in Appendix Tables \ref{tab:predicting_failure_historical} and \ref{tab:predicting_failure_modern_era}.

\begin{table}[htpb]
   \caption{\textbf{AUC Metric for Predicting Bank Failures with Fundamentals}  \label{tab:AUC} }
        \begin{minipage}{1\textwidth}
        \begin{center}
        \footnotesize
        \begin{tabular}{lccccccc}
        \toprule

         \cmidrule(lr){2-7}
         Prediction horizon $h$  & \multicolumn{5}{c}{1 year}  &\multicolumn{1}{c}{3 years}  &\multicolumn{1}{c}{5 years} \\             \cmidrule(lr){2-6}    \cmidrule(lr){7-7} \cmidrule(lr){8-8} \\

         \midrule
            & (1) & (2) & (3) & (4) & (5) & (6) & (7) \\ \midrule
              \multicolumn{8}{c}{\textbf{Panel A: Historical Sample (1863-1934)}} \\ \midrule 
       \input{./output/tables/05_tab_auc_historical} \\ \midrule \input{./output/tables/05_tab_auc_oos_historical} \\ \midrule
         
                   \multicolumn{8}{c}{\textbf{Panel B: Modern Sample (1959-2024)}} \\ \midrule       
             \input{./output/tables/05_tab_auc_modern_era} \\ \midrule
         \input{./output/tables/05_tab_auc_oos_modern_era} \\ \midrule
                  \multicolumn{8}{c}{\textbf{Specification details}} \\ \midrule
         
Insolvency & \checkmark & & \checkmark  & \checkmark  & \checkmark  
   & \checkmark  & \checkmark  \\
Noncore funding & & \checkmark & \checkmark  & \checkmark & \checkmark  & \checkmark  & \checkmark   \\
Insolvency $\times$ Noncore funding  & & & \checkmark  & \checkmark  & \checkmark  & \checkmark  & \checkmark    \\
Growth \& Aggregate conditions & & & & \checkmark   & \checkmark  & \checkmark    & \checkmark  \\
Deposit outflow before failure   &  & & & & >7.5\%  & & \\
Age controls & \checkmark  & \checkmark  & \checkmark  & \checkmark  & \checkmark  & \checkmark  & \checkmark   \\
        \bottomrule
        \end{tabular}
        \end{center}
        {\footnotesize Notes: This table reports the area under the receiver operating characteristic curve (AUC) across different specifications, samples, and horizons using in-sample and pseudo-out-of-sample classification. The corresponding regression coefficients underlying the models are reported in \Cref{tab:predicting_failure_historical} (for Panel A) and \Cref{tab:predicting_failure_modern_era} (for Panel B). Pseudo-out-of-sample AUCs are obtained by estimating the regression model with training data from 1863-1873 (Panel A), and 1959-1969 (Panel B), and iteratively expanding the sample for subsequent years.  We drop banks without information on deposit outflows before failure in column (5). This restricts the sample period to 1880-1934 in Panel (A), as data on deposits at failure are not available before 1880. Similarly, column (5) in Panel B is restricted to 1993-2024, as deposits in failure are not available before 1993. \Cref{tab:AUC_logit} in the Appendix shows the AUC when using logit instead of linear probability models.}
        \end{minipage}
 \end{table}%

Bank failures are highly predictable based on the AUC metric. The in-sample AUC for the full specification in column (4) ranges from 86\% in the historical sample to 95\% in the modern sample. On their own, measures of insolvency and noncore funding both predict failures. Insolvency is relatively more predictive in the modern sample, while noncore funding has a higher AUC in the historical sample. The interaction between solvency and funding also boosts the predictive performance slightly. In the modern sample, where the predictability is extremely high, insolvency alone captures most of the predictive content of fundamentals. 

The stronger predictive power of noncore funding relative to the insolvency measures in the historical sample has at least two potential explanations. First, funding pressures may play a more important role in pre-FDIC bank failures. Second, reliance on noncore funding can be more informative about a bank's distance to default than the insolvency metric. Creditors of weaker banks were more likely to lend through noncore funding \citep{White1983,Calomiris1997}. Furthermore, the informational content of book equity was lower, as banks were subject to less rigorous accounting standards and less strict provisioning rules in the historical sample.

There are several reasons for the  stronger overall predictive performance in the modern sample. First, as just discussed, the quality of the accounting data is higher in the modern sample. Second, in the historical sample, national banks with unit-branches were less diversified, implying that idiosyncratic shocks accounted for more failures. This makes these failures potentially harder to predict. Third, in the modern sample, bank failures are preceded by larger lending booms, which often imply predictable losses down the road. Finally, in the modern context, bank failures are, by and large, a supervisory decision \citep{Correia2025supervising}. Frictions in the supervisory process, in turn, may delay bank failures, thereby increasing predictability.

The pseudo-out-of-sample performance is nearly as strong as the in-sample predictive performance. The high predictability also extends to longer horizons. In columns (6) and (7) we assess the predictability of bank failure over three and five-year horizons. At the five-year horizon, the in-sample AUC is nearly 80\% for the historical sample, and it is even higher in the modern sample.

\begin{figure}
\centering
\caption{\textbf{ROC Curves for Bank Failure Prediction} \label{fig:ROC}}

\subfloat[1863-1934 Sample]{\includegraphics[trim=10cm 0cm 10cm 0cm, clip, width=0.5\textwidth]{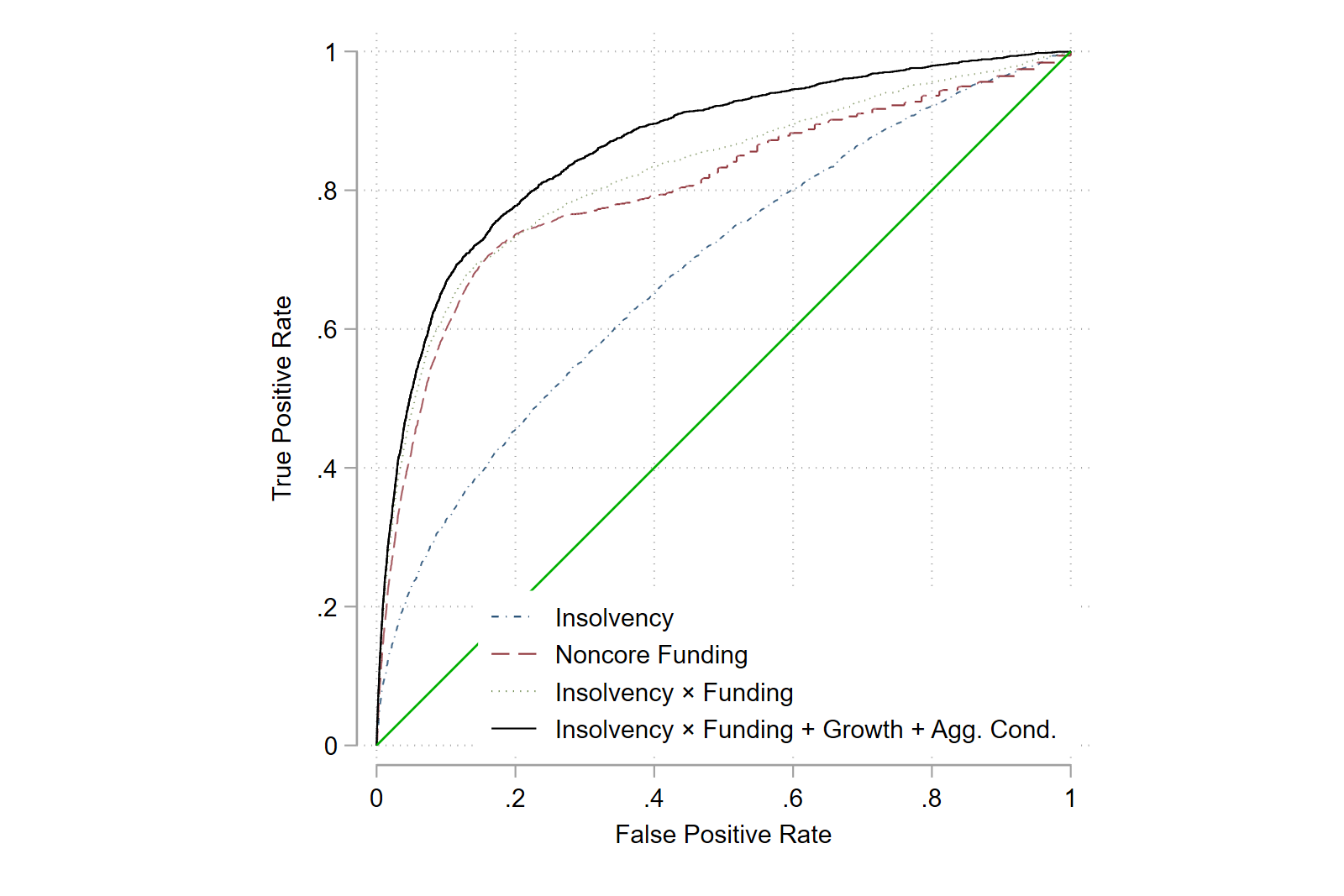}}
\subfloat[1959-2024 Sample ]{\includegraphics[trim=10cm 0cm 10cm 0cm, clip, width=0.5\textwidth]{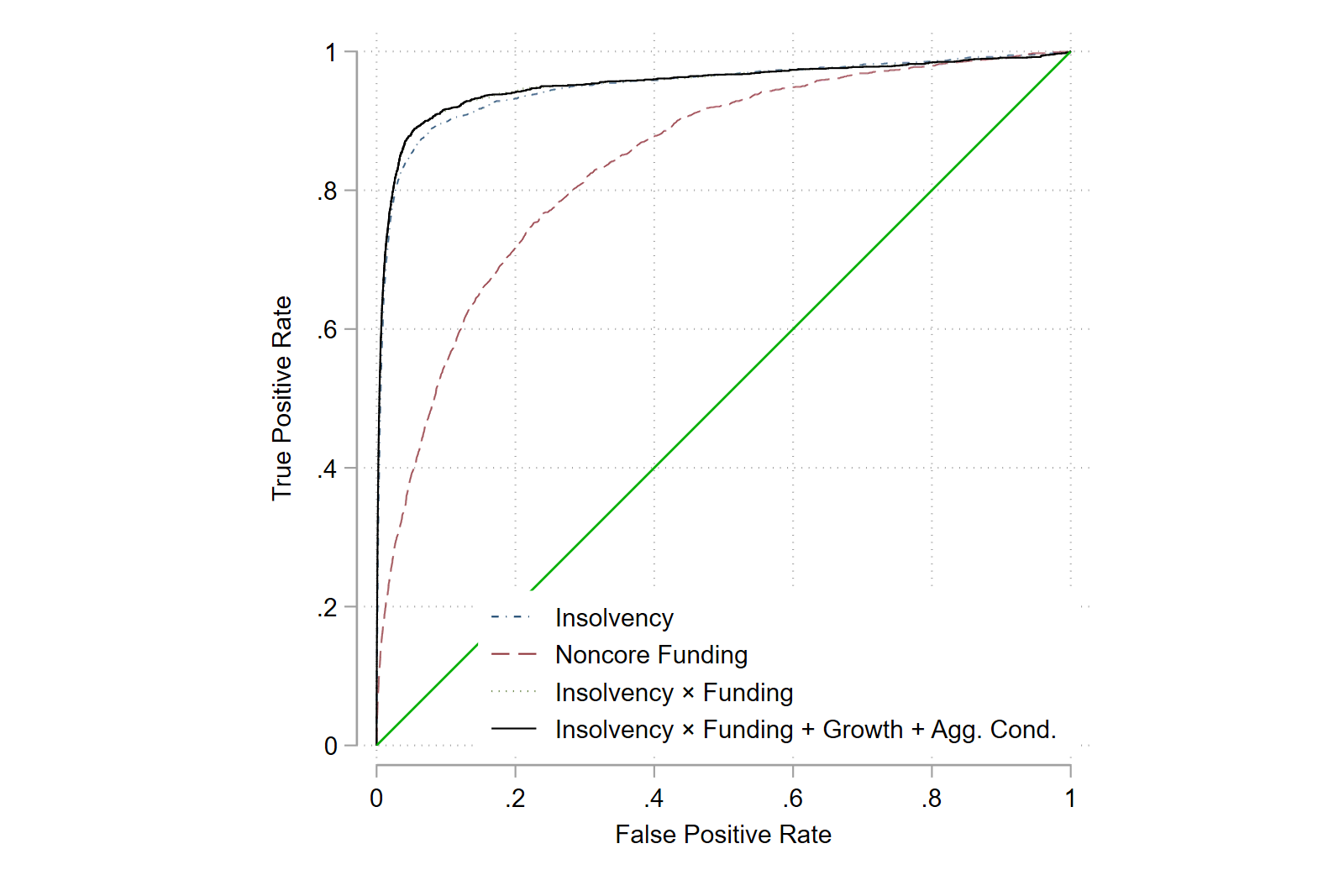}}

\begin{minipage}{\textwidth}
\footnotesize
Notes: This figure plots the receiver operating characteristic (ROC) curve for the estimates based on columns (1) through (4) of \Cref{tab:AUC}. 
\end{minipage}
\end{figure}

The high AUC statistics imply that bank failures can be classified with a high degree of accuracy. \Cref{fig:ROC} presents a visualization of the tradeoff between true positives and false positives---the ROC curve---across specifications for the historical and modern samples. \Cref{tab:true_false_positives} reports true positive rates, false positive rates, true negative rates, and false negative rates for various classification thresholds. The ROC curves imply that a forecaster willing to accept a 10\% false positive rate can achieve a true positive rate of 67\% in the historical sample and 92\% in the modern sample, again illustrating the strong predictability of bank failures.

\subsection{Additional Predictability Results and Robustness}

The estimated coefficients for the prediction models reported in \Cref{tab:predicting_failure_historical}  and  \Cref{tab:predicting_failure_modern_era} reveal several other interesting results. For subperiods of 1863-1934 sample, we also consider richer specifications using more detailed balance sheet line items that are only reported in specific periods (see \Cref{tab:predicting_failure_NBEra}, \Cref{tab:predicting_failure_early_fed}, and \Cref{tab:predicting_failure_GD}). These regression estimates reveal that higher surplus profit to equity, higher loans to assets, lower equity to assets, and higher reliance on noncore forms of funding, such as bills payable and rediscounts, all significantly predict higher future rates of failure.

Bank asset growth is also significantly associated with failure. The relation is non-monotonic and changes with the horizon. In the short-term, banks with low asset growth have the highest probability of failure. In contrast, at longer horizons of three to five years, the highest probability of failure is for banks that grow \textit{quickly}, as well as banks that contract the most.

Aggregate conditions are also informative. Low aggregate GDP growth over the past three years is associated with a higher probability of failure in the National Banking Era and Early Fed Era. This is consistent with \cite{Gorton1988} and \cite{Calomiris1991}, who find that bank failures and panics in the National Banking Era were more likely following negative macroeconomic news.

Using richer, period-specific specifications generally leads to slightly better predictive performance (see \Cref{tab:AUC_historical_detailed}), but most of the predictability is summarized by the simpler baseline models summarized in \Cref{tab:AUC}. \Cref{tab:AUC_historical_detailed} also shows that AUCs are substantial within subsamples of the historical sample, such as in the National Banking Era before the creation of the Federal Reserve (pre-1904), the Early Fed Era (1914-1928), and the Great Depression (1929-1933). Further, AUCs are similar across the bank size distribution (see \Cref{tab:AUC_size}). Bank failures are predictable based on weak fundamentals for both small and large banks. The predictive performance is also similar across linear probability and logit models (see \Cref{tab:AUC_logit}).

Finally, in addition to evaluating predictive performance using the area under the receiver-operating characteristic curve (AUC), we also assess the precision-recall (PR) curve. The precision-recall curve traces the relationship between the proportion of predicted bank failures that are correct (true positives relative to true positives plus false positives) against recall (true positives relative to true positives plus false negatives, also known as the true positive rate). Precision-recall is informative when predicting a rare outcome, such as bank failures.  \Cref{tab:PR_AUC} summarizes the area under the precision-recall curve (PR-AUC) across models and samples. Note that the PR-AUC has a different interpretation than the standard AUC. In particular, it is informative to compare the ratio of the PR-AUC to the unconditional probability of failure, as this base rate would be the PR-AUC of a naive model that classified all bank-year observations as failures.

Studying the precision-recall curve confirms the patterns discussed above. The PR-AUC is about 13 times higher than the failure rate for out-of-sample predictions from the historical sample, indicating moderately strong predictability (panels A and B, column 4). In the modern sample, the PR-AUC is over 70 times larger than the failure rate, indicating extremely strong classification performance.

%% file: output/tables/05_tab_auc_historical.tex
AUC (in-sample)     &       0.684&       0.802&       0.823&       0.864&       0.855&       0.799&       0.739

%% file: output/tables/05_tab_auc_oos_historical.tex
AUC (out-of-sample) &       0.774&       0.826&       0.846&       0.851&       0.839&       0.810&       0.775\\
\cmidrule(lr){1-8} N&      294574&      294253&      294247&      290088&      262636&      290088&      290088\\
No of Banks         &       12536&       12535&       12535&       12428&       11851&       12428&       12428\\
Mean of dep. var.   &         .79&         .79&         .79&          .8&         .45&         2.5&         4.1

%% file: output/tables/05_tab_auc_modern_era.tex
AUC (in-sample)     &       0.949&       0.846&       0.954&       0.953&       0.932&       0.889&       0.831

%% file: output/tables/05_tab_auc_oos_modern_era.tex
AUC (out-of-sample) &       0.942&       0.792&       0.946&       0.945&       0.922&       0.869&       0.805\\
\cmidrule(lr){1-8} N&      633407&      633404&      633404&      590645&      215320&      590645&      590645\\
No of Banks         &       23107&       23106&       23106&       22348&       13900&       22348&       22348\\
Mean of dep. var.   &         .32&         .32&         .32&         .34&        .032&         .97&         1.5

%% file: 06_runs.tex
\section{Failures With Bank Runs} 

\label{sec:predicting_runs}

In this section, we discuss the role of bank runs in failures. We define bank failures with runs as those featuring a large deposit outflow immediately before failure. As discussed in \Cref{sec:conceptual}, large deposit outflows are a necessary, but not sufficient, condition for bank runs to be the cause of failure. We first show that large deposit outflows were common in bank failures before the FDIC. We then show that failures with bank runs are strongly predicted by weak bank fundamentals.

\subsection{Deposit Outflows in Failing Banks}

How large are deposit outflows before bank failures, and how does this vary with the presence of deposit insurance? We calculate deposit outflows immediately before failure as the growth in deposits from the last call report to the time of bank failure. For the historical period, the OCC reports deposits at suspension for all bank failures during 1880-1934. For the modern sample, we obtain deposits at failure during 1993-2024 from the FDIC Failure Transaction Database.

\begin{figure}[h!]
\caption{\textbf{Deposit Outflow Immediately before Bank Failure}} 
\label{fig:deposit_before_failure}
\centering

  {\includegraphics[width=0.85\textwidth]{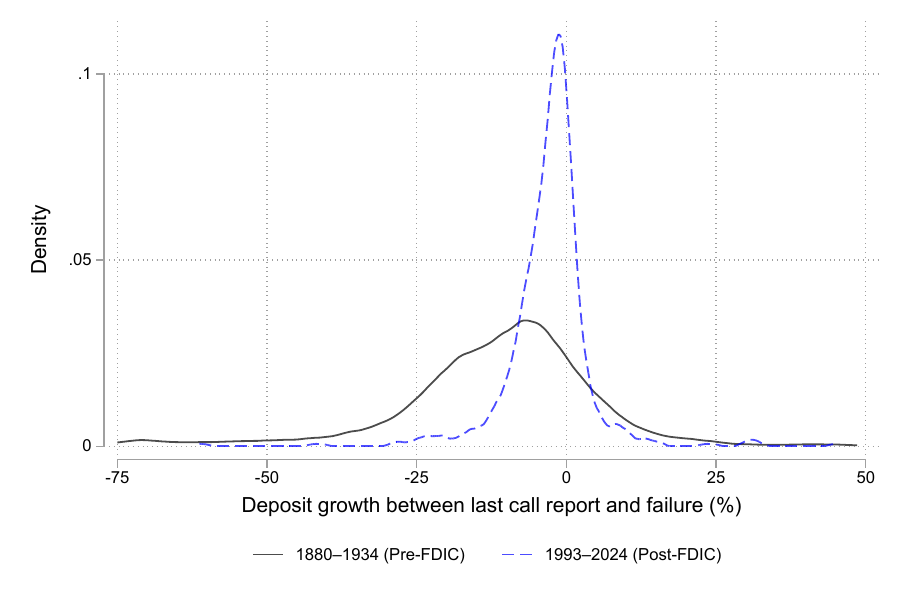}}

   \begin{minipage}{\textwidth}
\footnotesize
Notes: This figure shows the distribution of the growth in deposits between the last call report from before failure and the deposits reported in failure. Deposit growth is trimmed at -75\% and 50\%. We include failures from 1934 in the pre-FDIC sample. Even though the FDIC was founded in 1933, many receiverships in 1934 were associated with suspensions in 1933 (see \Cref{sec:historical_context}).
\end{minipage}
\end{figure}

\Cref{fig:deposit_before_failure} visualizes the distribution of deposit growth immediately before failure for the pre- and post-FDIC samples. \Cref{tab:deposit_outflows} reports details on the distribution of deposit growth immediately before failure for various eras. On average, banks experience a 14\% decline in deposits immediately before failure in the 1880-1934 sample. Moreover, 25\% of all pre-FDIC failures were preceded by deposit outflows of more than 20\%. Deposit outflows in failing banks were highest during the Great Depression. Before the banking holiday, deposits declined by an average of 21\% between the last call and failure. Thus, large deposit outflows were quite common before the FDIC became operational. At the same time, large deposit outflows were not universal. In 26\% of failures, deposits declined by less than 2.5\%. 
  
In contrast to the historical sample, average outflows before failure are much smaller in the modern sample. Deposit outflows immediately before failure averaged 2.5\% for failures during 1993-2024, a sample that includes the 2008 Global Financial Crisis. Deposit outflows exceeding 20\% are rare in the modern sample and only occur in 3\% of all failures. These patterns are in line with deposit insurance insulating a large share of depositors from losses, reducing the incentives for deposit withdrawals before failure. Indeed, \Cref{fig:funding} shows that failing banks in the modern sample see a net inflow of insured deposits.

\begin{table}[ht]
\footnotesize
\begin{center}
\begin{threeparttable}
\caption{\textbf{Deposit Growth in Failing Banks Immediately before Failure }} 
\label{tab:deposit_outflows}
    
\begin{tabular}{lcccccccc}
\toprule
Era & Average & \multicolumn{6}{c}{Share of failures with deposit growth falling within... } & $N$ \\

\cmidrule(lr){3-8} 
 & & <-30\%  & [-30,-20\%)  & [-20,-7.5\%)& [-7.5,-2.5\%) & [-2.5,0\%) &  >0\%  & \\ \midrule
  \multicolumn{9}{c}{Panel A: Pre versus Post-FDIC}\\
 \midrule
\input{./output/tables/06_deposit_outflows_general}  \\  \midrule
 \multicolumn{9}{c}{Panel B: By Era}\\
 \midrule
\input{./output/tables/06_deposit_outflows}  \\  
\bottomrule
\end{tabular}

\tiny
\begin{tablenotes} \item
\footnotesize
Notes: This table reports the percent change in nominal deposits from the last call report before failure to the time of failure. From 1880 through 1934, deposits in failure are as reported in the OCC annual reports table on national banks in receivership, which records deposits ``at date of suspension.''  After 1992, we use deposits in failure as reported in the FDIC's Failure Transaction Database. $N$ indicates the number of bank failures for which we match both deposits before failure and at suspension. In Panel B, we split the Depression sample into failures before and after the banking holiday in March 1933 due to the different nature in failures across those episodes \citep[see, e.g.,][]{Jaremski2023}. Deposit growth is clipped at 100\%.
\end{tablenotes}
\normalsize
	\end{threeparttable}
	\end{center}
\end{table}

\subsection{Predictability of Failures with Large Deposit Outflows}

Before the introduction of deposit insurance, bank failures were commonly associated with bank runs. This suggests that the bank runs are relevant for understanding bank failures. As discussed in \Cref{sec:conceptual}, there are at least three possibilities. First, non-fundamental bank runs could cause the failure of solvent banks \citep{Diamond1983}. Second, panic runs could cause the failure of weak but solvent banks \citep{Goldstein2005,He2012}. Or, third, fundamental runs could be the trigger for the failure of insolvent banks.\footnote{A fourth possibility that is difficult to test is that runs simply occur in response to news of imminent bank failure.}

To make progress in distinguishing between failures from non-fundamental runs and runs based on weak fundamentals, we next investigate whether failures with bank runs are less connected to weak fundamentals than other bank failures. \Cref{fig:cond_prob} reveals that failures with large deposit outflows are related to weak fundamentals. In particular, \Cref{fig:cond_prob} shows that the conditional probability of failure with large deposit outflows over the next three years is increasing in measures of insolvency and noncore funding, similar to all failures. We classify failures with large deposit outflow as those with outflows of more than 7.5\% between the last call report and failure. The cutoff is necessarily arbitrary, but the results are robust to different cutoff choices. In the historical sample, moving from healthy fundamentals (below the 50\textsuperscript{th} percentile) to high insolvency or funding vulnerability is associated with an increase in the probability of failure similar to the increase for all failures. While failures with large deposit outflows are rare in the modern sample, these failures are also associated with weaker fundamentals.

To reinforce this point, we estimate \Cref{eq:pred} separately for failures with large deposit outflows. Comparing columns (4) and (5) in \Cref{tab:AUC} reveals that the predictive performance of fundamentals is nearly identical for bank failures with large deposit outflows as for all failures. In the historical sample, the in-sample AUC is 86\% both for failures with large deposit outflows and all failures. \Cref{tab:AUC_historical_detailed} shows that the AUCs for predicting failures with runs is slightly higher than for all failures using richer period-specific  models for the National Banking Era (1880-1904) and the Early Federal Reserve Era (1914-1928), while the AUC for predicting failures with runs is slightly lower for the Great Depression. On balance, throughout the sample before deposit insurance, failures with runs are approximately as predictable as other failures.

Thus, the failures associated with large deposit outflows---failures that likely involved runs---are not wholly unexpected events that are disconnected from fundamentals. The high predictability of failures with large deposit outflows cuts against the view that failures were often caused by non-fundamental runs that brought down solvent banks randomly. This pattern also holds for the pre- Federal Reserve and pre-FDIC sample, when non-fundamental runs were in principle possible. Instead, it suggests that runs either caused the failure of weak but solvent banks or occurred in \textit{ex ante} insolvent banks. This finding generalizes insights from existing empirical studies that have focused on studying specific panic episodes \citep[see, e.g.,][]{Wicker1996,Calomiris1997,Calomiris2003a} and establishes that weak fundamentals are typically a necessary condition for a bank to fail, even in the absence of a government safety net.

\subsection{Depositor Inattentiveness before Deposit Insurance}

The predictability of bank failures indicates that weak banks have a high in-sample predicted probability of failure. In Figures \Cref{fig:cond_prob} and \Cref{fig:interaction}, we saw that the predicted probability of failure exceeded 25\% for the most vulnerable banks. Appendix \Cref{tab:oos_predicted} reports the distribution of out-of-sample predicted probabilities of failure for failing banks in the year before failure for the historical pre-FDIC sample. More than 48\% of pre-FDIC failures have a predicted probability of failure over the next three years in excess of 20\% in the year before failure, compared to the unconditional three-year probability of 2.5\%, using the relatively richer, period-specific regression models. Further, more than 34\% of all failures are associated with a predicted probability of failure over three years exceeding 40\%, a very high likelihood of failure for an individual bank. 

The high predicted failure probabilities suggest that depositors are slow to react to the increased risk of bank failure. This, in turn, points to a role for behavioral frictions such as inattentive depositors or neglect of downside risk \citep[e.g.,][]{GSV2012,Jiang2023,LlambiasOrdonez2024}. Importantly, high predicted probabilities do not rule out the importance of coordination or the role of panic-based runs in causing the failure of weak but solvent banks. However, they do suggest that bank runs potentially happen later than theoretical benchmarks would suggest.

To illustrate the potential implication of the high predicted probabilities, recall that the information we use to estimate the probability of failure is public and therefore available to contemporary depositors. Moreover, depositor loss rates before the establishment of the FDIC were substantial, averaging 34\% for pre-FDIC bank failures (see \Cref{tab:depositor_losses}).\footnote{This is a conservative estimate that does not account for the time value of money, as depositors often had to wait several years before obtaining this recovery rate (see \Cref{fig:depositor_recovery}).} Appendix \Cref{tab:oos_predicted} presents a simple calculation of the excess returns that a risk-neutral and a risk-averse depositor would require to be compensated for holding deposits in these failing banks. Depending on the degree of risk aversion assumed, we find that between one-fifth and one-half of all failing banks would have faced a required annual excess return above 5\% on their unsecured debt financing.  While we do not have data on deposit rates, \cite{Cox1967} shows, based on a sample of national banks in 1929, that essentially no bank paid such high interest rates on deposits. We speculate that such high levels of interest rates were not likely to be offered by weak banks, but rather that the high predicted probabilities of failure are a result of depositor inattention or other information frictions.

\section{Fundamentals and Aggregate Waves of Bank Failures}
\label{sec:waves}

Individual bank failures, including those accompanied by runs, are predictable based on past fundamentals. Does the predictability at the micro level extend to forecasting aggregate waves of failures during systemic banking crises?

While fundamentals may predict individual bank failures, the connection between fundamentals and failures during systemic banking crises may differ for two reasons. First, fundamentals could become less predictive of failures during crises in which many banks fail. For example, panics may decouple bank failures from fundamentals. Increased uncertainty during crises may lead creditors to withdraw even from healthy banks, breaking the cross-sectional link between weak fundamentals and failure \citep{Chari1988,Gorton1988,Allen2000}.

We find no evidence that fundamentals are less predictive of bank failures during crises. The AUC for predicting bank failures is similar during times of major banking crises as for the overall sample (see \Cref{tab:auc_by_era} in the Appendix). Therefore, fundamentals perform well in ranking which banks are likely to fail during crises, cutting against failures driven by indiscriminate panics.

Second, crises may feature \textit{excess failures} beyond what fundamentals predict due to amplification mechanisms. For example, crises can feature chain reactions where bank failures lead to losses for other banks through interdependent claims \citep{Allen2000,Acemoglu2015} and fire sales that weaken all banks \citep{Gertler2015}. Amplification can also occur through contagion that triggers funding pressure and runs on weak banks. These amplification mechanisms can increase the fundamental threshold at which banks fail, leading more banks to fail than would otherwise occur.

We examine whether deteriorating fundamentals can forecast the aggregate rate of bank failures, including spikes in bank failures during systemic banking crises. To do this, we obtain the pseudo-out-of-sample predicted probability of failure for each bank and year $\hat p_{b,t+1|t}$ from estimation of \Cref{eq:pred} using data only up to year $t$. As the baseline, we use the model in column (4) from \Cref{tab:AUC}, namely the model with $\text{Insolvency}_{bt}$, $\text{Noncore Funding}_{bt}$, their interaction, $\text{Asset Growth}_{bt}$, and $\text{Aggregate Conditions}_{t}$. We then compute the average predicted failure rate
$ \overline p_{t|t-1} = \sum_{b \in B_{t-1}} w_{b,t-1} \hat p_{b,t|t-1},$ where $w_{bt}$ is the weight on bank $b$ at time $t$ and $B_t$ is the set of all banks in year $t$. As in \Cref{tab:AUC}, we use an initial training sample of 10 years and estimate $\overline p_{t|t-1}$ separately for the 1863-1934 and 1959-2024 samples due to differences in data availability. We weight banks equally. Results are similar when weighting banks by size. We present baseline results from linear probability models, but show robustness to using logit models in the appendix.

\begin{table}[ht]
\begin{center}
\begin{threeparttable}
\caption{\textbf{Fundamentals Predict Aggregate Rate of Bank Failures} }\label{tab:predicting_aggregate}
\footnotesize
    \begin{tabular}{lcccc}
\toprule
Dependent variable  & \multicolumn{4}{c}{Aggregate Failure Rate}  \\
    \cmidrule(lr){2-5}   
\input{./output/tables/06_aggregate_predicted_actual_regs}
\\
\bottomrule
\end{tabular}
\begin{tablenotes} 
\item 
\footnotesize
Notes: This table presents time series regressions of the annual aggregate failure rate in year $t$ on the average predicted failure rate $\overline p_{t|t-1}$.  The average predicted failure rate is constructed out-of-sample using an expanding sample that only incorporates information up to year $t-1$. The predicted failure rate is based on the model in column (4) of \Cref{tab:AUC} estimated with a linear probability model. Appendix \Cref{tab:predicting_aggregate_robustness} shows the results from using alternative variables and logit estimation. Predicted probabilities are constructed using the first 10 years of the sample as training data.  Newey-West standard errors in parentheses with truncation parameter $S=1.3T^{1/2}$ following \cite{Lazarus2018HAR}. *, **, and *** indicate significance at the 10\%, 5\%, and 1\% level, respectively. 
\end{tablenotes}
\end{threeparttable}
\end{center}
\end{table}

\begin{figure}[h!]
\centering
\caption{\textbf{Fundamentals Predict Aggregate Waves of Bank Failures } \label{fig:aggregate}}

\includegraphics[width=0.85\textwidth]{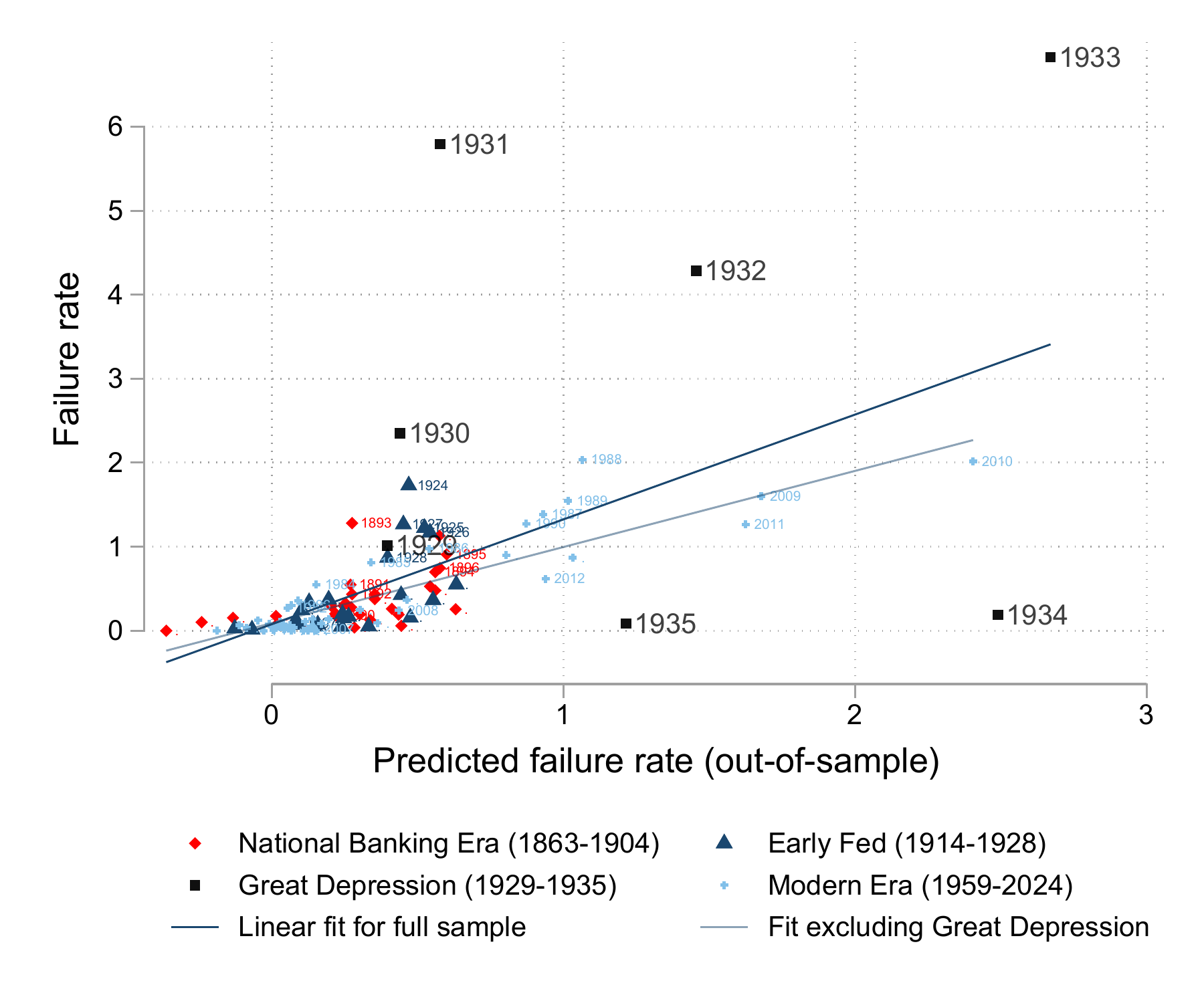}

\begin{minipage}{\textwidth}
\footnotesize 
Notes: This figure plots the realized aggregate failure rate against the predicted aggregate failure rate, $\overline p_{t|t-1}$. The predicted aggregate failure rate for year $t$ is constructed using only information up to year $t-1$, so the prediction is pseudo out-of-sample using an initial training sample of ten years.  The predictions for each sample period are based on the linear probability model in column (4) of \Cref{tab:AUC}.  \Cref{tab:predicting_aggregate} reports the regression version of this figure for the full sample and by era. Negative predicted probabilities are possible due to the use of a linear probability model. \Cref{fig:aggregate_granular} shows the results using logit estimation and with richer period-specific models. 
\end{minipage}
\end{figure}

\Cref{fig:aggregate} plots the realized aggregate failure rate against the out-of-sample predicted aggregate failure rate, $\overline p_{t|t-1}$. \Cref{tab:predicting_aggregate} presents estimates of the corresponding time-series regression:
$$ \text{FailureRate}_t = \alpha + \beta \overline p_{t|t-1} + u_t.$$ There is strong out-of-sample predictability of aggregate bank failures based on past fundamentals. The $R^2$ for the full sample is 40\%. Using richer period-specific models implies an $R^2$ of 59\% (see \Cref{tab:predicting_aggregate_robustness}). Deteriorating fundamentals matter not only for individual bank failures; they are also key to understanding widespread bank failures during major banking crises in the U.S. Banking crises are thus, to a significant extent, baked into weak fundamentals, rather than fully unexpected jumps to bad equilibria. This evidence reinforces existing evidence using aggregate data that banking crises are predictable based on variables that forecast subsequent bank losses, such as business failures \citep{Gorton1988}, credit and asset price growth \citep[e.g.,][]{Schularick2012, Greenwood2022,MullerVerner2023}, and declining bank equity prices \citep{Baron2021}.

The predicted failure rate is similar in magnitude to the actual failure rate in many years. 
However, it generally underpredicts the actual failure rate during the Great Depression, consistent with potential excess failures during this major crisis. The estimated slope $\hat \beta $ in  \Cref{tab:predicting_aggregate} is larger than one in the full sample and in the pre-1934 sample.\footnote{\Cref{tab:predicting_aggregate_robustness} show that the estimated slope is larger than one in simpler models with fewer predictors but close to one using richer period-specific models.}  Interestingly, the coefficient estimate is not significantly different from one in the modern sample (column 3) or when we exclude the Great Depression (column 4). Excess failures are thus mainly driven by the Great Depression, which featured record bank failures from 1930 to 1933.\footnote{While there are excess failures in the main Depression years 1930 through 1933, our model overpredicts failure in 1934 and 1935. While the over-prediction in 1935 may be directly linked to government interventions and an increased safety net that suppresses failure, the over-prediction in 1934 is in part also related to the timing of the Banking Holiday of 1933 and the timing of the subsequent appointment of receivers \citep{Jaremski2023}. In particular, many banks that suspended in early 1933 reported their last call report in 1932 but the receiver was only formally appointed in 1934, see also \Cref{fig:failures_and_suspensions}. This implies that our model that predicts failure one year ahead is subject to a mechanical sample selection, as some banks that fail in 1934 are dropped due to the lack of financial statements in 1933. 

Moreover, there is less evidence of excess failures for other major episodes, such as the panics of the National Banking Era (with the exception of 1893), the Savings and Loan Crisis of the 1980s, and the 2008 Global Financial Crisis.}

The predictability of the aggregate failure rate is especially high in the modern sample. The estimated coefficient in column (3) of \Cref{tab:predicting_aggregate} is close to unity, the constant is close to zero, and the $R^2$ is 81\%. This is likely partly due to improvements in the accounting data, which more quickly reflect bank losses. It may also reflect a change in the nature of bank failures. In the post-FDIC era, the timing of failure is partly determined by government supervisors, since deposit insurance can blunt market forces that would force a bank failure \citep{Walter2004,Correia2025supervising}. Therefore, modern bank failures occur later in crises. For example, during the 2008 financial crisis, the highest rate of failures occurred in 2010, followed by 2011 and 2009.\footnote{During the Savings and Loan Crisis of the 1980s, regulatory forbearance significantly delayed failure. Since Prompt Corrective Action was introduced in 1991, a critically undercapitalized bank must be closed or raise new capital within 90 days, accelerating failure for troubled banks.} In contrast, in the historical sample, the timing of failure was determined by market forces, such as a run or bank owners seeking to limit their losses.

%% file: output/tables/06_deposit_outflows_general.tex
1880-1934 (Pre-FDIC)&-13.71&0.13&0.12&0.33&0.16&0.07&0.19&2729\\
1993-2024 (Post-FDIC)&-2.55&0.01&0.02&0.13&0.30&0.30&0.24&546

%% file: output/tables/06_deposit_outflows.tex
1880-1913 (NB Era)&-14.40&0.18&0.16&0.32&0.13&0.04&0.17&499\\
1914-1918 (Early Fed)&-15.89&0.13&0.14&0.38&0.18&0.06&0.11&672\\
1929-1933 (Depr., pre-Hld.)&-21.19&0.17&0.16&0.44&0.14&0.05&0.04&962\\
1933-1934 (Depr., post-Hld.)&1.39&0.01&0.02&0.10&0.21&0.13&0.53&596\\
1993-2006&-5.64&0.08&0.11&0.33&0.21&0.08&0.19&63\\
2007-2024&-2.14&0.00&0.01&0.10&0.32&0.33&0.25&483

%% file: output/tables/06_aggregate_predicted_actual_regs.tex
                    &         (1)   &         (2)   &         (3)   &         (4)   \\
\cmidrule(lr){1-5} Predicted failure rate, \( \overline p_{t|t-1} \) &        1.25***&        1.58***&        0.97***&        0.91***\\
                    &      (0.24)   &      (0.29)   &      (0.15)   &      (0.12)   \\
Constant            &        0.08   &        0.07   &        0.06   &        0.09***\\
                    &      (0.05)   &      (0.13)   &      (0.04)   &      (0.03)   \\
\cmidrule(lr){1-5} N&         117   &          61   &          55   &         111   \\
$R^2$               &        0.40   &        0.38   &        0.81   &        0.59   \\
Sample              &        Full   &   1874-1934   &   1970-2024   &Exclude 1929-1934   

%% file: 07_recovery_rates.tex
\section{Recovery Rates and Asset Quality in Failure}
\label{sec:recovery}

Thus far, we have established that bank failures---including failures involving runs---are strongly connected to weak bank fundamentals. This evidence casts doubt on the role of non-fundamental runs in explaining most bank failures and crises. In this section, we study recovery rates on assets held by failing banks at failure. Under certain assumptions, asset recovery rates allow us to shed additional light on the relative importance of bank runs and insolvency in explaining bank failures. 

\subsection{Recovery Rates and Asset Quality} 

\Cref{tab:recovery_rates} reports the recovery rate on assets in bank receiverships for the 1863-1934 sample. We define the recovery rate, $R$, as the ratio of the cash the receiver collected to the book value of total assets available to the receiver. Total assets are defined as assets at suspension and assets received after suspension. Recovery rates on assets are low in the pre-FDIC sample, averaging 52\%.\footnote{The average recovery rate is even lower when excluding the failures that occurred after the banking holiday in the Great Depression. Bank failures during the banking holiday were the product of a government decision, which led to the closure of some healthier banks that likely had higher quality assets \citep{Jaremski2023}. \Cref{fig:recovery-rate} plots the average recovery rate over time.} In 43\% of failures, recovery rates are below 50\%. Another 49\% have recovery rates between 50\% and 75\%. Only 8\% of failed banks have recovery rates above 75\%.

\begin{table}[ht!]
\small
\begin{center}
\begin{threeparttable}
\caption{\textbf{Asset Recovery Rates in Failure, 1863-1934 } \label{tab:recovery_rates}
} 
\centering
\footnotesize
   \begin{tabular}{lccccccc}
\toprule
Era  &
\multicolumn{6}{c}{Recovery Rate, $R$} & $N$  \\ \cmidrule(lr){2-7}
  & Average & \multicolumn{5}{c}{Share of failures with recovery rate within...} & \\ \cmidrule(lr){2-2} \cmidrule(lr){3-7} 
&   & <25\% & [25\%-50\%) & [50\%-75\%) & [75\%-95\%) & $\geq$95\% & \\   \midrule
\input{./output/tables/04_assets_in_failure_collected} \\
\input{./output/tables/04_assets_in_failure_all_collected} \\
		\bottomrule
\end{tabular}
 
\begin{tablenotes} \item
Notes: This table reports asset recovery rates for failed banks. The sample covers failed national banks from 1863 to 1934. Data are collected from the OCC's annual report to Congress; tables on ``National banks in charge of receivers'' (various years). The recovery rate on assets is the total collected funds in receivership relative to the sum of total assets held at suspension and collected after suspension. This represents the share of assets that the receiver was ultimately able to recover to compensate claimholders. Note that the receiver also collected funds from shareholders due to double-liability, which increased the overall amount of available funds to distribute to debt holders. \Cref{fig:recovery-rate} in the appendix plots the average recovery rate by year.

\end{tablenotes}
\normalsize
\end{threeparttable}
\end{center}
\end{table}

What explains these low recovery rates? We first study the OCC's assessments of asset quality made immediately upon a bank's closure. As part of the initial examination process during receivership, the OCC divided the book value of assets into three categories: \textit{Estimated Good}, \textit{Estimated Doubtful}, and \textit{Estimated Worthless}. \Cref{tab:quality} provides statistics on these assessments for all failures in the 1863-1934 sample.  On average, the OCC estimated 36\% of assets to be good, 47\% to be doubtful, and 18\% to be worthless. Thus, receivers on average assessed 65\% of assets as either ``doubtful'' or ``worthless,'' indicating a pessimistic view of failed banks' asset quality at the time of bank failure.

The assessed shares of good, doubtful, and worthless assets are strongly related to actual recovery rates. Column (1) in \Cref{tab:pred_recovery_rate} reports a regression of the realized recovery rate on the initial assessment of asset quality. It shows that each dollar of good, doubtful, and worthless assets predicts a recovery rate of \$0.89, \$0.54, and \$0.08, respectively, with an $R^2$ of 94\%. While the OCC assessments may contain biases, this indicates that they provide useful information about asset quality. Taken together, the low assessed asset quality and low recovery rate on failed banks' asset holdings suggest that unrealized losses relative to the book value of assets were likely an important determinant of failure.\footnote{\cite{James1991} studies 412 bank failures between 1985 and 1988. He finds that asset losses averaged 30\% for failing banks. \cite{James1991} argues that a significant portion of these losses reflect past unrealized losses, rather than liquidation discounts.  Focusing on bank failures between 1986-2007, \cite{Bennett2015} find that the average loss amounted to 33.2\% of total assets. Further, \citet{Granja2017} show that in the aftermath of the GFC, the average FDIC loss on a failed bank was around 28\% of assets, with a substantial part of these losses resulting from frictions in the market for failed banks. Our evidence is broadly consistent with these papers, although we find that the recovery rates were lower in the pre-FDIC sample.} 

\begin{table}[ht!]
\small
\begin{center}
\begin{threeparttable}
\caption{\textbf{OCC Assessment of Asset Quality in Failure, 1863-1934 } \label{tab:quality}
} 
\centering
\footnotesize
   \begin{tabular}{lcccc}
\toprule
Era  &
\multicolumn{3}{c}{Assets at suspension} & $N$   \\ \cmidrule(lr){2-4}
 &  \shortstack{Estimated \\ Good} & \shortstack{Estimated \\ Doubtful} & \shortstack{Estimated \\ Worthless} &  \\  \midrule
\input{./output/tables/04_assets_in_failure_assessed}
\input{./output/tables/04_assets_in_failure_all_assessed} \\

\bottomrule
    \end{tabular}
 	
\begin{tablenotes} \item
  
Notes: This table reports estimates of the share of good, doubtful, and worthless assets at the time of suspension. The sample covers failed national banks from 1863 to 1934. Data are collected from the OCC's annual report to Congress; tables on ``National banks in charge of receivers,'' (various years). Good, doubtful, and worthless assets at suspension are normalized by total assets at suspension. 

\end{tablenotes}
\normalsize
\end{threeparttable}
\end{center}
\end{table}

\Cref{tab:pred_recovery_rate} further shows that recovery rates are related to several other bank-level proxies of asset quality. We measure these proxies in the last call report before failure. The recovery rate is lower in smaller banks, in line with less well-diversified banks suffering larger losses in failure. It is also lower in less well-capitalized banks, suggesting that poor asset performance depresses both pre-failure capitalization and recovery in failure. Similarly, it is lower in banks that report a higher share of non-performing loans and a higher loans to liquid asset ratio before failure, indicating that banks with more bad loans tend to experience larger asset losses in failure. Finally, a larger boom-bust cycle before failure is associated with a lower recovery rate in failure. In particular, the recovery rate is lower for banks that have rapid asset growth from ten to seven years before failure, consistent with rapid asset growth going hand-in-hand with more risk at the margin. It is also lower for banks that have the slowest growth in the last three years before failure.

The low recovery rate on assets in the pre-FDIC sample implies substantial loss rates for depositors in this period compared to the modern period. Appendix \Cref{tab:depositor_losses} presents estimates on the loss rates for uninsured depositors for bank failures in the pre- and post-FDIC samples. In the pre-FDIC sample, 81\% of failures involved losses for depositors, and the average unconditional depositor recovery rate was 66\%. Moreover, depositors often experienced a substantial delay before receiving their funds. On average, the depositor recovery rate in the initial year is only about 35\% (see \Cref{fig:depositor_recovery}). In contrast, in the 1992-2022 sample, only 20\% of failures involved losses for uninsured depositors, and the average unconditional loss rate is 6\%.

\begin{table}[ht!]
\centering
\begin{threeparttable}
\caption{\textbf{Predictors of Asset Recovery Rate in Receivership}}
\label{tab:pred_recovery_rate}
\scriptsize
\begin{minipage}{1.0\textwidth}
\begin{tabular}{lccccc}
\toprule
Dependent variable & \multicolumn{5}{c}{Recovery Rate, $R$} \\
\cmidrule(lr){2-6} 
\input{./output/tables/07_predicting_recovery_rates_paper} \\ 
\bottomrule
\end{tabular}

\begin{tablenotes} 
\item 
\footnotesize
Notes: This table reports results from estimating regressions of the following form: 
${R}_b = \alpha + \beta X_{b} +\epsilon_b.$ The sample is receiverships in 1863-1934. The recovery rate, $R_b$, is calculated as the ratio of the total funds collected in receivership over total assets held at suspension and additional assets received after suspension. $X_b$ is a set of bank characteristics observed before or in receivership, including characteristics such as the share of assets assessed as \emph{good}, \emph{doubtful} or \emph{worthless} by the receiver at the onset of the receivership and balance sheet characteristics from before receivership. NPL is proxied by Other Real Estate Owned, available for the 1889-1904 subsample. We also include indicators for the quintile of the loan growth distribution (across all banks within the same calendar year) in both the boom phase (10 to 3 years before failure) and the bust phase (3 years before failure to last call before failure). The regression in column (1) is estimated without a constant.  *, **, and *** indicate significance at the 10\%, 5\%, and 1\% level, respectively. 

\end{tablenotes}
\end{minipage}
\end{threeparttable}
\end{table}%

\subsection{The Share of Fundamentally Insolvent Banks} 

As we outlined in \Cref{sec:conceptual}, a central yet challenging question is whether runs commonly cause the failure of weak but solvent banks or whether runs mainly trigger the failure of \textit{ex ante} insolvent banks. To address this question, we spell out a simple framework that we take to the granular data on pre-FDIC bank receiverships.

Suppose a failed bank has book assets $A$ and debt $D$, where debt includes both deposits and non-deposit borrowing. Further, assume the bank has unrealized losses on assets before entering failure of $\lambda$ and incurs additional losses in receivership of $\rho$. $\rho>0$ captures the idea that bank closure itself may cause additional losses. These additional losses can result from a wedge between the value of the assets in the bank and the value for the next best user. For instance, the value of bank assets can be tied to the human capital of the manager \citep[e.g.,][]{Diamond1984,HartMoore1994,Diamond2001}, or the receiver may be inefficient in managing a failed bank's assets.  The recovery rate we observe in receivership, $R$, is thus a combination of unrealized asset losses from before failure and losses in receivership: $R = (1-\lambda)(1-\rho)$. Further, let $v$ be the franchise value as a fraction of current book assets. 

The bank is insolvent irrespective of the run if the value of its assets and franchise falls short of its debt:
\begin{align*}
(1-\lambda)(1+v) A = \frac{R}{1-\rho}(1+v) A < D.
\end{align*}
If this inequality does not hold, the failed bank was solvent in the absence of the run. Let $\ell=D/A$ denote the bank's leverage. The insolvency condition can be written as
\begin{align}
    \frac{1+v}{1-\rho} < \frac{\ell}{R}. \label{eq:insolvent}
\end{align}
This condition captures the intuition that, all else being equal, loss rates ($1-R$) and leverage ($\ell$) cannot be too high for a run to cause the failure of an \textit{ex ante} solvent bank. If the recovery rate $R$ is high relative to leverage $\ell$, it would be plausible that the run destroyed the bank's franchise value, but the bank was solvent absent the run. In contrast, very low recovery rates combined with high leverage would suggest that the bank was insolvent, irrespective of whether a run took place or not.

Our newly digitized micro-data on bank receivership allows us to estimate $\ell$ and $R$ for each bank. Specifically, we calculate $\ell$ as the amount of claims proved by depositors plus secured and preferred liabilities paid at failure, relative to total assets (see \Cref{sec:data_receiverships} for details). As above, we calculate the recovery rate $R$ as total collections from assets relative to total assets. 

The loss in receivership, $\rho,$ and franchise value, $v$, on the other hand, are not directly observable. The loss in receivership $\rho$ is likely to be positive if the bank manager is better than the receiver in collecting on bank assets. At the same time, unlike non-financial firms, which hold mostly assets that are considerably more valuable inside a firm than outside a firm, banks largely hold assets that can be separated and repossessed, such as securities and loans. Moreover, OCC receivers were generally considered to be experts in collecting on failed banks' assets, could not sell assets without the approval of the Comptroller and a court order, and were charged with liquidating assets in an orderly fashion \citep{upham1934closed}. OCC receivers often held assets for several years in order to avoid fire sales. The median receivership lasts for six years (see \Cref{tab:receivership_length}). Therefore, the loss in receivership is not necessarily large. Banks enter failure with very low-quality assets, which suggests managers of these failed banks may not have been highly skilled at screening and collecting from borrowers.

There is also a range of plausible values for the franchise value, $v$. Franchise value varies across banks and over time based on technology, regulation, the competitive environment, and market conditions \citep[e.g.,][]{Keeley1990}. Estimates of franchise value based on Tobin's \textit{q} or valuations of bank income and expenses range from 0\% to 20\% \citep[e.g.,][]{Demsetz1996,MaScheinkmann2020,Hirtle2024,Demarzo2024}. These estimates are, of course, based on specific equilibria and may not apply to failing banks. Indeed, we view these estimates as likely upper bounds for failing banks, given that we find these banks generally undergo a deteriorating business model before failure.

\begin{table}[htbp]
\centering
\caption{Share of Fundamentally Insolvent Banks by $\rho$ and $v$, 1863-1934} 
\label{tab:rho_v} 
\begin{minipage}{1\textwidth}
\begin{center}
\begin{tabular}{l|ccccccc}
\toprule
$\rho$ & $v = 0$ & $v = 2.5\%$ & $v = 5\%$ & $v = 7.5\%$ & $v = 10\%$ & $v = 15\%$ & $v = 20\%$ \\
\midrule
\input{./output/tables/07_recovery_rho_v}  \\ 
\bottomrule
\end{tabular}
        \end{center}
 {\footnotesize Notes: This table reports the share of failed banks that are fundamentally insolvent, calculated as $$\frac{1}{N}\sum_b   \mathbb{I}\left[\frac{1+v}{1-\rho} < \frac{\ell_b}{R_b} \right].$$ $\rho$ and $v$ are the unobservable loss in receivership and the franchise value as a fraction of current book assets, respectively. $\ell_b$ and $R_b$ are the observable ratio of debt to assets and recovery rate on assets, respectively. The calculations are based on the sample of national banks placed in charge of receivers from 1863 through 1934. }
         \end{minipage}
\end{table}

\Cref{tab:rho_v} presents the share of fundamentally insolvent banks, defined as the share of banks for which condition \eqref{eq:insolvent} holds. Given the range of possible values of the receivership loss rate and franchise value, we report the shares for various combinations of $\rho$ and $v$. When $\rho=v=0$, the framework implies that the share of fundamentally insolvent banks is 81\%. Assuming a $\rho=0.1$ and $v=0.05$ implies 60\% of failed banks were fundamentally insolvent. Interestingly, when we separately examine failures by those with and without large deposit outflows immediately before failure, we find that a greater share of banks that failed with a run were likely to be insolvent (see \Cref{tab:rho_v_by_run}), consistent with runs happening in the weakest banks.

\begin{figure}[ht]
\caption{\textbf{Share of Banks Subject to Run that Were Not Fundamentally Insolvent}}
\label{fig:run_solvent}
\centering

\includegraphics[width=0.9
\textwidth]{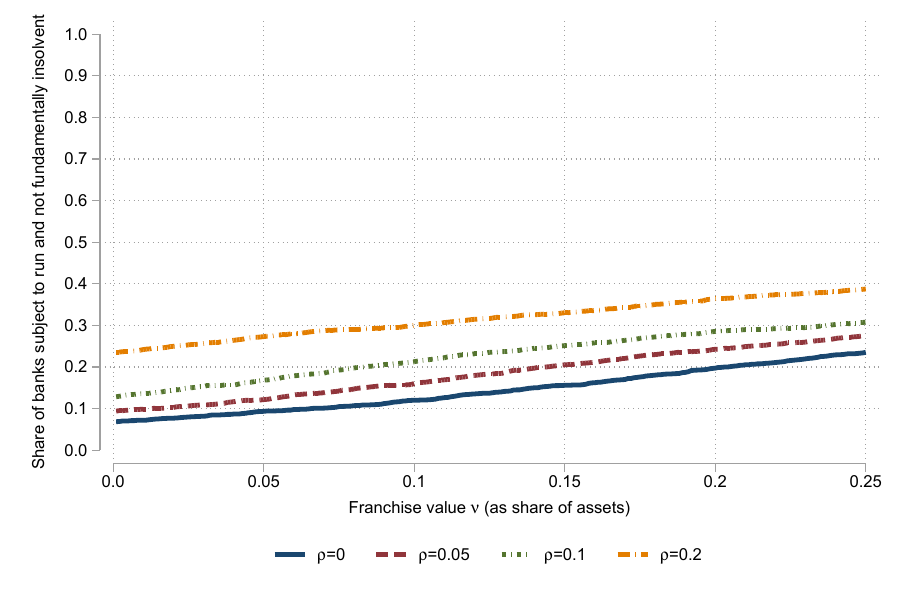}

\begin{minipage}{\textwidth}
\footnotesize
Notes:  This figure plots the share of banks that are subject to a deposit outflow of more than 7.5\% between their last call report and suspension and that are not fundamentally insolvent, defined as $$\frac{1}{N}\sum_b   \mathbb{I}\left[\frac{1+v}{1-\rho} \geq \frac{\ell_b}{R_b} \right] \times Run_b.$$ $\rho$ is the loss in receivership, and $v$ is the franchise value as a fraction of current book assets, $\ell$ is the ratio of debt to assets, $R$ is the recovery rate on assets, and $Run_b$ is an indicator variable that equals one for banks that experienced a decline in deposits between the last call report and failure of greater than 7.5\%. We restrict the sample to national banks placed in charge of receivers between 1880 and 1934 for which ``deposits at suspension'' is reported by the OCC.
\end{minipage}
\end{figure}

What is the share of fundamentally solvent banks that could have failed because of a run? To answer this question, we make use of two testable implications of bank run theories, discussed in \Cref{sec:conceptual}. First, for a bank run to represent a plausible cause of failure,  a failing bank needs to experience significant deposit outflows before failure. Second, the failed bank should not have been fundamentally insolvent.

\Cref{fig:run_solvent} presents the share of failed banks subject to a run that were \textit{not} fundamentally insolvent, i.e., $\frac{1+v}{1-\rho} \geq \frac{\ell_b}{R_b}$. As before, we define a bank as being subject to a run if it loses more than 7.5\% of its deposits immediately before failure. We report this share for various values of $\rho$ and $v$. 

For $\rho=v=0$, less than 8\% of bank failures could have plausibly been caused by a run on an \textit{ex ante} solvent bank. In the language of \cite{Goldstein2005} from \Cref{sec:conceptual}, less than 8\% of banks fail with $\theta \in (\underline{\theta},\theta^*)$. The higher the assumed potential value of the bank absent a run, the greater is the share of banks that could have failed because of the run. With $\rho=v=0.05$, this share rises to 13\%. For $\rho=v=0.1$, which would imply that value destruction because of failure itself exceeds 20\%, the share is 22\%. While the exact share of fundamentally solvent banks that failed because of a run depends on  assumptions about the potential value destroyed by the run, the calculations summarized in \Cref{fig:run_solvent} suggest that runs on otherwise solvent banks account for a modest, although not negligible, share of failures before the introduction of deposit insurance. Put differently, even when assuming that bank failure induced large value reduction, the majority of pre-FDIC failures involved either fundamentally insolvent banks or banks that were not subject to large deposit outflows.

\subsection{Cause of Failure Assigned by OCC} 

Our empirical evidence suggests that asset losses that drove banks to insolvency were an important driver of pre-FDIC bank failures. While runs were common in failing banks, in most cases, these runs were likely the trigger of failure of \textit{ex ante} insolvent banks, rather than the original cause of failure for otherwise solvent banks.

\begin{figure}[ht]
\caption{\textbf{Causes of Bank Failures as Classified by the OCC, 1863-1937}}
\label{fig:causes_of_failures}
\centering 
\includegraphics[width=0.7\textwidth]
{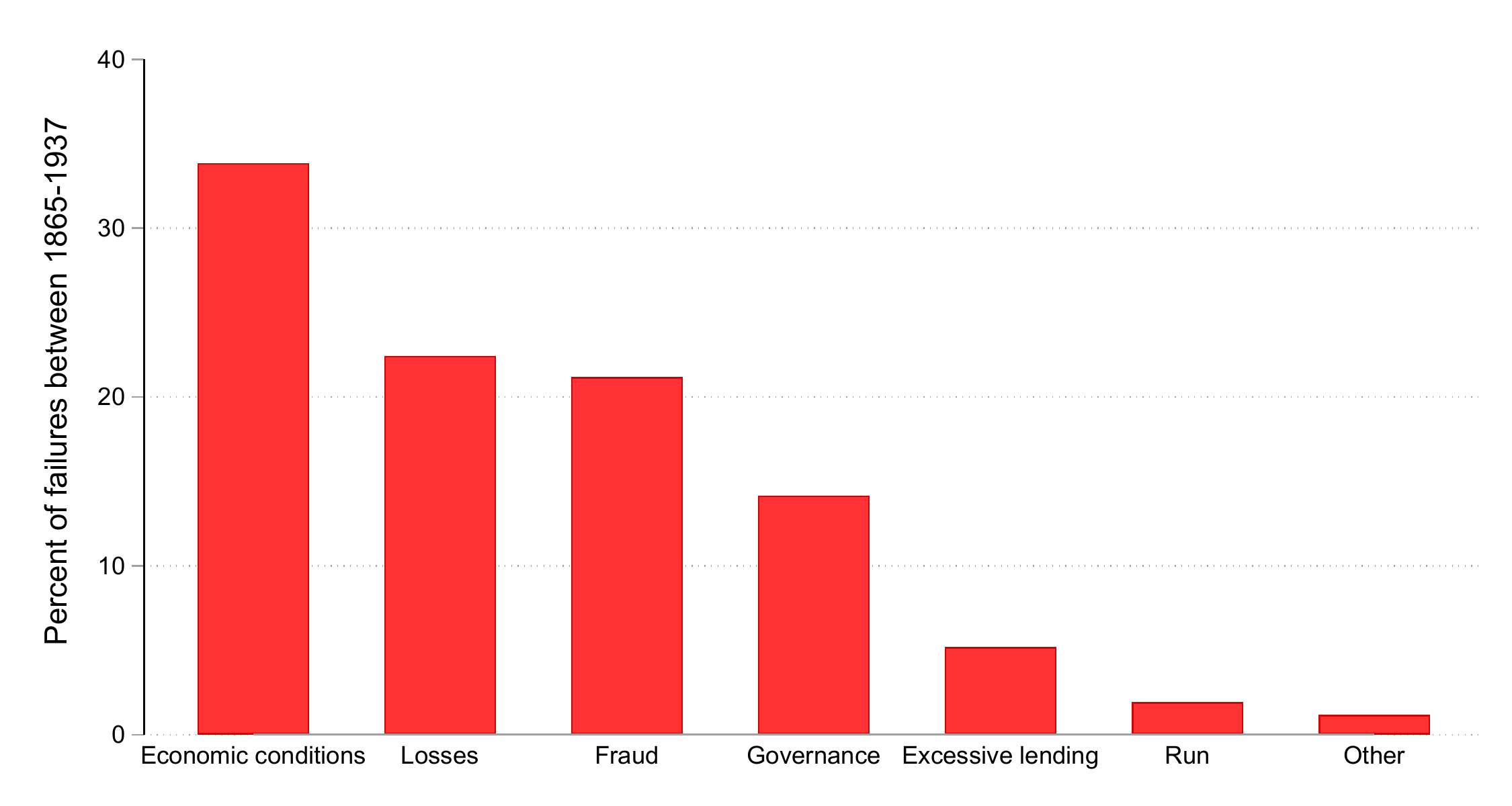}
\begin{minipage}{\textwidth}
\footnotesize
Notes: Causes of failure are as classified by the OCC in the tables of national banks in charge of receivers from the OCC annual report to Congress for various years. We categorize the detailed list of failure reasons as described in \Cref{sec:data_receiverships}. The classification of the causes of bank failures by the OCC is essentially complete for failures from 1863-1928, partially complete for failures from 1929-1931 and 1934-1937, and entirely missing for failures in 1932 and 1933 (see \Cref{fig:failure_reasons_classification_across_time}). 
\end{minipage}
\end{figure}

How does this interpretation align with contemporary views of pre-FDIC failures? As a final piece of evidence, \Cref{fig:causes_of_failures} reports the distribution of the ``cause of failure'' provided by the OCC for most failures that occurred between 1863 and 1937. We summarize the causes of failure assigned by the OCC into seven broad categories: economic conditions, excessive lending, losses, fraud, governance issues, run, and other factors.\footnote{Systematic classification of the cause of bank failures by the OCC is essentially complete for 1863-1928, but it is only partial for failures that occurred during 1929-1931 and is not available for banks that failed in 1932 or 1933. See \Cref{sec:data_receiverships} for details on the sample coverage and how we classify the detailed causes of failure into broad categories.} While the OCC classification may contain errors, it nevertheless provides insight into what examiners on the ground saw as the main cause of failure for each bank.

\Cref{fig:causes_of_failures} shows that the most common cause of failure provided by the OCC is ``economic conditions.'' This category includes failures attributed to deflation, crop loss, and a local financial depression. The second most common category is ``losses.'' The third most common category is ``fraud,'' which also often occurs in response to losses. Other common causes are ``governance issues'' and ``excessive lending,'' which refers to a bank with excessive exposure to one counterparty. The most common causes of failure are thus related to deteriorating asset quality and poor fundamentals. 

Bank runs, in contrast, account for less than 2\% of all failures according to the OCC, despite the preponderance of failures that involved large deposit outflows in this period. ``Run'' covers instances where the bank was closed by a run, heavy withdrawals, and a lack of public confidence, as well as cases where the bank was closed by directors in anticipation of a run or due to rumors of a run. Contemporary observers rarely saw runs as the principal cause of failure \citep[see also][]{Fed1936,Calomiris1991}.

%% file: output/tables/04_assets_in_failure_collected.tex
1863-1913 (NB Era)&0.45&0.15&0.46&0.34&0.05&0.01&533\\
1914-1918 (Early Fed)&0.48&0.06&0.49&0.41&0.02&0.01&732\\
1929-1933 (Depr., pre-Holiday)&0.50&0.07&0.39&0.52&0.02&0.01&1031\\
1933-1934 (Depr., post-Holiday)&0.65&0.01&0.09&0.69&0.21&0.00&605 

%% file: output/tables/04_assets_in_failure_all_collected.tex
All&0.52&0.07&0.36&0.49&0.07&0.01&2901 

%% file: output/tables/04_assets_in_failure_assessed.tex
1863-1913 (NB Era)&0.35&0.40&0.26&533\\
1914-1918 (Early Fed)&0.35&0.40&0.26&732\\
1929-1933 (Depr., pre-Holiday)&0.31&0.56&0.15&1031\\
1933-1934 (Depr., post-Holiday)&0.46&0.47&0.08&605\\

%% file: output/tables/04_assets_in_failure_all_assessed.tex
All&0.36&0.47&0.18&2901 

%% file: output/tables/07_predicting_recovery_rates_paper.tex
                    &         (1)   &         (2)   &         (3)   &         (4)   &         (5)   \\
\cmidrule(lr){1-6} \textit{Good} at suspension&        0.89***&               &               &               &               \\
                    &      (0.01)   &               &               &               &               \\
\textit{Doubtful} at suspension&        0.54***&               &               &               &               \\
                    &      (0.01)   &               &               &               &               \\
\textit{Worthless} at suspension&        0.08***&               &               &               &               \\
                    &      (0.02)   &               &               &               &               \\
Size (log(assets))  &               &        2.93***&               &               &               \\
                    &               &      (0.29)   &               &               &               \\
Surplus/equity      &               &               &       15.49***&               &               \\
                    &               &               &      (1.77)   &               &               \\
Noncore funding/assets&               &               &       -2.50   &               &               \\
                    &               &               &      (3.27)   &               &               \\
Loans/assets        &               &               &      -28.62***&               &               \\
                    &               &               &      (2.12)   &               &               \\
NPL/loans           &               &               &               &      -25.38***&               \\
                    &               &               &               &      (9.52)   &               \\
Q1 of loan growth (t-10,t-3) &               &               &               &               &       -1.38   \\
                    &               &               &               &               &      (1.15)   \\
Q2 of loan growth (t-10,t-3) &               &               &               &               &       -0.45   \\
                    &               &               &               &               &      (1.17)   \\
Q4 of loan growth (t-10,t-3) &               &               &               &               &       -0.73   \\
                    &               &               &               &               &      (1.18)   \\
Q5 of loan growth (t-10,t-3) &               &               &               &               &       -3.44***\\
                    &               &               &               &               &      (1.14)   \\
Q1 of loan growth (t-3, t-1) &               &               &               &               &       -7.34***\\
                    &               &               &               &               &      (1.10)   \\
Q2 of loan growth (t-3, t-1) &               &               &               &               &       -3.37***\\
                    &               &               &               &               &      (1.20)   \\
Q4 of loan growth (t-3, t-1) &               &               &               &               &       -2.10   \\
                    &               &               &               &               &      (1.34)   \\
Q5 of loan growth (t-3, t-1) &               &               &               &               &       -0.43   \\
                    &               &               &               &               &      (1.36)   \\
\cmidrule(lr){1-6} N&        2682   &        2710   &        2708   &         418   &        1986   \\
$R^2$               &       0.936   &       0.035   &       0.098   &       0.017   &       0.038   

%% file: output/tables/07_recovery_rho_v.tex
0\%&0.81&0.77&0.75&0.72&0.69&0.62&0.56\\
5\%&0.74&0.71&0.68&0.65&0.62&0.55&0.49\\
10\%&0.67&0.64&0.60&0.57&0.54&0.48&0.43\\
20\%&0.50&0.48&0.45&0.42&0.40&0.36&0.31

%% file: 09_conclusion.tex
%
%

\section{Discussion}
\label{sec:conclusion}

This paper studies failing banks using novel data spanning 1863-2024 on more than 37,000 banks in the United States. Throughout the sample, bank failures are preceded by gradually deteriorating solvency and increasing reliance on expensive noncore funding. Worsening fundamentals are often preceded by rapid lending growth and driven by a realization of credit risk. As a result, bank failures are highly predictable. Failures with bank runs, a common phenomenon before deposit insurance, are as predictable as failures without runs. Aggregate waves of bank failures are also predictable by weak bank fundamentals. Most pre-FDIC bank failures featured poor asset quality, with an average asset recovery rate of 52\%. 

Our findings have important implications for theories of bank failures and banking crises. We generalize the notion that weak bank fundamentals are a necessary condition for failure and show that non-fundamental runs on healthy banks are a rare cause of failure, even before deposit insurance. More than that, we find that most pre-FDIC bank failures involved \textit{ex ante} insolvent banks. Runs on weak but solvent banks likely account for only a modest share of failures, barring strong assumptions about the value destruction of receiverships. 

One explanation for why runs on solvent banks are an uncommon cause of failure is that private sector arrangements may often resolve liquidity shortages, even in the absence of government interventions. For instance, to the extent that interbank market participants can distinguish whether distress results from liquidity or solvency concerns, interbank lending can provide insurance against temporary liquidity pressures \citep{Blickle2022}. For the historical U.S. banking system, solvent banks facing runs could suspend convertibility until they could secure additional liquidity, often through arrangements such as clearinghouses, rather than engage in mutually destructive asset fire sales \citep{Gorton1985suspension,Gorton1985}. Therefore, while runs on healthy banks did occur, these generally did not trigger failure.

At the same time, our evidence emphasizes that runs on failing banks were frequent before the adoption of deposit insurance. Runs were likely an important mechanism for closing insolvent banks, determining the timing of many failures. Absent supervisory intervention, many pre-FDIC bank failures can be thought of as requiring a double trigger of both insolvency and a bank run that made a bank illiquid. Furthermore, our findings on the dynamics in failing banks point to the importance of a feedback loop in which solvency deterioration leads to increased reliance on expensive funding that further erodes solvency. These ``slow run'' dynamics are not captured by most bank run models.  Overall, our findings suggest that the \textit{solvency view} goes a long way to understanding the primary cause of bank failures, but the runnable nature of bank liabilities, central to the \textit{bank runs view}, can be important for understanding when and how banks fail.

Our evidence offers a nuanced perspective on the wide-ranging views of the role of depositor discipline \citep{CalomirisKahn1991,Diamond2001}. Before deposit insurance, failures involving large deposit outflows were common. This suggests that depositor behavior was often important for triggering failure. In contrast, in the modern era, deposit outflows are small, and insured deposits even flow into failing banks. This indicates important changes in the extent to which depositors discipline banks due to changes in regulation, as also suggested by \cite{PuriJF}. At the same time, large depositor losses and the significant predictability of bank failures also suggest that depositors are slow to react to information about bank fundamentals, potentially consistent with theories of inattention. Hence, depositor discipline before deposit insurance was imperfect.

Our results also speak to theories of banking crises based on asymmetric information. An older literature models bank runs as potentially arising from asymmetric information in which depositors may react to signals of poor fundamentals or to changes in beliefs triggered by noisy signals such as rumors  \citep[see, e.g.,][]{Gorton1988,Chari1988}. In this view, even solvent banks can fail if depositors, lacking full information, panic and withdraw funds based on rumors. Our evidence that most banks that fail have weak fundamentals and are insolvent, as well as the predictability of failures, suggests that runs triggered by asymmetric information are unlikely to result in failure but may be resolved through other mechanisms. 

Our results further inform research on the ``information view'' of financial crises \citep{DangGortonHolmstrom2012,Dang2020information}. In this view, banking crises happen when creditors revise their assessment of bank assets after receiving signals about the state of the banking system or the economy. These revisions, in turn, cause liabilities to become more fragile, as counterparties shorten maturities and demand collateral. Sufficiently bad signals trigger system-wide runs out of bank debt and into cash. Consistent with the information view, we find that bank failures and runs are preceded by losses that lead banks to have more fragile funding. However, our findings of significant predictability of aggregate waves of failure based on weak fundamentals suggest that information or behavioral frictions in belief formation may also be key to understanding the nature of financial crises.

Our findings emphasizing the central role of fundamentals and bank solvency have important policy implications. The predictability of bank failures implies a role for \textit{ex ante} interventions to prevent bank failures or mitigate their damage \citep{Gennaioli2018}. The fact that bank failures are predictable supports the active use of prompt corrective action measures, such as limiting dividend payouts and the use of non-core funding for poorly capitalized banks. More generally, our findings emphasize the importance of requiring financial intermediaries to be well-capitalized. Our findings also imply that \textit{ex post} interventions during a crisis must address fundamental solvency issues. When a crisis is driven by poor fundamentals, liquidity support without addressing insolvency is insufficient for mitigating the adverse consequences of banking sector distress, as argued by recent empirical and theoretical work \citep{Baron2024, Amador2024}.

Finally, our conclusions are subject to important caveats. Our empirical analysis focuses on bank failures. We do not study bank runs that do not ultimately result in bank failure. Panic-based runs could force otherwise healthy banks or banking systems to suspend convertibility of deposits into cash. Such suspensions can cause uncertainty and disrupt the payments system, with adverse real economic effects, even if no bank failures follow \citep{Sprague1910}. In short, while we emphasize the important role of fundamental insolvency for bank failures, our paper should not be read as saying that runs and panics do not matter for understanding the severity of banking crises.

%% file: 99_historical_context.tex
\clearpage
%
%

\section{Evolution of the U.S. Banking System and Bank Failures}
\label{sec:historical_context}

\begin{figure}[htpb]
\caption{\textbf{Failing Banks: 1863-2024}}
\label{fig:bank_failures_across_time_appendix}
\centering

  \includegraphics[width=0.9\textwidth]{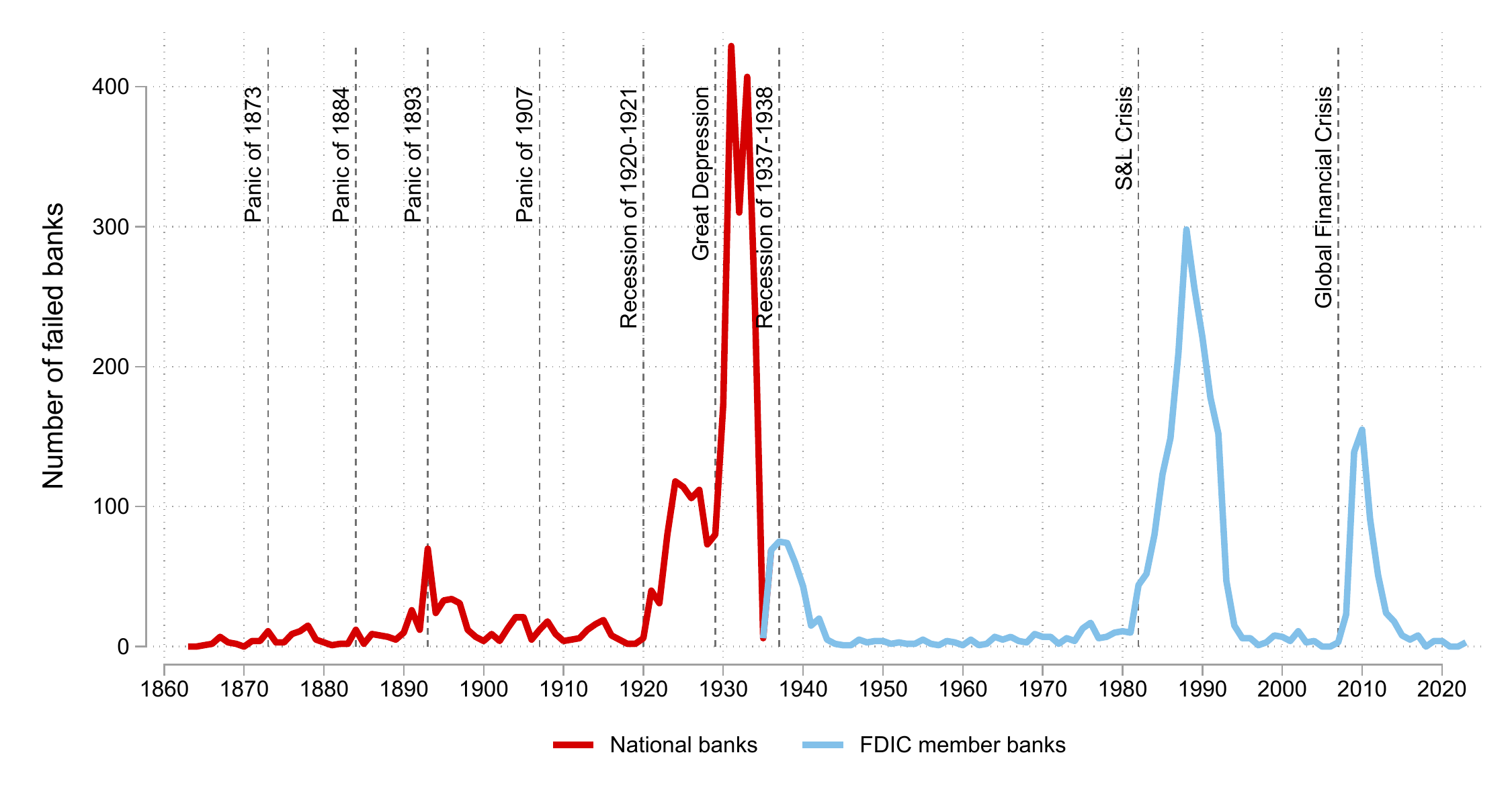}   

\begin{minipage}{\textwidth}
\footnotesize
Notes: This figure plots the number of failed banks by year.  Vertical lines indicate selected major banking crises and economic downturns. The red line plots the number of failing national banks, defined as national banks placed into receivership. The blue line plots the number of banks classified as failed by the FDIC. We restrict our sample of FDIC member banks to National Member Banks, State Member Banks, and State Nonmember Banks and exclude Savings Associations, Savings Banks, and Savings and Loans.
\end{minipage}

\end{figure}

This section charts the evolution of bank failures and banking regulation in the U.S. since 1863. \Cref{fig:bank_failures_across_time} in the paper shows the failure rate throughout our sample. \Cref{fig:bank_failures_across_time_appendix} plots the number of failures.  \Cref{tab:history} summarizes the key institutional and regulatory features by era.


\paragraph{National Banking Era} 
Our sample begins at the start of the National Banking Era, which spans the period between the Civil War and the founding of the Federal Reserve System, roughly 1863 to 1913. It was preceded by what is now referred to as the Free Banking Era, during which banks could charter under state laws after fulfilling a simple set of regulatory requirements \citep[see, e.g.][]{Rolnick1983}. Since the United States did not have a central bank for most of the nineteenth century, these state-chartered banks were the main issuers of notes in circulation. To be able to issue bank notes, banks had to cover their note issuance with purchases of state-issued government bonds. This changed during the Civil War, when the federal government needed to finance the war. To increase demand for federal government bonds, Congress passed two laws (the National Currency Act in 1863 and the National Banking Act in 1864) that allowed banks to be chartered under federal law—thus the name: national banks. Like state banks before them, national banks were allowed to issue bank notes when backed by government bonds. Currency issued by state banks was taxed at a high rate that encouraged banks to adopt national charters and purchase federal government bonds. The National Banking Act also established a regulatory and supervisory authority, the Office of the Comptroller of the Currency (OCC). The OCC published national bank balance sheets every year in an annual report to Congress, as discussed in the data section. 

Other than issuing currency, national banks operated very much as banks do today, by taking deposits and making loans. However, there was very little government interference. For instance, there was no insurance for deposits. Moreover, as there was no central bank, there was also no lender of last resort to help banks in a crisis. The Treasury performed quasi-central bank operations toward the end of the National Banking Era, but the interventions were small \citep{FriedmanSchwartz}. Thus, in this period, we can be reasonably confident that bank behavior was not driven by the anticipation of government support. Moreover, national banks were restricted to operating as unit banks, which meant that each bank could only operate a single branch serving a single location. Finally, capital regulation during the National Banking Era did not restrict the leverage ratio but reflected entry barriers \citep{Carlson2022}. At the founding of a bank, the bank charter would determine the dollar-amount of capital paid in to the bank with a minimum amount determined by the population of the bank's location. Thereafter, banks were largely able to choose their leverage freely subject to market conditions, though banks did face restrictions on dividend payouts based on their surplus. National banks were subject to double liability. In the event of failure, a receiver would levy an additional assessment on the bank shareholders' equal to the par value of subscribed capital to cover losses to depositors \citep{Grossman2010}. National banks were subject to double liability until 1937. National banks also faced portfolio restrictions limiting their capacity to lend against real estate collateral \citep{White1983}.

The National Banking Era witnessed a series of banking crises. The banking crisis chronology of \cite{Baron2021} records banking crises featuring widespread bank failures and panic-runs in 1873, 1884, 1890, 1893, and 1907.
For the National Banking Era, \Cref{fig:bank_failures_across_time} shows that the rate of failure of national banks was highest around the Panic of 1893.

While there was limited government intervention in response to banking panics and crises, banks responded to funding pressures and runs through various private sector arrangements. In many cases, banks would suspend convertibility into cash, either by fully or partially restricting withdrawals. Suspension was usually undertaken by a group of banks through local clearinghouse associations. In response to panics, clearinghouses would also halt the publication of individual bank balance sheets and instead publish an aggregate balance sheet for all members \citep{Gorton1985}. Further, clearinghouses would act as quasi-central banks by issuing loan certificates, a joint liability of all members, which banks could use in the clearing process. In certain panics (1893 and 1907), clearinghouses issued loan certificates directly to the public, expanding the money supply. Clearinghouses would also audit banks that were in distress to evaluate their solvency, and insolvent banks were expelled from the clearinghouse. Clearinghouse associations operating at the city level provided emergency liquidity to member banks during panics \citep{Jaremski2018}. The New York Clearinghouse (NYCH) lent to troubled banks throughout the country in certain cases \citep{Gorton1985}. The NYCH performed these functions effectively in the Panic of 1873, but not in later panics, as member banks struggled to internalize the collective interest of such interventions \citep{Wicker1996}.
 
\begin{table}
\caption{Evolution of the U.S. Banking System}
\label{tab:history}
\footnotesize
\centering
\input{historical_context.tex}

  \begin{minipage}{\textwidth}
  {\footnotesize
 Notes: *There was no deposit insurance for national banks until the founding of the Federal Deposit Insurance Corporation (FDIC) in 1933. However, some states had already implemented deposit insurance schemes for state-chartered banks before 1933 \citep[see][]{calomiris2019stealing}. ** Local branching was permitted for state banks in selected chartered states. National banks were not allowed to branch until the McFadden Act of 1927. This Act allowed national banks to branch in states in which state-chartered institutions were permitted to branch.} 
\end{minipage}
\end{table}

\paragraph{Early Federal Reserve \& Great Depression} The recurring financial crises of the National Banking Era led to the creation of a central bank through the Federal Reserve Act of 1913. The Federal Reserve could serve as a lender of last resort and had the responsibility to supervise member banks \citep[see, e.g.,][]{carlson2025young}. 

The McFadden Act 1927 liberalized restrictions on national banks. Before the Act, national banks were prohibited from opening branches. The Act allowed national banks to branch in states where state banks were permitted to branch, a step toward liberalization of geographic restrictions \citep[see, e.g.,][]{Rajan2016}. The McFadden Act also liberalized rules for Federal Reserve member banks to lend against real estate and expanded lending limits to single borrowers.  Moreover, the McFadden act rechartered the Federal Reserve into perpetuity, removing the risk that the charter would be revoked, as had occurred with the First and Second Banks of the United States.


The 1920s witnessed a rise in banking failures. Failures were concentrated in agricultural states and primarily affected small, rural banks. The rise in bank failures was driven by a sharp decline in agriculture and land prices in agrarian states, as well as rising urbanization that weakened the position of rural banks \citep{FriedmanSchwartz}. \Cref{fig:bank_failures_across_time} shows that the failure rate of national banks reached a new high in the 1920s, even before the Great Depression.

The Great Depression further exacerbated distress among banks, and several scholars have argued that the banking crisis, in turn, contributed to the severity of the downturn 
\citep[e.g.,][]{FriedmanSchwartz,Bernanke1983}. The wave of bank failures prompted a decades-long debate about whether failures were mainly liquidity-based due to depositor runs \citep{FriedmanSchwartz} or driven by fundamentals such as rising losses \citep{Calomiris2003a}. \citet{Richardson2009} exploit that the Atlanta and St. Louis Federal Reserve banks followed different lender of last resort policies and find that Fed liquidity reduced bank failures and boosted lending, pointing to a role for liquidity-based failures. This also highlights that lender of last resort facilities were not uniformly available, even with a central bank, especially as the discount window became stigmatized.


The Great Depression prompted a wave of banking reforms. Deposit insurance was introduced in 1933 and then made permanent in 1934 with the creation of the FDIC. State-level deposit insurance systems had existed before, but these became inoperative by Great Depression \citep{calomiris2019stealing}. However, state-level deposit insurance schemes did not apply to national banks. 
Great Depression banking reform also imposed a range of limits of banking activities \citep{Kroszner2014}. The Glass-Steagall Act 
prohibited commercial banks from engaging in investment banking activities (corporate bond and equity underwriting). It also imposed limits on interest rates that banks could pay on deposits, known as Regulation Q \citep{Gilbert1986}. 

The Great Depression brought an end to shareholder double liability. The Banking Act of 1933 allowed for the issuance of shares without double liability, and the Banking Act of 1935 allowed national banks to remove double liability from their shares in 1937 \citep{TuftsTufts2001}. Double liability was unpopular among bank shareholders following the high rates of failure during the Depression. It was also seen as ineffective in preventing bank failures and unnecessary with the advent of deposit insurance \citep{Grossman2001}.

\paragraph{Some notes on the Dual Banking System}
In our analysis in the main paper, we rely entirely on data on national banks for the period prior to the founding of the FDIC. As noted in the main text, the main reason for focusing on national banks is the availability of consistent records provided by the OCC  on both balance sheets and bank failures. However, it is important to highlight that the US banking system featured several types of financial institutions that were not chartered under federal law but state law. National banks always coexisted alongside state banks, trusts, and private banks, with the relative importance of each type of institution varying over time. 

\begin{figure}[htpb]
\caption{\textbf{Number of Banks by Type and the Market Share of National Banks}}
\label{fig:descriptives_nb_shares}
\centering

  \subfloat[Number of banks]{\includegraphics[width=0.7\textwidth]{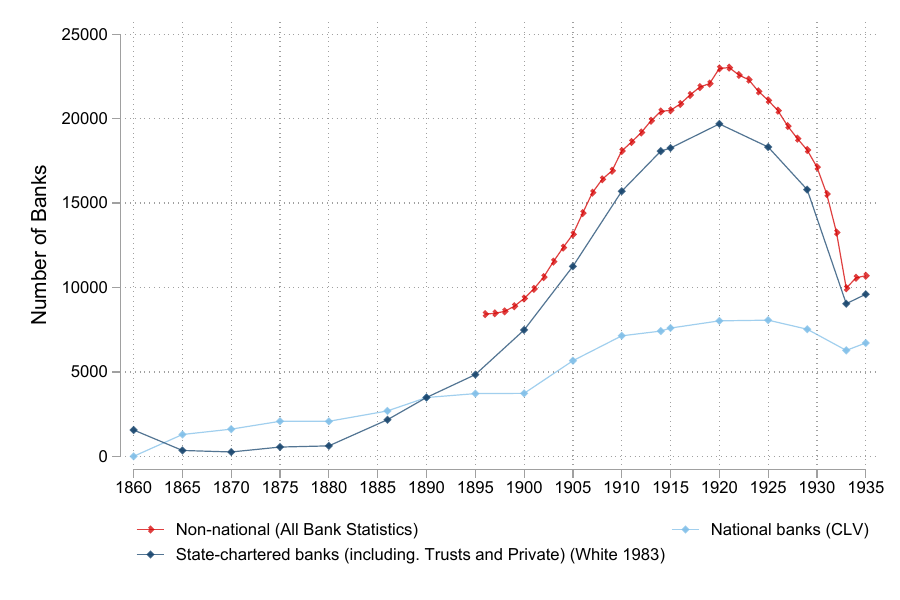}}

  \subfloat[Market share of national banks based on total assets]{\includegraphics[width=0.7\textwidth]
{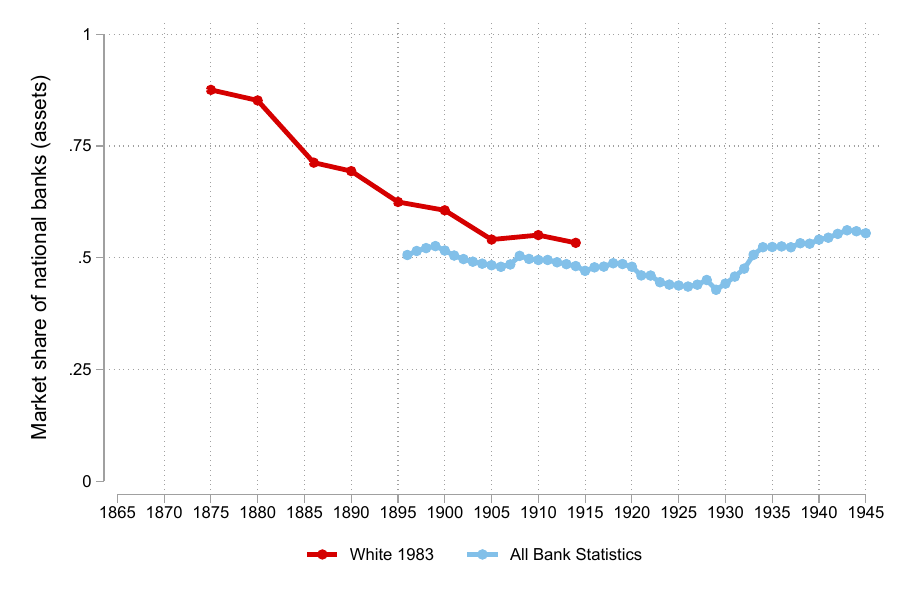}}

   \begin{minipage}{\textwidth}
\footnotesize
Notes: Data on state banks, trusts, and private bank are as indicated in the legend either taken from  \citet{White1983} and ``All Bank Statistics'' digitized by \citet{Flood1998}. State bank assets are available from 1875 onwards in \citet{White1983}; assets of trusts and private bank from 1886 onwards in ``All Bank Statistics''. Number of national banks in panel (a) are taken from the sample used in this paper.
\end{minipage}
\end{figure}

For instance, Figure \ref{fig:descriptives_nb_shares} plots the number of national banks and state-chartered institutions (panel (a)) and their market share of total banking assets (panel (b)). 
National banks had a very large  market share of the entire banking market ranging of around 80\%  at the onset of the National Banking Era. This large market share is related to the fact that the National Banking Act imposed a tax on notes issued by state banks, making state bank charters very unattractive. However, starting in the 1880s, the rise of deposits as form of money  slowly eroded the note-issuing advantage of national banks. Thus, over time the market share of national banks started to shrink, reaching 45\% by the 1930s. More generally, state-chartered institutions tended to be active in smaller markets in which national banks, which faced considerable stricter regulatory requirements, were not viable. Hence, state banks tended to be smaller in size, but there tended to be more state banks than national banks. This naturally implies that national banks tended to have larger, more financially sophisticated depositors than state banks. 

\begin{figure}[ht]
\caption{\textbf{Bank Failures and Suspensions by Bank Type}}
\label{fig:failures_and_suspensions}
\centering

{\includegraphics[width=1.0\textwidth]{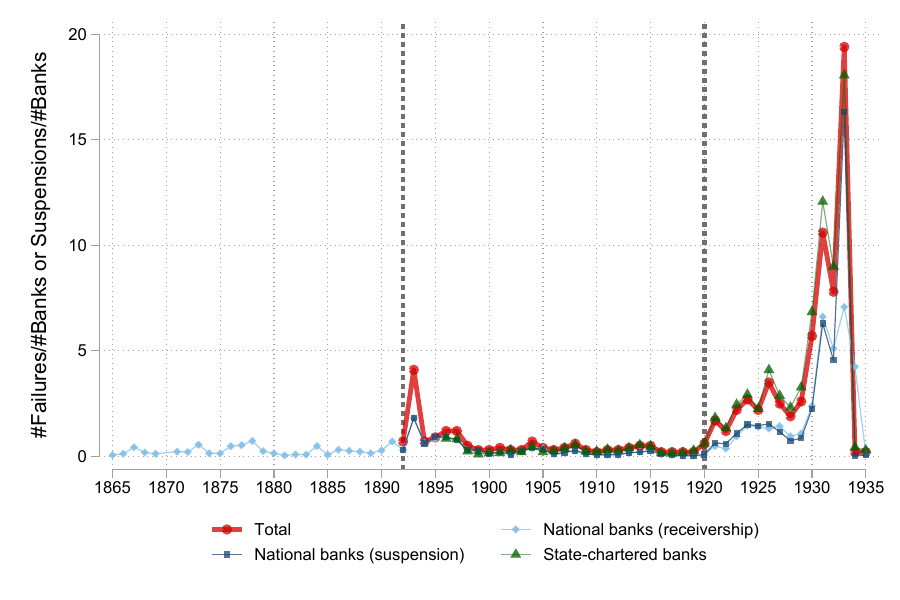}}

   \begin{minipage}{\textwidth}
\footnotesize
Notes: This figure combines various sources to construct a rate of bank failure from 1863 through 1935 for national banks, state-chartered banks, and all banks combined. From 1863-1891, we use the U.S. Comptroller of the Currency Annual Report list of failed banks. From  1892-1920,  the number of bank suspensions are from Chapter X of the ``Historical statistics of the United States, Colonial Times to 1957'' (available \hyperlink{https://www2.census.gov/library/publications/1960/compendia/hist_stats_colonial-1957/hist_stats_colonial-1957-chX.pdf}{here}). For 1921-1934, data are the Board of Governor's from ``Bank Suspensions, 1892 - 1935'' (available \hyperlink{https://fraser.stlouisfed.org/title/bank-suspensions-1892-1935-403}{here}).
Note that differences in sources between 1864 and 1892, 1892-1920, and 1921-1934  complicate comparisons across time. 
\end{minipage}
\end{figure}

While \Cref{fig:bank_failures_across_time} in the main text plots the failure rate (receivership) of national banks, \Cref{fig:failures_and_suspensions} plots the suspension and receivership rates for national banks and suspension rates for state-chartered institutions. Observe that before 1892, there is no reliable source of state bank suspensions and receivership. After 1892, it became possible to construct a series for both.  \Cref{fig:failures_and_suspensions}  shows that failure rates co-moved broadly, with state banks facing slightly higher failure rates than national banks. However, note that the counts of state-chartered institutions changed across sources (differing in the inclusion of trusts, mutual banks, and private institutions), making it \textit{de facto} impossible to construct a consistent time series of failure rates across all bank types. Hence, the levels of failure rates before and after 1920 are not comparable across time. 


\clearpage

\paragraph{Boring Banking} We refer to the era from 1959 through 1982 as the ``Boring Banking'' Era. The term ``Boring Banking'' is inspired by Paul Krugman, who wrote in the New York Times on April 9, 2009: ``Thirty-plus years ago, when I was a graduate student in economics, only the least ambitious of my classmates sought careers in the financial world. Even then, investment banks paid more than teaching or public service - but not that much more, and anyway, everyone knew that banking was, well, boring.''
During this era, failure rates were low. 
Banks' activities were restricted by the Depression-era regulations. Furthermore, the 1956 Bank Holding Company Act allowed states to restrict entry by out-of-state banks and holding companies, which effectively prohibited interstate banking. 
There was no explicit capital requirements. Instead, capital regulation was conducted by supervision, and supervision focused not just on capital ratios but on a broader range of quantitative and qualitative factors \citep{Haubrich2020}.

Rising inflation and interest rates led to outflows of deposits from commercial banks and into money market funds that were not subject to interest rate ceilings. This led to a phasing out of interest rate ceilings on deposits with the 1980 Depository Institutions Deregulation and Monetary Control Act \citep{Kroszner2014}.

\paragraph{Deregulation and Savings \& Loan (S\&L) Crisis} 

The period of low bank failure rates came to an end in the late 1970s as the number of bank failures started to edge up. Bank failures further increased throughout the 1980s. While the failures in the S\&L crisis were highest among thrifts, 
commercial banks also saw high failure rates during 1980s (see \Cref{fig:bank_failures_across_time}). The S\&L crisis is often dated to 1984 based on the failure of Continental Illinois, which represented the largest bank failure in U.S. history at the time. Failures in the 1980s were driven by a combination of high interest rates, the severe recessions over 1980-1982, losses on oil and gas loans, and losses from exposure to the Latin American debt crisis.


In response to rising bank failures and a trend of declining bank capital ratios discussed below, the 1980s witnessed the formal introduction of modern regulatory capital ratios that require a minimum amount of equity finance as a share of total assets. In the early 1980s, both the OCC and the Federal Reserve communicated a simple leverage ratio requirement of 5\% equity finance and noted that banks falling short of this cutoff would be considered undercapitalized. Both the Federal Reserve and the OCC published numerical capital ratios in 1981.  According to \citet{TarulloBasel}, the agencies  in effect adopted a minimum requirement of capital-to-assets of 5\%. The FDIC only published guidelines on ``minimum acceptable levels'' of primary capital. The original published requirements excluded the 17 largest banks (those with \$15B or more in assets) but by June 1983, these banks were also included. The International Lending Supervision Act (ILSA) of 1983 then formally required regulatory agencies to explicitly set capital ratios. By 1985, Federal Reserve, OCC, and the FDIC had formalized and published final regulations similar to those of the original 1981 guidelines. 
Rising inflation and interest rates led to outflows of deposits from commercial banks and into money market funds that were not subject to interest rate ceilings. This also led to a phasing out of interest rate ceilings on deposits with the 1980 Depository Institutions Deregulation and Monetary Control Act \citep{Kroszner2014}.

Following this period of formalizing capital regulation, capital requirements based on risk-weighted assets also became increasingly popular. In the 1950s, the Federal Reserve developed its Analyzing Bank Capital (ABC) model, which was an early method to construct a capital ratio based on risk-weighted assets. The S\&L Crisis also led Congress to pass the FDIC Improvement Act in 1991. A key provision of this Act was the introduction of Prompt Corrective Action (PCA), which requires supervisors avoid exercising forbearance and to take increasingly severe actions when a bank is deemed to be undercapitalized. 

At the same time, there was a move toward levelling the international playing field for banks. To this end, the Basel I accord was finalized in 1988 and implemented in the U.S. in 1991. Basel I introduced a minimum capital requirement of 8\% based on risk-weighted assets. Risk-weight varied from 0\% for supposedly risk-free assets such as cash up to 100\% for the most illiquid and risky forms of bank lending such as corporate debt. Further, to address various practical issues around the implementation of Basel I was revised and followed by Basel II in 2007. The Basel II framework left the overall required amount of capital unchanged but allowed for the possibility of banks opting into using their own internal risk models rather than the simple risk weights provided in Basel I. Moreover, Basel II attempted to address issues around off-balance sheet exposures that allowed for an effective circumvention of capital requirements. 

\paragraph{Global Financial Crisis and Beyond}
Finally, the Global Financial Crisis (GFC) initiated
additional drastic changes in regulation and supervision of financial institutions. Basel III and the Dodd-Frank Act led to both more stringent and more complicated capital requirements. Capital ratios were increased relative to Basel II and the definition of what constitutes capital was tightened. Capital requirements became differentiated by bank, with the tighter requirements for the largest banks, the Global Systemically Important Banks. Basel III also introduced a capital conservation buffer, limiting bank payouts when capital falls close to the minimum capital ratios, and a counter-cyclical capital buffer (CCyB), which is set at the discretion of the Board of Governors of the Federal Reserve. 

The aftermath of the GFC also saw the rise of stress testing. A stress test assesses whether banks are sufficiently capitalized to withstand adverse scenarios. Effectively, the stress test represents a form of a forward-looking, bank-specific capital requirement.  In early 2009, at the height  of the crisis, the Supervisory Capital Assessment Program (SCAP)---subsequently replaced by the Comprehensive Capital Analysis and Review (CCAR)---represented the first stress
testing effort. SCAP aimed to ensure that the 19 largest banks had sufficient capital coming
out of the crisis to absorb losses under poor economic conditions. The Dodd-Frank Act formalized regular stress tests for the largest banks (DFAST) in 2013. Under CCAR, each bank must propose a capital distribution plan incorporated into the stress test, whereas DFAST uses a standardized capital distribution plan that holds
dividends at their current level and sets net repurchases to zero. DFAST also requires banks to run
(and disclose) stress tests using the same set of inputs (i.e., the Fed’s scenarios and the standard
capital distribution plan) but with their own, internally developed model.

%% file: historical_context.tex
\begin{tabular}{l|c|c|c|c|c}
  \toprule
  
  Era & Years & \shortstack{Deposit \\ insurance} & \shortstack{Central \\ bank} & Capital regulation & \shortstack{Geographic \\ restrictions} \\[.1cm] \midrule 
  National Banking Era & 1863-1913 & No & No & \$ by pop  & Unit-branch** \\[.1cm] 
  Early Federal Reserve & 1914-1928 & No* & $\checkmark$ & \$ by pop & Unit-branch** \\[.1cm]
  Great Depression & 1929-1935 & No* & \checkmark & \$ by pop & Local branching \\[.1cm]
  Boring Banking & 1959-1982 & $\checkmark$ & $\checkmark$ &  Supervisory Discretion& Local branching  \\[.1cm]
  Deregulation and S\&L & 1982-2006 & $\checkmark$ & \checkmark & \shortstack{Leverage ratio in 1985 \\ Basel I in 1989 }& Limited until 1994  \\[.1cm]
  Global Financial Crisis & 2007-2015 & $\checkmark$ & $\checkmark$ &  Basel II/III + DFAST & No \\[.1cm]
  Post-crisis & 2015- & $\checkmark$ & $\checkmark$ &  Basel II/III + DFAST  & No \\
  \bottomrule
\end{tabular}

%% file: 99_appendix_figures.tex

\clearpage
\section{Additional Results}

\subsection{Additional Results on Dynamics in Failing Banks} 

This subsection discusses additional results and robustness of the findings in \Cref{sec:facts}.

\paragraph{Additional results on declining solvency} The patterns in \Cref{fig:losses_post} suggest that failures are mainly associated with realized credit risk. How does the net interest margin (NIM) evolve in the run-up to failure? In Appendix \Cref{fig:II_IE_NIM} we show that failing banks see both rising interest income (indicating higher risk taking) \textit{and} rising interest expenses (in line with higher reliance on expensive forms of funding, as discussed below).

Further note that restricting our sample to the 1970s and 1980s and thus the period of the Volcker shock, we do not find evidence that failing banks experienced deteriorating net interest margins. This is consistent with \cite{Wright1996FRB}, who find that \textit{thrifts} saw falling NIM in early 1980s, while commercial banks had much more stable NIM.

It is important to highlight, however, that the resulting stable NIM does not suggest that the realization of interest rate risk cannot also be a source of trouble for failing banks. Making inference from NIM to interest rate exposure is problematic for various reasons \citep[see, e.g.,][]{begenau2015banks,begenau2022unstable}. While we believe our evidence is indicative that realized credit risk in the recent history of the U.S. has been a common cause of failure, we do not preclude that realized interest rate risk is not also a potentially important driver of bank failures. 

\begin{figure}[ht]
\caption{\textbf{Funding of Failing Banks: Outcomes in Natural Logarithms}}
\label{fig:funding_levels}
\centering

  \subfloat[1863-1934: Deposits and Noncore Funding]{\includegraphics[width=0.65\textwidth]
{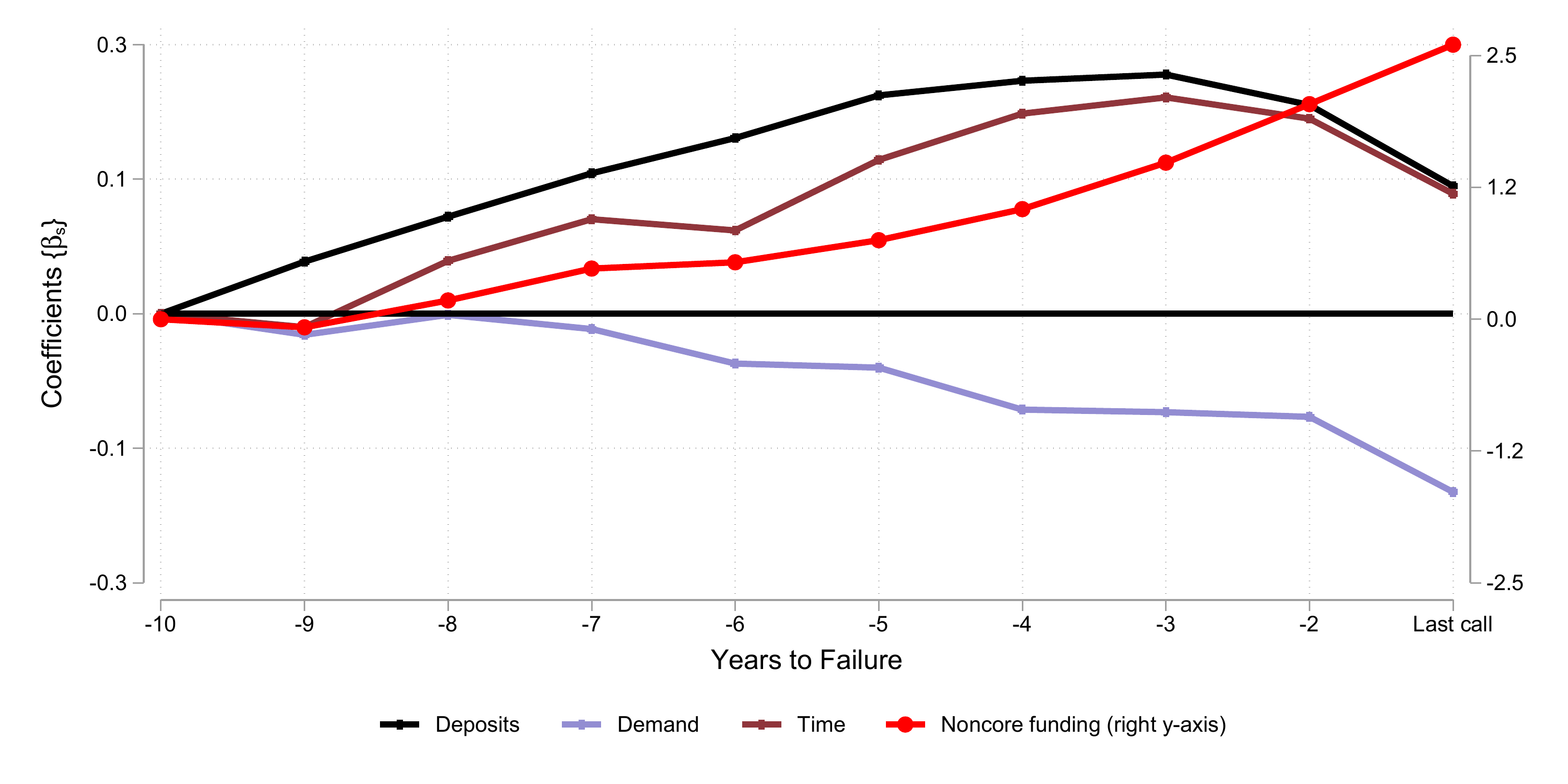}}

  \subfloat[1959-2024: Time, Demand, and Brokered Deposits, and Wholesale Funding]{\includegraphics[width=0.65\textwidth]{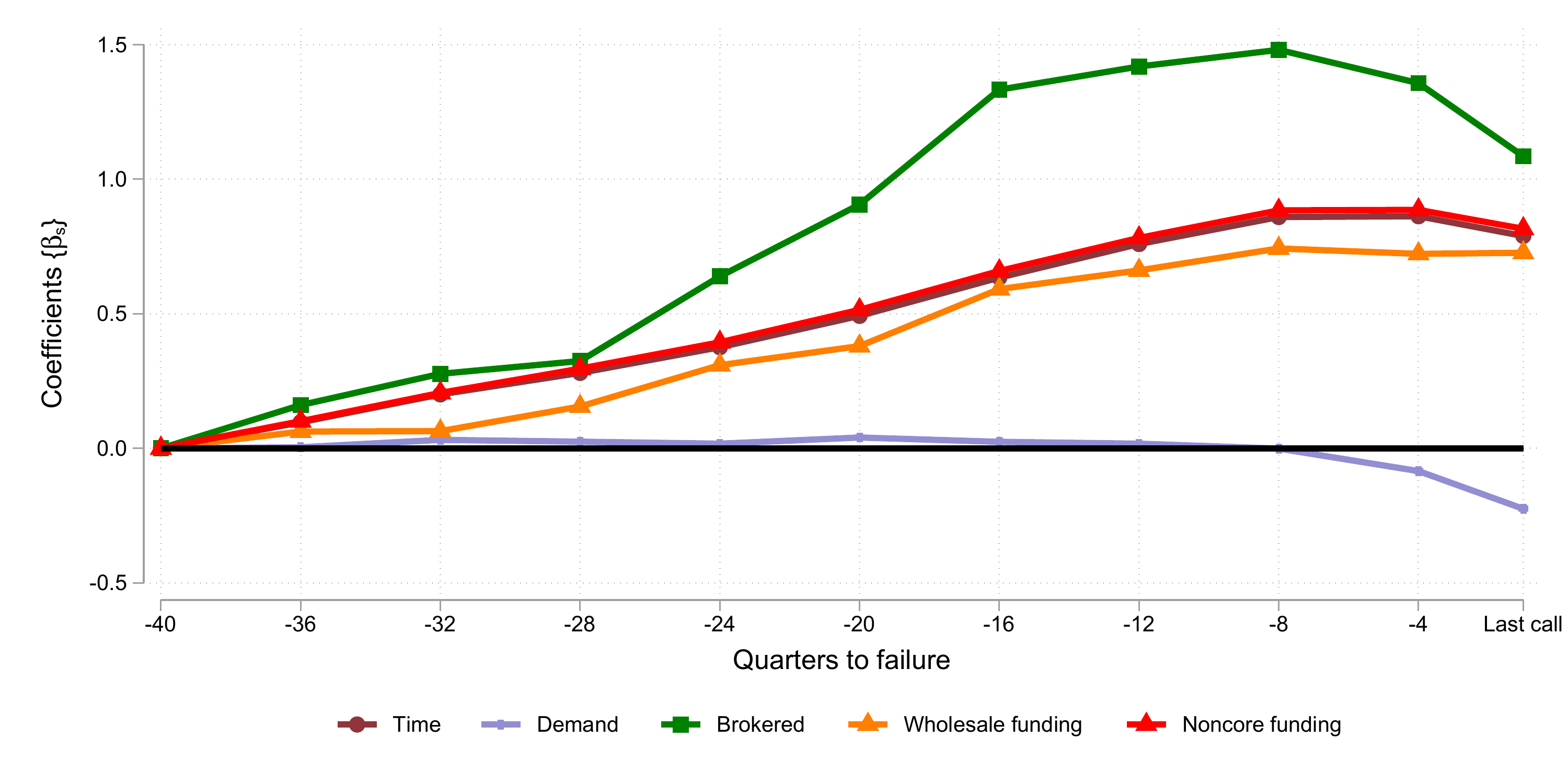}}
  %

   \begin{minipage}{\textwidth}
   \footnotesize
Notes: The figure shows the sequence of coefficients from a regression of the following form:
\[y_{b,t} = \alpha_b + \sum_{j=-10,j\neq -10}^{0} \beta_j \times \mathbf{1}_{j=t} +\epsilon_{b,t}  \] 
where $y_{bt}$ is the natural logarithm of the line item indicated in the figure legends and $\alpha_b$ is a set of bank fixed effects. The sample is restricted to failing banks and to the ten years before they fail. In panel (a), the sample is restricted to data from 1863 though 1934 and in panel (b) to data from 1959 through 2024. 

In panel (a) noncore funding is defined as the  total assets net of total deposits (including unsecured interbank deposits), equity, and national bank notes.  In panel (b),  noncore funding is the sum of wholesale funding and time deposits, where wholesale funding is the amount reported in the call report line item ``other borrowed money'' which pools various sources of bank wholesale funding, such as advances from Federal Home Loan Banks (FHLBs), other types of wholesale borrowings in the private market, and credit extended by the Federal Reserve. Note that for the historical sample, we do not include time deposits in noncore funding, as this item is only reported separately during 1915-1928.

\end{minipage}
\end{figure}

\paragraph{Additional results on funding dynamics}  Appendix \Cref{fig:funding_levels} presents the dynamics of liabilities in logs, as opposed to as a share of assets. The figure shows that, in both the historical and modern sample, wholesale funding also rises at a similar pace in percentage terms as other forms of noncore funding. However, the rise is from a lower initial share of assets.

\paragraph{Additional results on asset boom and bust}

\begin{figure}[htpb]
\caption{\textbf{Asset Growth in Failing Banks for Failures Occurring during Financial Crises versus Normal Times}}
\label{fig:dynamics_inside_outside_crises}
\centering
{\includegraphics[width=0.85\textwidth]
{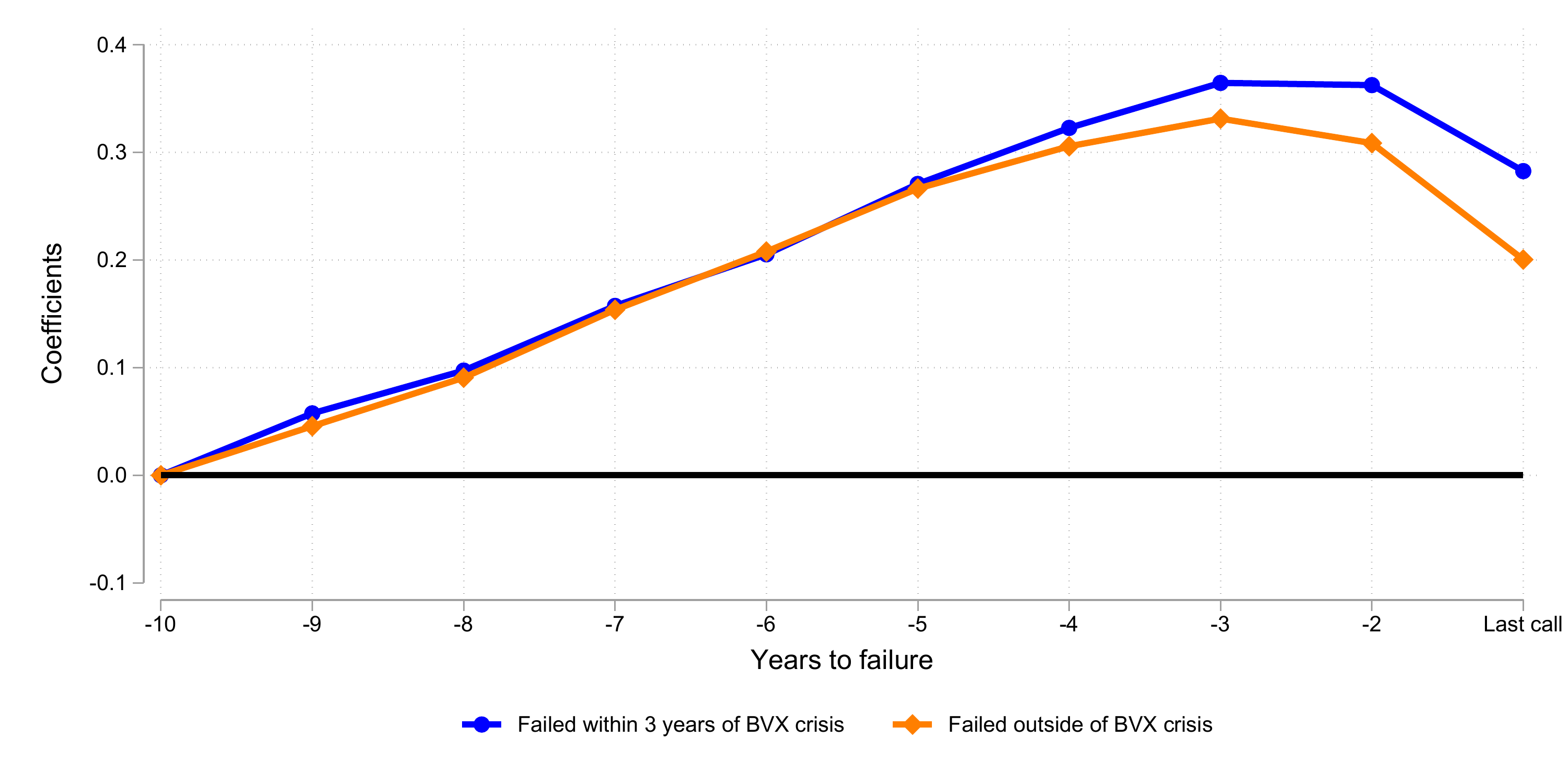}}
\begin{minipage}{\textwidth}
\footnotesize
Notes: Both panels show the sequence of coefficients from a regression of the following form:
\[y_{b,t} = \alpha_b + \sum_{j=-10,j\neq -10}^{0} \beta_j \times \mathbf{1}_{j=t} +\epsilon_{b,t}  \] 
where $y_{bt}$ is either bank $b$' assets, deposits, or loans and $\alpha_b$ is a set of bank fixed effects. The sample is restricted to failing banks only and to the ten years before they fail. Financial crises are defined according to \citet{Baron2021}


\end{minipage}
\end{figure}

\begin{figure}[ht]
\caption{\textbf{Failure Probability in the Cross-Section of Asset Growth}}
\label{fig:nonmonotonic}
\centering
\includegraphics[scale=1.0]{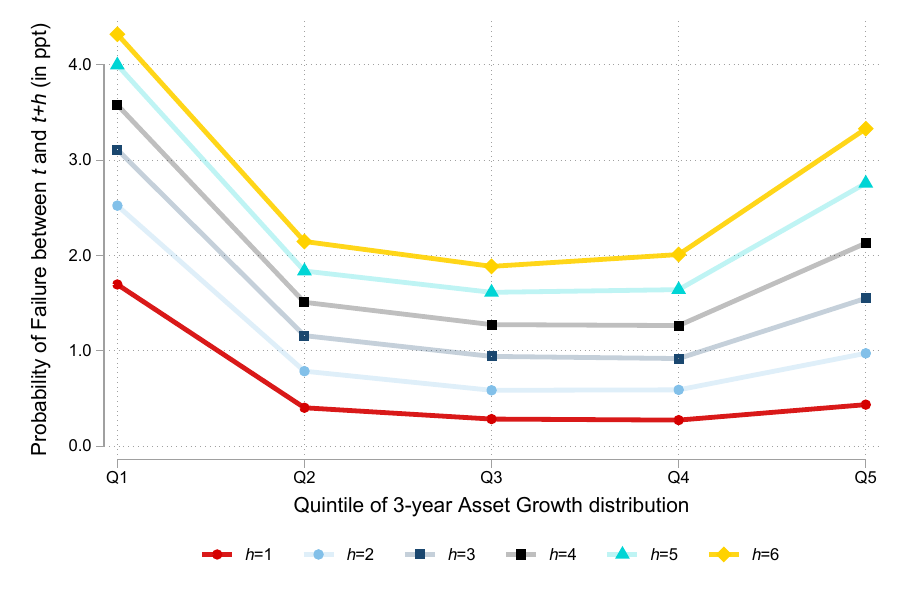}
\begin{minipage}{\textwidth}
\footnotesize
Notes: This figure plots the frequency of failure at the one to six year horizons across quintiles of the three-year asset growth distribution.  Appendix \Cref{fig:nonmonotonic_pre_post} shows this figure separately for the pre- and post-FDIC samples.
\end{minipage}
\end{figure}

The boom-bust pattern in \Cref{fig:assets_failing_banks} is not simply driven by the fact that bank failures are more common at the end of a boom-bust cycle. First, the boom-bust pattern is similar for banks failing outside of major banking crises (see \Cref{fig:dynamics_inside_outside_crises}). Second, rapid asset growth predicts subsequent failure in the cross-section of banks (see \Cref{fig:nonmonotonic}). In contrast, at short horizons, banks with lowest growth are most likely to fail. The relation between asset growth and future failure is stronger in the 1959-2024 sample. For the historical sample, there is a strong relation between low growth and failure within one to three years, but a weaker relation between rapid growth and failure in five to six years (see Appendix \Cref{fig:nonmonotonic_pre_post}).

\Cref{fig:assets_failing_banks_subsamples} shows the estimates across finer subsamples. Asset growth prior to failure is especially large in the period leading up to the 2008 Global Financial Crisis, followed by the 1959-1981 and 1982-2006 periods. There are several potential explanations for why the boom-bust pattern has become stronger in the modern era. First, in the historical period, bank expansions were constrained by geographic restrictions, limiting the growth of individual banks. Second, in recent decades, banks have greater access to more elastic non-core sources of funding, such as brokered deposits and funding in the Eurodollar market.\footnote{Accounts of major bank failures in the 1970s and 1980s begin to stress rapid growth financed by non-core funding as an important factor. For example, Franklin National Bank of New York and Continental Illinois were both the largest bank failures to date at the time of their failures. These banks both underwent rapid growth financed by wholesale funding, especially from the Eurodollar market \citep{Continental}.} Third, in the historical period, national banks faced restrictions on lending against real estate, making them less exposed to real estate booms and busts, an important driver of large lending booms. Finally, the anticipation of government interventions and deposit insurance after the Great Depression may have increased risk-taking \citep{calomiris2019stealing}.

\begin{figure}[htbp]
\caption{\textbf{Liquid and Illiquid Assets in Failing Banks}}
\label{fig:illiquid_liquid}
\centering

    \subfloat[1863-1935 ]{\includegraphics[width=0.7\textwidth]{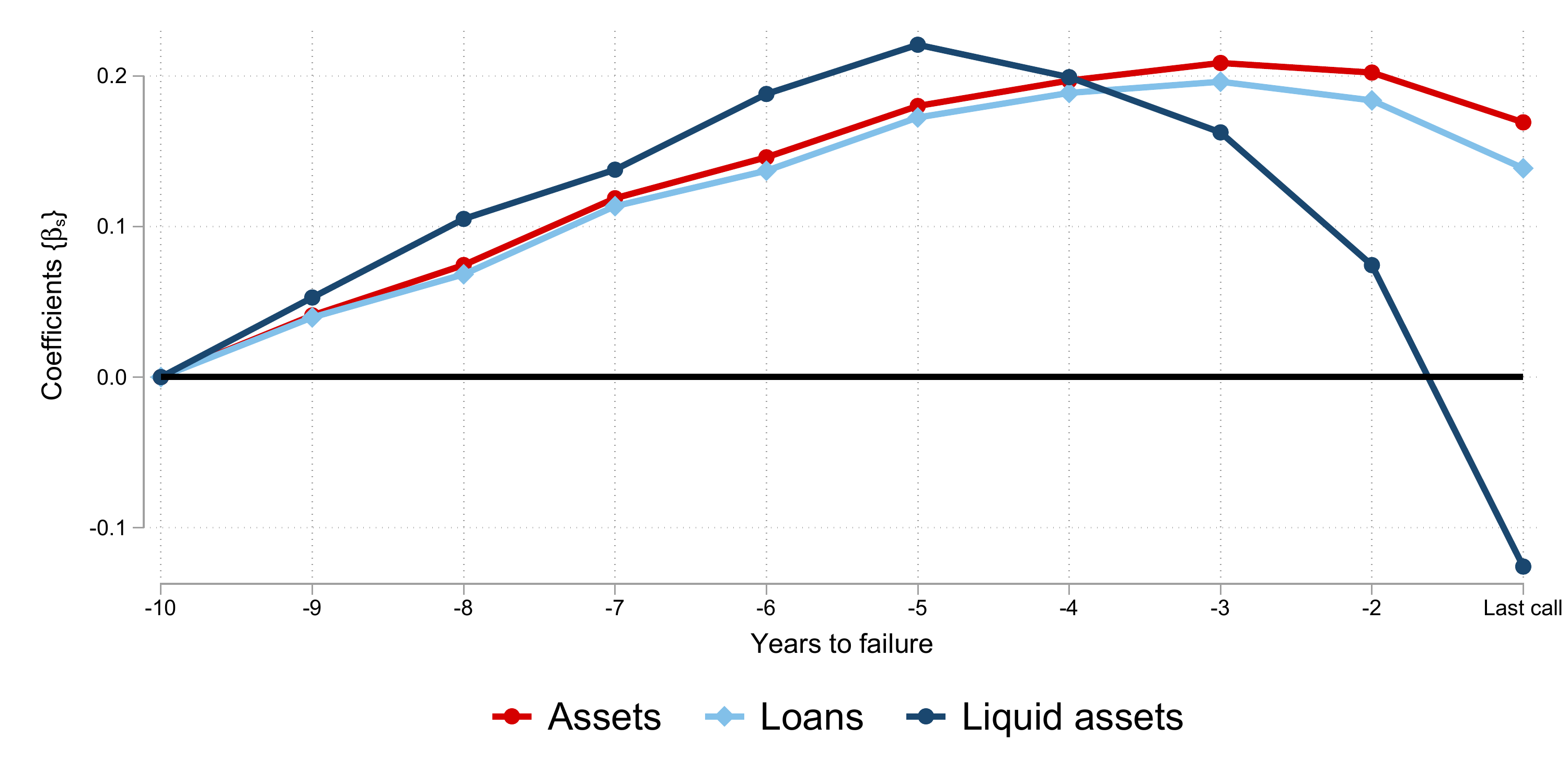}}
    
  \subfloat[1959-2024]{\includegraphics[width=0.7\textwidth]{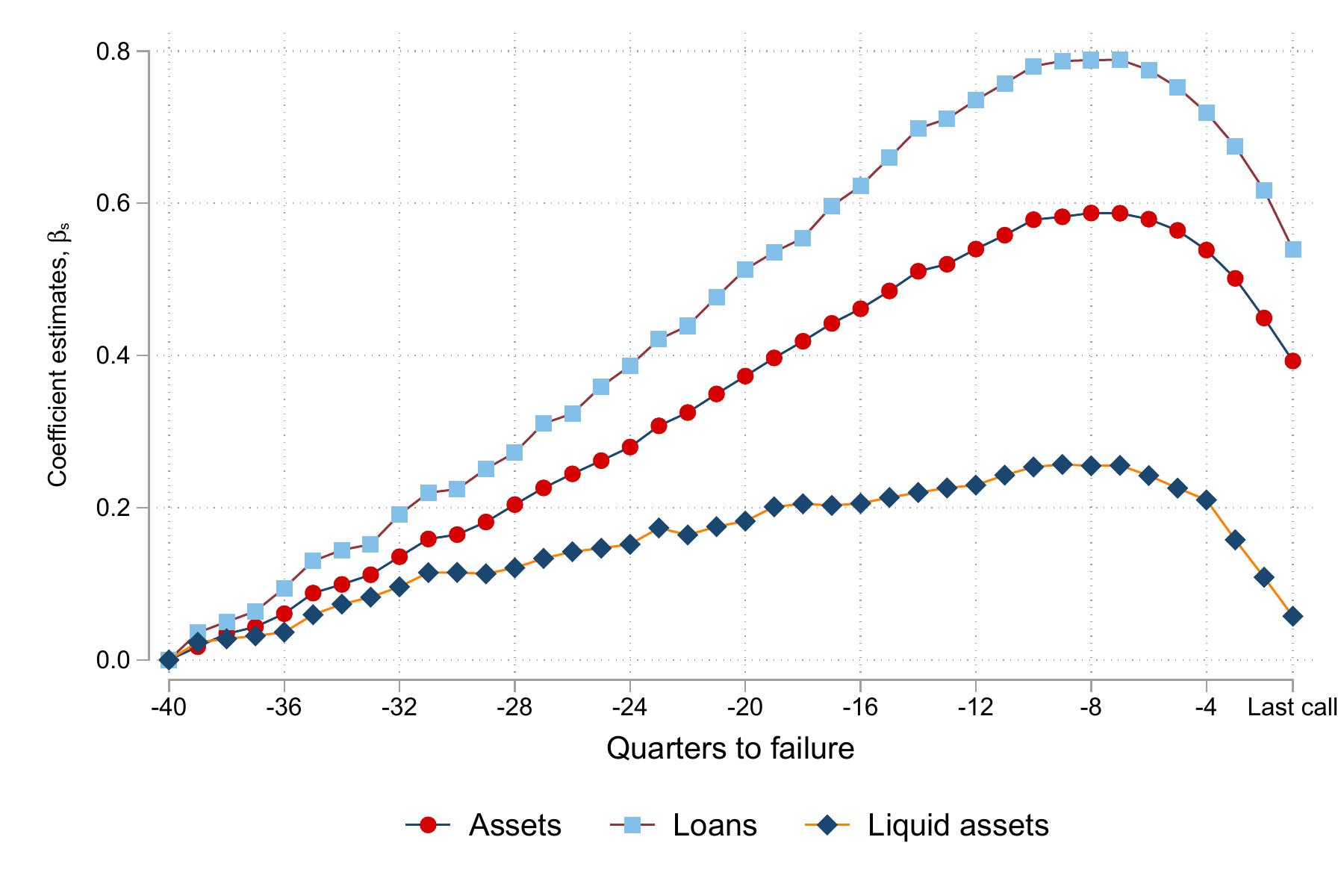}}
   
\begin{minipage}{\textwidth}
\footnotesize
Notes: This figure plots the sequence of coefficients from estimating \Cref{eq:dynamic}  with the logarithm of either assets, loans, or liquid assets (all deflated by the CPI) as the dependent variable for different samples. The specification includes a set of bank fixed effects. The sample is restricted to failing banks and to the ten years before they fail. From 1863 through 1941, we define liquid assets as the sum of currency, checks, legal  tender, interbank claims, bonds to secure deposits and bonds on hand, and bills of national banks and state banks, as well as FRS reserves after 1913. From 1959 onwards, liquid assets are defined as currency and reserves held, balances with other banks, cash items in collection, and security holdings (both government-issued and private label).
\end{minipage}
\end{figure}

\begin{figure}[htpb]
\caption{\textbf{Asset Growth in Failing Banks is Driven by Real Estate and C\&I Lending}}
\label{fig:loans}
\centering
\includegraphics[width=0.8\textwidth]{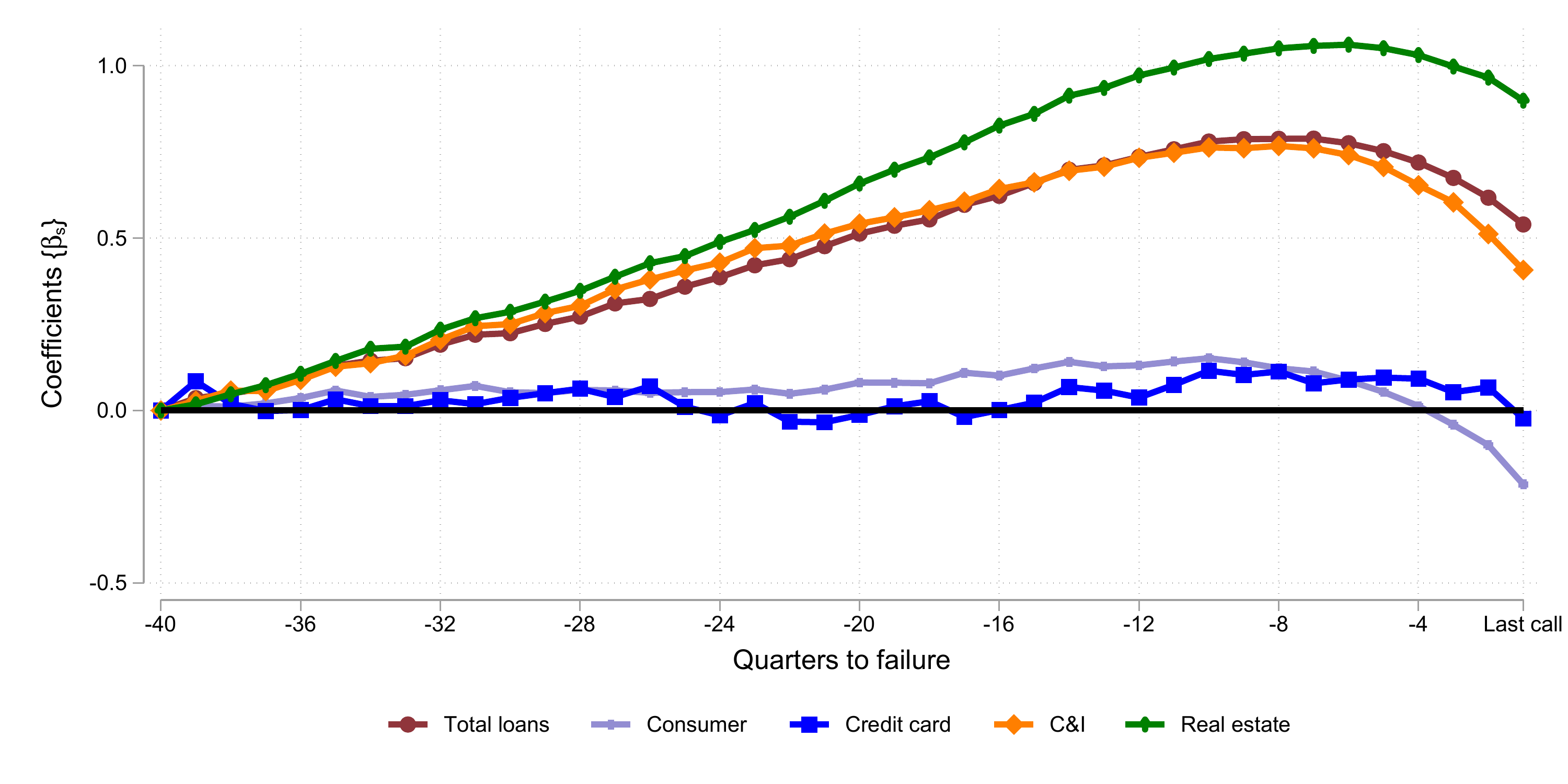}
   \begin{minipage}{\textwidth}
\footnotesize
Notes: This figure presents the sequence of coefficients from a regression of the following form
\[y_{b,t} = \alpha_b + \sum_{j=-10,j\neq -10}^{0} \beta_j \times \mathbf{1}_{j=t} +\epsilon_{b,t},  \] %
where $y_{bt}$ is a type of bank loan. The sample is restricted to failing banks and to the ten years before they fail. The estimates are based on the post-1959 sample. Data on loan types is not available for the pre-1935 sample. 
\end{minipage}
\end{figure}

\Cref{fig:illiquid_liquid} shows that rapid asset growth is concentrated in illiquid loans. In contrast, liquid assets such as cash and securities rise more slowly than total assets. An implication of the rapid credit expansion in failing banks is that their asset holdings tilt increasingly towards illiquid loans that are associated with higher credit risk in the decade before failure. For the modern sample, \Cref{fig:loans} shows that failing banks see the strongest boom in real estate lending (loans secured by real estate), followed by C\&I lending. On the other hand, credit card and consumer lending are flat in real terms in the run-up to failure.


\clearpage

\begin{figure}[ht]
\caption{\textbf{Interest Income, Expenses and NIM: 1959-2024}}
\label{fig:II_IE_NIM}
\centering
\includegraphics[width=0.8\textwidth]{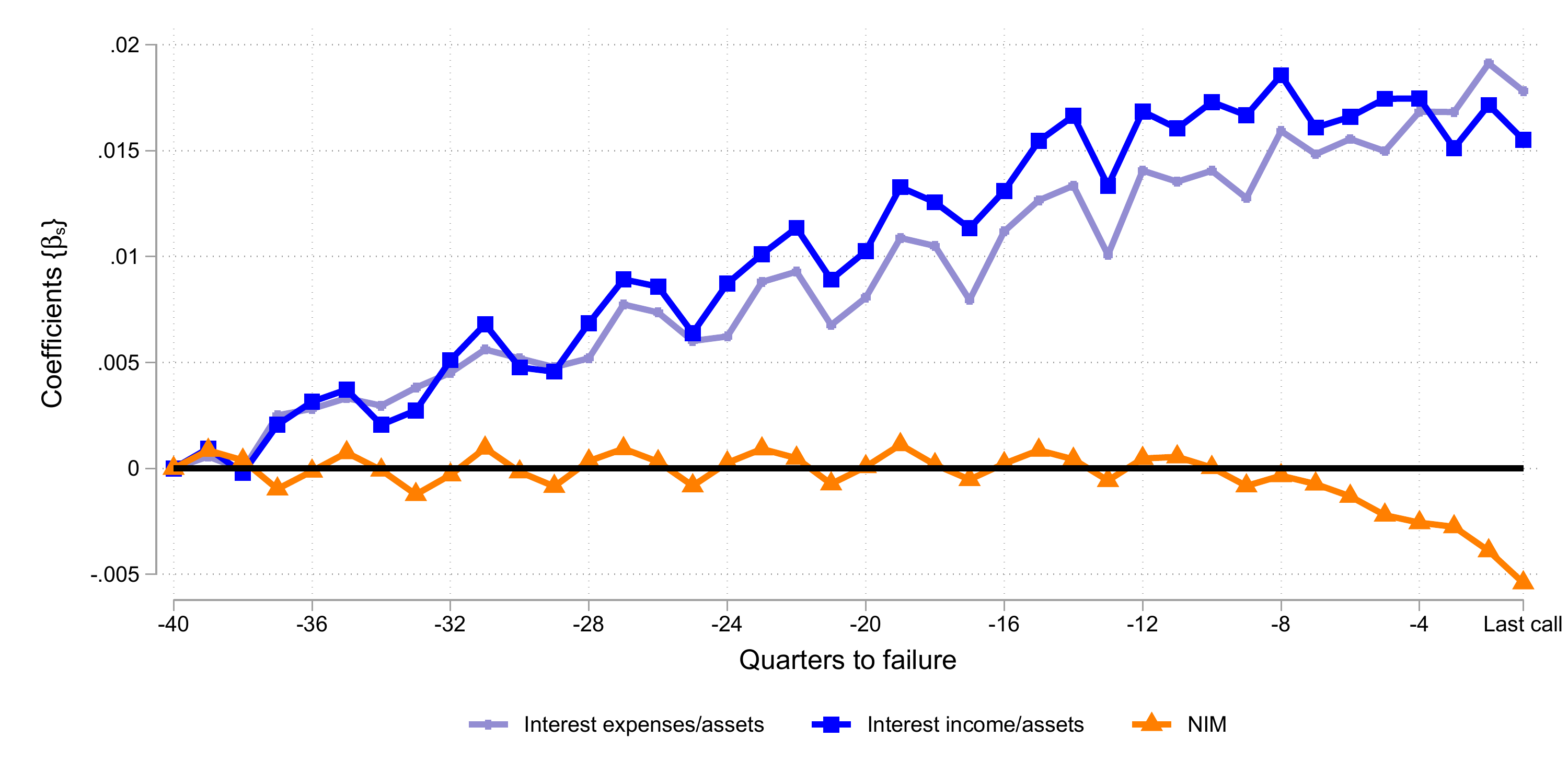}
\begin{minipage}{\textwidth}
\footnotesize
Notes: The figure shows the sequence of coefficients from a regression of the following form:
\[y_{b,t} = \alpha_b + \sum_{j=-10,j\neq -10}^{0} \beta_j \times \mathbf{1}_{j=t} +\epsilon_{b,t}  \] 
where $y_{bt}$ is the ratio indicated in the figure legends, and $\alpha_b$ is a set of bank fixed effects. The sample is restricted to failing banks and to the ten years before they fail and banks that fail after 1959. The net interest margin (NIM) is defined as the difference between total interest income net of interest expenses normalized by total assets. \end{minipage}
\end{figure}

\begin{figure}[htpb]
\caption{\textbf{Assets in Failing Banks: 1863-2024, By Historical Subsamples}}
\label{fig:assets_failing_banks_subsamples}
\centering

\includegraphics[width=0.9\textwidth]{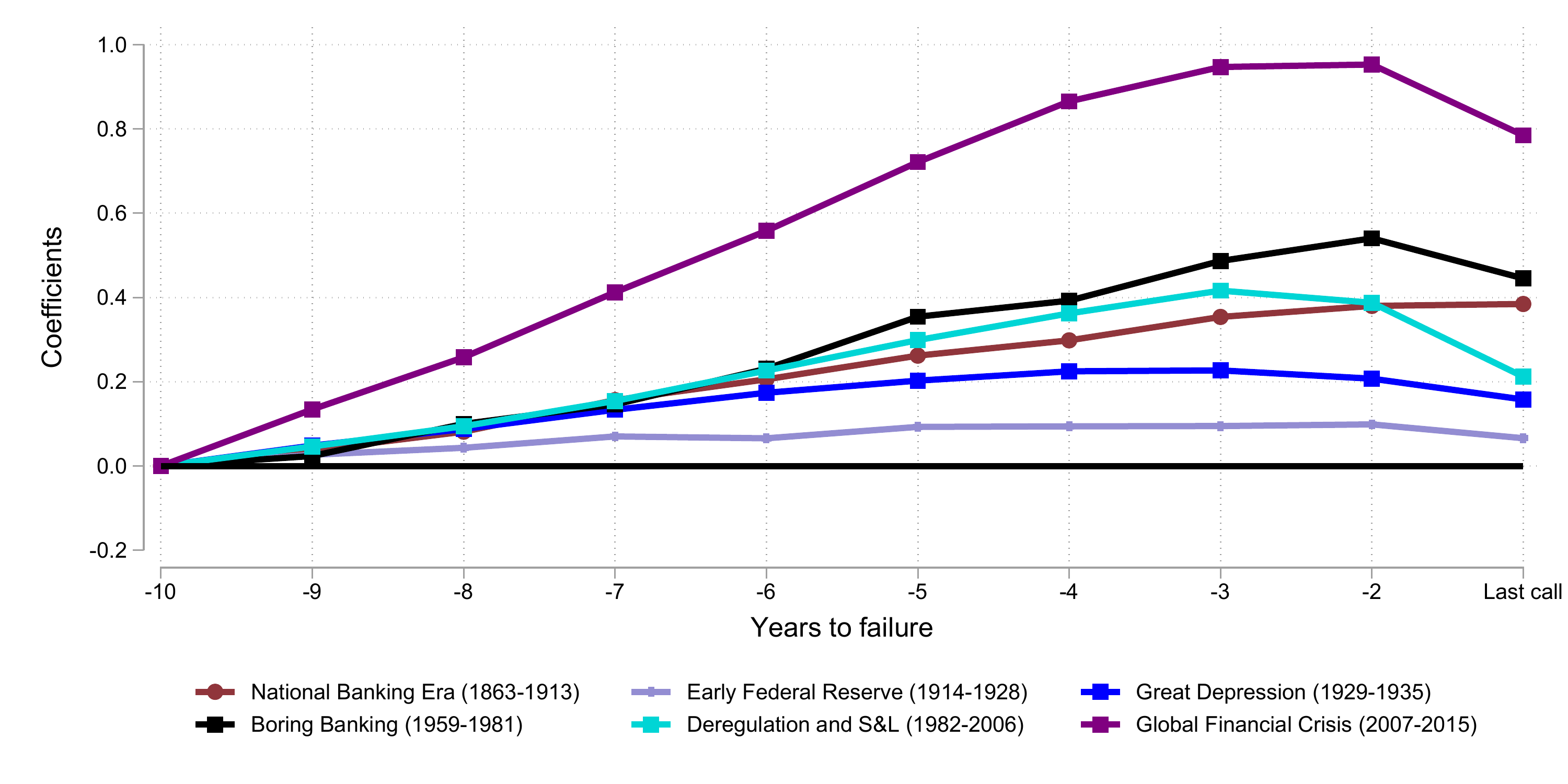}
   \begin{minipage}{\textwidth}
\footnotesize
Notes: This figure reports the sequence of coefficients from estimating \Cref{eq:dynamic} with log total assets (deflated by CPI) as the dependent variable for various subsamples. The regression includes a set of bank fixed effects. The sample is restricted to failing banks and to the ten years before they fail. The sub-samples indicated in the figure legends are selected based on the years in which a bank failed.
\end{minipage}
\end{figure}

\begin{figure}[htpb]
\caption{\textbf{Non-Monotonic Intertemporal Relation between Growth and Failure Probability}}
\label{fig:nonmonotonic_pre_post}
\centering

  \subfloat[Pre-1935]{\includegraphics[width=0.8\textwidth]{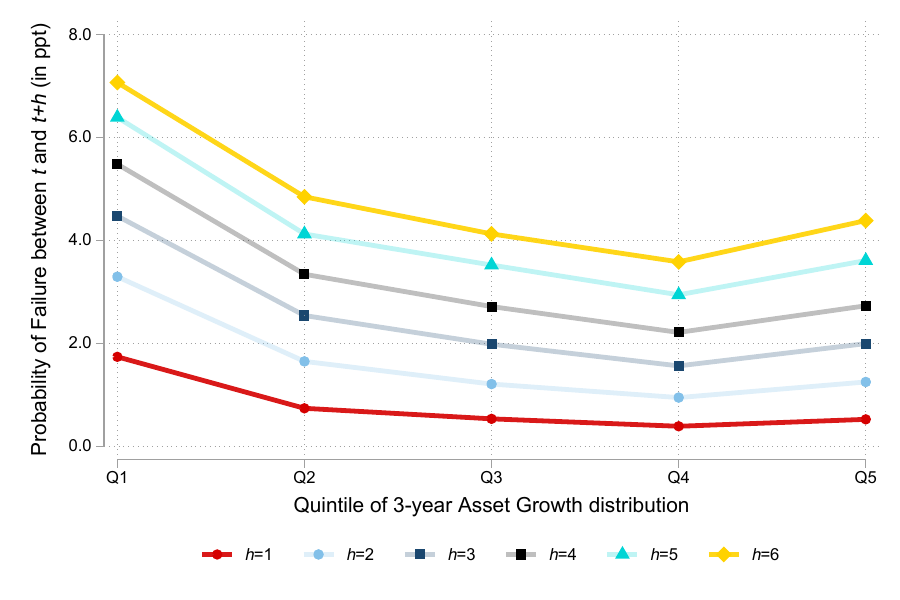}}

  \subfloat[Post-1935]{\includegraphics[width=0.8\textwidth]
{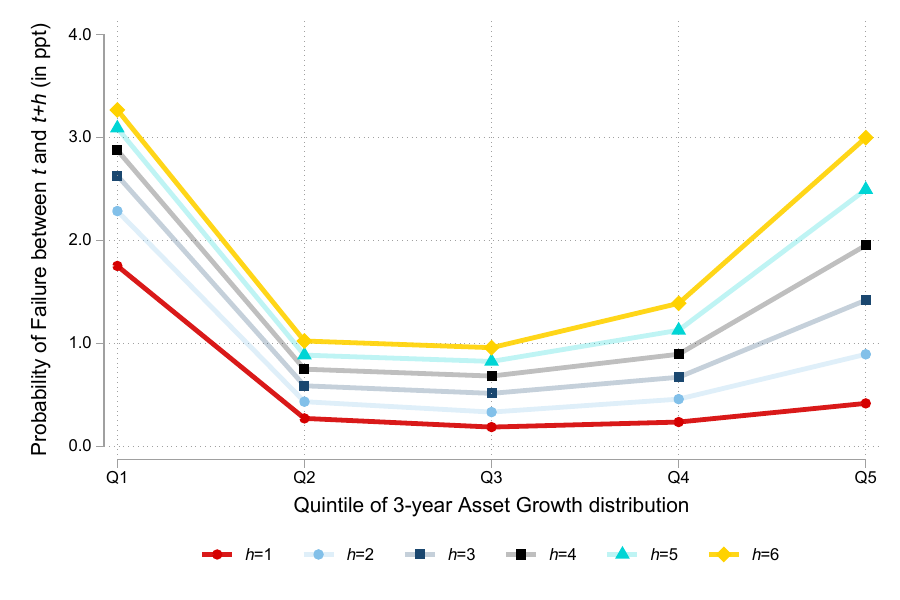}}

   \begin{minipage}{\textwidth}
\footnotesize
Notes:  This figure plots the frequency of failure at the one to six year horizons across quintiles of the three-year asset growth distribution. Panel (a) presents the results for the 1863-1935 sample, and panel (b) presents the results for the 1959-2024 sample.

\end{minipage}
\end{figure}

\begin{figure}[h!]
\centering
\caption{\textbf{Fundamentals Predict Aggregate Waves of Bank Failures: Robustness to Alternative Models } \label{fig:aggregate_granular}}

\subfloat[LPM, Baseline in \Cref{fig:aggregate}]{
\includegraphics[width=0.5\textwidth]{./output/figures/06_aggregate_predicted_actual.pdf}}
\subfloat[Logit, Same predictors as in \Cref{fig:aggregate}]{
\includegraphics[width=0.5\textwidth]{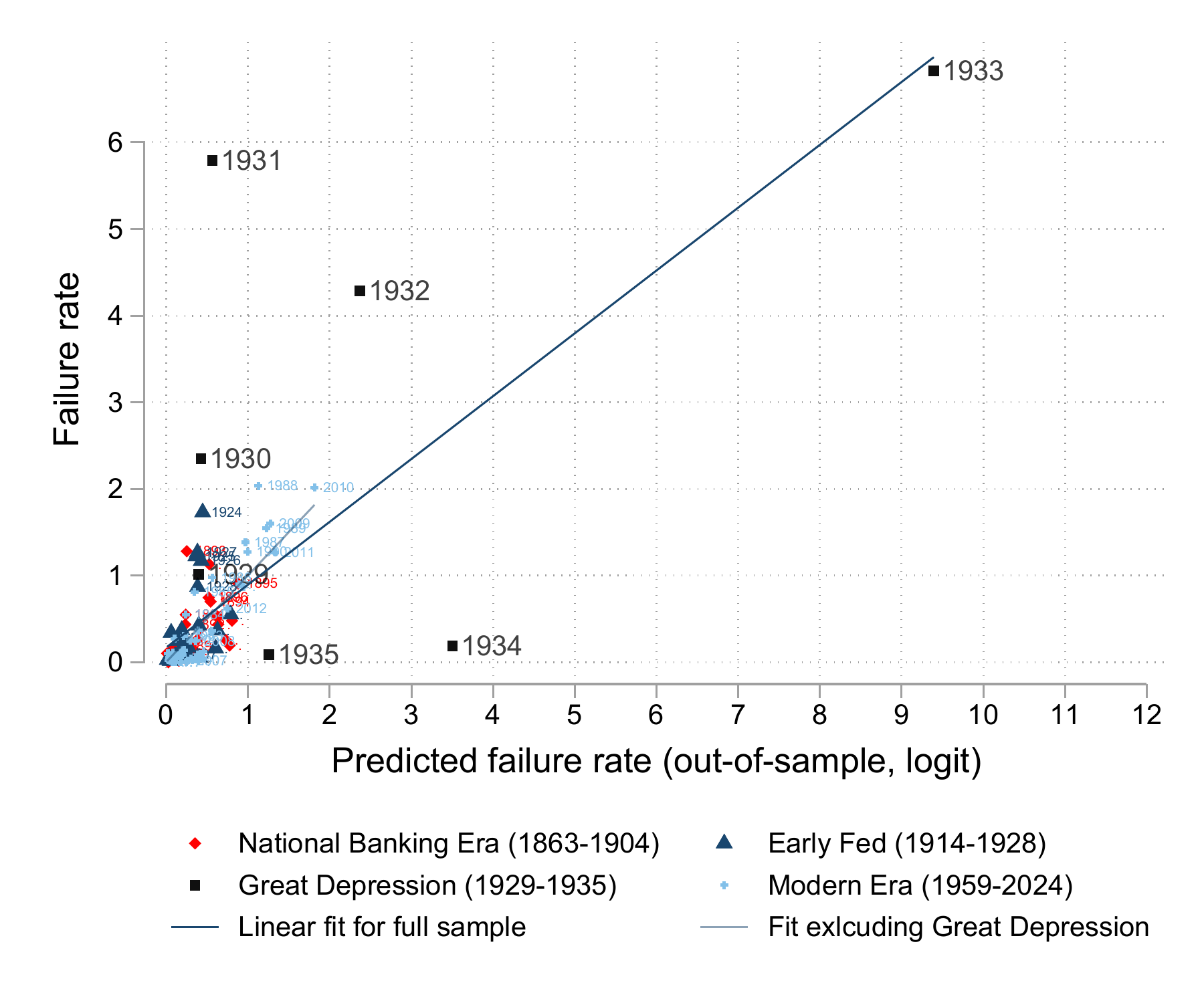}}

\subfloat[LPM, Richer period-specific models ]{
\includegraphics[width=0.5\textwidth]{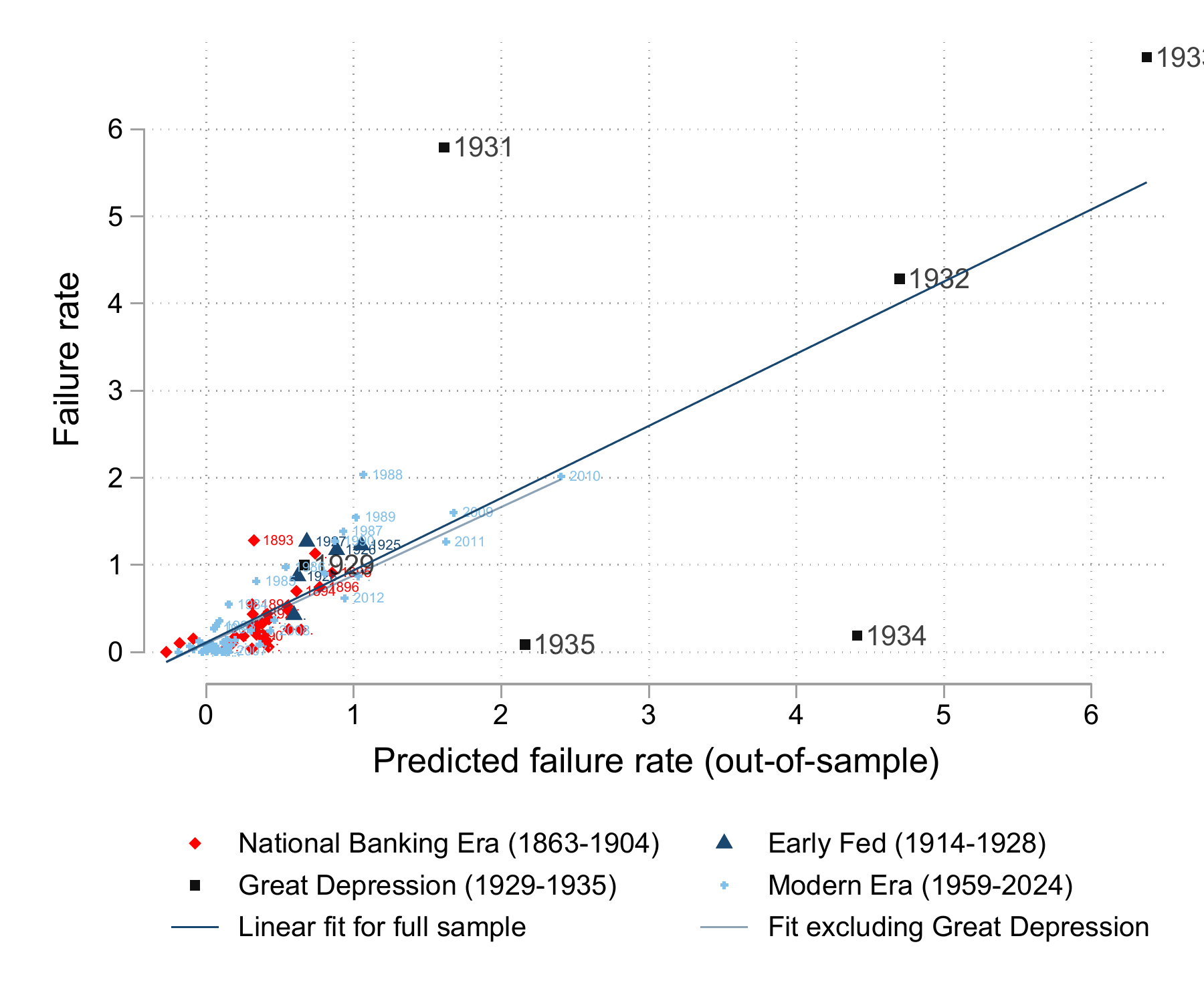}}


\begin{minipage}{\textwidth}
\footnotesize 
Notes: This figure plots the realized aggregate failure rate against the predicted aggregate failure rate, $\overline p_{t|t-1}$, based on various models of $\overline p_{t|t-1}$. Panels (a) and (c) use linear probability models, while panel (b) reports results from a logit model. Panels (a) and (b) use the baseline predictors from column (4) in \Cref{tab:AUC}. Panel (c) is based on the richer period-specific models in column (4) of \Cref{tab:predicting_failure_modern_era}, \Cref{tab:predicting_failure_NBEra}, \Cref{tab:predicting_failure_early_fed}, and \Cref{tab:predicting_failure_GD}. The predicted aggregate failure rate for year $t$ is constructed using only information up to year $t-1$, so the prediction is pseudo out-of-sample. Both measures start 10 years after the start of our data so that we have a sufficiently long training sample.  
\end{minipage}
\end{figure}

\begin{figure}[ht]
\caption{\textbf{Asset and Deposit Recovery Rate over Time Following Bank Failure: 1920-1939} }
\label{fig:depositor_recovery}
\centering

\includegraphics[width=0.8\textwidth]{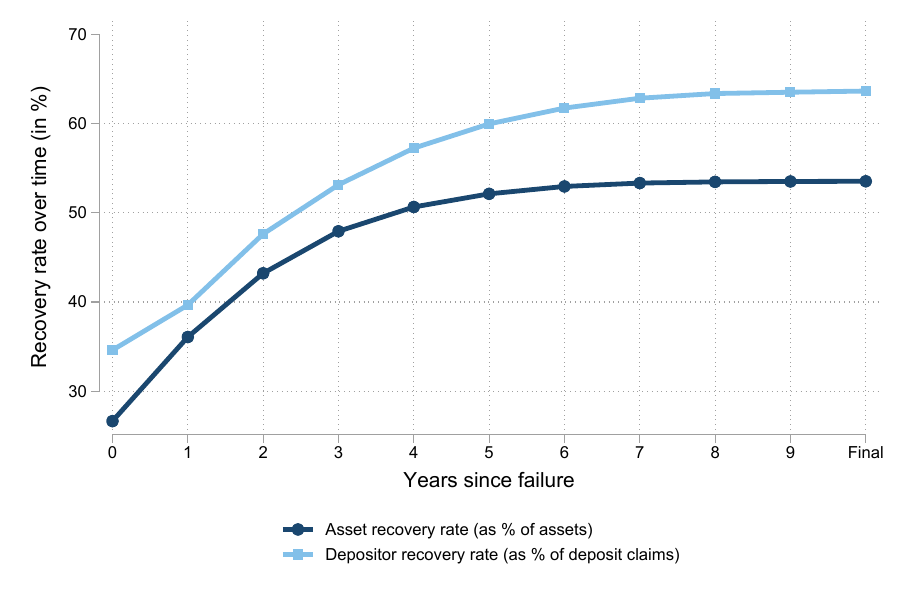}



\begin{minipage}{\textwidth}
\footnotesize
Notes: The figure shows the average asset and depositor recovery rate as a function of the time since failure. The asset recovery rate is calculated as ``collected from assets'' divided by the sum of ``assets at suspension'' and the maximum reported ``additional assets received after suspension.'' The depositor recovery rate is the ``dividend (in percent)'' directly reported by the OCC. The sample covers bank failures from 1920 to 1939, as this sample allows us to observe the dividend payments to depositors in each year from the suspension to when the bank is finally closed. Data are collected from the OCC's annual report to Congress, tables on ``National banks in charge of receivers'' (various years). 
\end{minipage}
\end{figure}





\begin{figure}
\caption{Asset Recovery Rate over Time}
    
    \centering
    \includegraphics[width=0.9\linewidth]{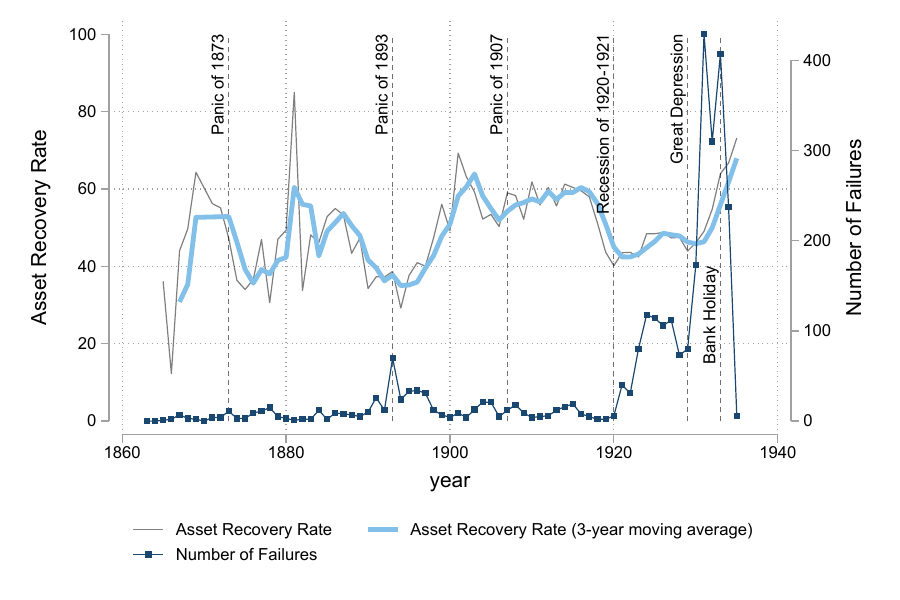}
    \label{fig:recovery-rate}
    
\begin{minipage}{\textwidth}
\footnotesize
Notes: This figure plots both the average and the three-year moving average of the recovery rate in receivership on the left y-axis and the number of total national bank failures on the right y-axis. The recovery rate is calculated as ``collected from assets'' divided by the sum of ``assets at suspension'' and the maximum reported ``additional assets received after suspension.''
\end{minipage}
\end{figure}

%% file: 99_appendix_tables.tex
\clearpage

\begin{table}[ht!]
  \centering
  \caption{\textbf{Summary Statistics: Bank-Level Data, 1863-1941}}\label{tab:summary_statistics}
        \begin{minipage}{1.0\textwidth}
        \begin{center}
        \scalebox{0.8}{
        \input{./output/tables/03_tab_sumstats_prewar.tex}}
        \end{center}
        {\footnotesize \textit{Notes}: This table reports summary statistics for the bank-level data based on the OCC's annual report. Data are at annual frequency. }
        \end{minipage}
\end{table}%

\begin{table}[ht!]
  \centering
  \caption{\textbf{Summary Statistics: Bank-Level Data, 1959-2024}}\label{tab:summary_statistics_modern}
        \begin{minipage}{1.0\textwidth}
        \begin{center}
        \scalebox{0.8}{
        \input{./output/tables/03_tab_sumstats_postwar.tex}}
        \end{center}
        {\footnotesize \textit{Notes}: This table reports summary statistics for the bank-level data based on the FFIEC Call Report. Net income, Loan Loss Provisions (LLP), and net interest income are based on annual, end-of-year data.  All other variables are quarterly. The net interest margin is calculated as the ratio of net interest income over total assets.}
        \end{minipage}
\end{table}%

\begin{table}[htpb]
   \caption{\textbf{AUC Metric for Predicting Bank Failures with Fundamentals Using Richer Period-Specific  Models}\label{tab:AUC_historical_detailed}}
        \begin{minipage}{1\textwidth}
        \begin{center}
        \footnotesize
        \begin{tabular}{lccccccc}
        \toprule

         Prediction horizon $h$  & \multicolumn{5}{c}{1 year}  &\multicolumn{1}{c}{3 years}  &\multicolumn{1}{c}{5 years} \\             \cmidrule(lr){2-6}  \cmidrule(lr){7-7}  \cmidrule(lr){8-8} \\

         \midrule
            & (1) & (2) & (3) & (4) & (5) & (6) & (7) \\ \midrule
            
         \multicolumn{8}{c}{\textbf{Panel A: National Banking Era (1863-1904)}} \\ \midrule 
       \input{./output/tables/05_tab_auc_NBEra} \\ \midrule \input{./output/tables/05_tab_auc_oos_NBEra} \\ \midrule
             \multicolumn{8}{c}{\textbf{Panel B: Early Federal Reserve (1914-1928)}} \\ \midrule       
             \input{./output/tables/05_tab_auc_early_fed} \\ \midrule
         \input{./output/tables/05_tab_auc_oos_early_fed} \\ \midrule
         \multicolumn{8}{c}{\textbf{Panel C: Great Depression (1929-1934)}} \\ \midrule       
             \input{./output/tables/05_tab_auc_GD} \\ \midrule
         \input{./output/tables/05_tab_auc_oos_GD} \\ \midrule

                  \multicolumn{8}{c}{\textbf{Specification details}} \\ \midrule
         
Insolvency & \checkmark & & \checkmark  & \checkmark  & \checkmark  
 & \checkmark  & \checkmark   \\
Noncore funding & & \checkmark & \checkmark  & \checkmark & \checkmark  & \checkmark  & \checkmark  \\
Insolvency $\times$ Noncore funding  & & & \checkmark  & \checkmark  & \checkmark  & \checkmark  & \checkmark  \\
Growth \& Aggregate conditions & & & & \checkmark   & \checkmark  & \checkmark  & \checkmark   \\
Deposit outflow before failure   &  & & & & >7.5\%  & & \\
Age controls & \checkmark  & \checkmark  & \checkmark  & \checkmark  & \checkmark  & \checkmark  & \checkmark  \\
        \bottomrule

        \end{tabular}
        \end{center}
        {\footnotesize Notes: This table reports the area under the receiver operating characteristic curve (AUC) across different specifications, samples, and horizons using in-sample and pseudo-out-of-sample classification. The corresponding regression coefficients underlying the models for Panel A can be found in \Cref{tab:predicting_failure_NBEra}, for Panel B in \Cref{tab:predicting_failure_early_fed}, and for Panel C in \Cref{tab:predicting_failure_GD}. Pseudo-out-of-sample AUCs are obtained by estimating the regression model with an initial training sample from 1863-1873 (Panel A), 1914-1919 (Panel B), and 1880-1904 (Panel C). In column (5) of Panel A, we restrict the sample to data from 1880 onward, and out-of-sample predictions are based on an initial training sample from 1880-1890.}
        \end{minipage}
 \end{table}%

\begin{table}[htpb]
\caption{\textbf{Regression Models Predicting Bank Failures: 1863-1934}}
\label{tab:predicting_failure_historical}
\begin{minipage}{1\textwidth}
        \begin{center}
             \scriptsize
        \begin{tabular}{lccccccc}
        \toprule
        Horizon $h$  & \multicolumn{5}{c}{Fail in next year } & 3 years & 5 years \\
        \cmidrule(lr){2-6} \cmidrule(lr){7-7}\cmidrule(lr){8-8} 
        Withdrawals before failure   &  & & & & $>$7.5\%& & \\
        \cmidrule(lr){2-6}  \cmidrule(lr){7-7} \cmidrule(lr){8-8}
        \input{./output/tables/05_tab_predicting_failure_historical} \\ 
        \bottomrule
        \end{tabular}
        \end{center}
        {\footnotesize Notes: This table presents OLS estimates of \Cref{eq:pred} with failure between $t$  and $t+h$ as the dependent variables for the 1863-1934 sample.  In addition to the reported  predictor variables, we also include the log of a bank's age. \citet{Driscoll1998} standard errors in parentheses with bandwidth $h=2$; *, **, and *** indicate significance at the 10\%, 5\%, and 1\% level, respectively. 
        }
        \end{minipage}
\end{table}%

\begin{table}[htpb]
\caption{\textbf{Regression Models Predicting Bank Failures: 1959-2024}}
\label{tab:predicting_failure_modern_era}
\begin{minipage}{1\textwidth}
 \begin{center}
     
        \scriptsize
         \begin{tabular}{lccccccc}
        \toprule
        Horizon $h$  & \multicolumn{5}{c}{Fail in next year } & 3 years & 5 years \\
      \cmidrule(lr){2-6} \cmidrule(lr){7-7}\cmidrule(lr){8-8} 
        Withdrawals before failure   &  & & & & $>$7.5\% & & \\
        \cmidrule(lr){2-6}  \cmidrule(lr){7-7} \cmidrule(lr){8-8}
        \input{./output/tables/05_tab_predicting_failure_modern_era} \\ 
        \bottomrule
        \end{tabular}
        \end{center}
        {\footnotesize Notes: This table presents OLS estimates of \Cref{eq:pred} with failure between $t$  and $t+h$ as the dependent variables for the 1959-2024 sample.  In addition to the reported  predictor variables, we also include the log of a bank's age. The sample in column (5) is restricted to the years from 1993-2024 due to the unavailability to deposits in failure before 1993. \citet{Driscoll1998} standard errors in parentheses with bandwidth $h=2$; *, **, and *** indicate significance at the 10\%, 5\%, and 1\% level, respectively. 
        }
        \end{minipage}
 \end{table}%

\begin{table}[htpb]
   \caption{\textbf{AUC Metric for Predicting Bank Failures with Fundamentals Using Logit Model} \label{tab:AUC_logit}}
        \begin{minipage}{1\textwidth}
        \begin{center}
        \footnotesize
        \begin{tabular}{lccccccc}
        \toprule

         Prediction horizon $h$  & \multicolumn{5}{c}{1 year}  &\multicolumn{1}{c}{3 years}  &\multicolumn{1}{c}{5 years} \\             \cmidrule(lr){2-6}    \cmidrule(lr){7-7} \cmidrule(lr){8-8} \\

         \midrule
            & (1) & (2) & (3) & (4) & (5) & (6) & (7) \\ \midrule
              \multicolumn{8}{c}{\textbf{Panel A: Historical Sample (1863-1934)}} \\ \midrule 
       \input{./output/tables/05_tab_auc_historical_glm} \\ \midrule \input{./output/tables/05_tab_auc_oos_historical_glm} \\ \midrule
         
                   \multicolumn{8}{c}{\textbf{Panel D: Modern Sample (1959-2024)}} \\ \midrule       
             \input{./output/tables/05_tab_auc_modern_era_glm} \\ \midrule
         \input{./output/tables/05_tab_auc_oos_modern_era_glm} \\ \midrule
                  \multicolumn{8}{c}{\textbf{Specification details}} \\ \midrule
         
Insolvency & \checkmark & & \checkmark  & \checkmark  & \checkmark  
 & \checkmark  & \checkmark    \\
Noncore funding & & \checkmark & \checkmark  & \checkmark & \checkmark  & \checkmark  & \checkmark   \\
Insolvency $\times$ Noncore funding  & & & \checkmark  & \checkmark  & \checkmark  & \checkmark  & \checkmark    \\
Growth \& Aggregate conditions & & & & \checkmark   & \checkmark  & \checkmark  & \checkmark   \\
Deposit outflow before failure   &  & & & & >7.5\%   & & \\
Age controls & \checkmark  & \checkmark  & \checkmark  & \checkmark  & \checkmark  & \checkmark  & \checkmark   \\
        \bottomrule
        \end{tabular}
        \end{center}
        {\footnotesize Notes: This table reports the area under the receiver operating characteristic curve (AUC) across different specifications, samples, and horizons using in-sample and pseudo-out-of-sample classification. The predictions are based on logit models with the same predictive variables in each column as in \Cref{tab:AUC}. Pseudo-out-of-sample AUCs are obtained by estimating the regression model on an initial training sample from 1863-1873 (Panel A) and 1959-1969 (Panel B) and iteratively expanding the sample for subsequent years.   We drop banks without information on deposit outflow before failure in column (5), restricting the sample period to 1880-1934 in Panel (A), as data on deposits at failure are not available before 1880. Similarly, column (5) in Panel B is restricted to the years from 1993-2024, deposits in failure are not available before 1993.}
        \end{minipage}
 \end{table}%

\begin{table}[htpb]
   \caption{\textbf{AUC Metric for Predicting Bank Failures with Fundamentals by Size} \label{tab:AUC_size}}
        \begin{minipage}{1\textwidth}
        \begin{center}
        \footnotesize
        \begin{tabular}{lccccc}
        \toprule

      \cmidrule(lr){2-6}
         Predictions horizon $h$ & \multicolumn{5}{c}{1 year} \\
         \cmidrule(lr){2-6}
         Size Quintile & 1st & 2nd & 3rd & 4th & 5th \\

         \midrule
            & (1) & (2) & (3) & (4) & (5) \\ \midrule
         \multicolumn{5}{c}{\textbf{Panel A: Historical Sample (1863-1934)}} \\ \midrule 
       \input{./output/tables/05_tab_auc_historical_by_size} \\ \midrule \input{./output/tables/05_tab_auc_oos_historical_by_size} \\ \midrule
             
      
                   \multicolumn{5}{c}{\textbf{Panel B: Modern Era (1959-2024)}} \\ \midrule       
             \input{./output/tables/05_tab_auc_modern_era_by_size} \\ \midrule
         \input{./output/tables/05_tab_auc_oos_modern_era_by_size} \\ \midrule
                  \multicolumn{5}{c}{\textbf{Specification details}} \\ \midrule
         
Insolvency & \checkmark  & \checkmark  
 & \checkmark  & \checkmark  & \checkmark   \\
Noncore funding & \checkmark  & \checkmark  
 & \checkmark  & \checkmark & \checkmark   \\
Insolvency $\times$ Noncore funding  & \checkmark  & \checkmark  
 & \checkmark  & \checkmark & \checkmark   \\
Age controls & \checkmark  & \checkmark  
 & \checkmark  & \checkmark  & \checkmark   \\
        \bottomrule
        \end{tabular}
        \end{center}
        {\footnotesize Notes: This table reports the area under the receiver operating characteristic curve (AUC) for column (4) of \Cref{tab:predicting_failure_historical} (Panel A) and \Cref{tab:predicting_failure_modern_era} (Panel B) when splitting the sample into five different size categories. Size categories correspond to quintiles of the assets distribution, calculated within a given year. Pseudo-out-of-sample AUCs are obtained by estimating the regression model with an initial training sample of 10 years and iteratively expanding the sample for subsequent years.}
        \end{minipage}
 \end{table}%

\begin{table}[htpb]
\caption{\textbf{Predicting Bank Failures: Richer Period-Specific Models for 1863-1904}}
\label{tab:predicting_failure_NBEra}
\begin{minipage}{1\textwidth}
        \begin{center}
             \scriptsize
        \begin{tabular}{lccccccc}
        \toprule
        Horizon $h$  & \multicolumn{5}{c}{Fail in next year } & 3 years & 5 years \\
        \cmidrule(lr){2-6} \cmidrule(lr){7-7} \cmidrule(lr){8-8}
        Withdrawals before failure   &  & & & & $>$7.5\% & & \\
        \cmidrule(lr){2-6}  \cmidrule(lr){7-7} \cmidrule(lr){8-8}
        \input{./output/tables/05_tab_predicting_failure_NBEra} \\ 
        \bottomrule
        \end{tabular}
        \end{center}
        {\footnotesize Notes: This table presents OLS estimates of \eqref{eq:pred} with failure between $t$  and $t+h$ as the dependent variables for the 1863-1904 sample.  In addition to the reported predictor variables, we also include the log of a bank's age. \citet{Driscoll1998} standard errors in parentheses with bandwidth $h=2$; *, **, and *** indicate significance at the 10\%, 5\%, and 1\% level, respectively. 
        }
        \end{minipage}
 \end{table}%

\begin{table}[htpb]
\caption{\textbf{Predicting Bank Failures: Richer Period-Specific Models for 1914-1928}}
\label{tab:predicting_failure_early_fed}
\begin{minipage}{1\textwidth}
        \begin{center}
             \scriptsize
        \begin{tabular}{lccccccc}
        \toprule
        Horizon $h$  & \multicolumn{5}{c}{Fail in next year } & 3 years & 5 years \\
        \cmidrule(lr){2-6}  \cmidrule(lr){7-7} \cmidrule(lr){8-8}
        Withdrawals before failure   &  & & & & $>$7.5\% & & \\
        \cmidrule(lr){2-6}  \cmidrule(lr){7-7} \cmidrule(lr){8-8}
        \input{./output/tables/05_tab_predicting_failure_early_fed} \\ 
        \bottomrule
        \end{tabular}
        \end{center}
        {\footnotesize Notes: This table presents OLS estimates of \eqref{eq:pred} with failure between $t$  and $t+h$ as the dependent variables for the 1914-1928 sample. In addition to the reported  predictor variables, we also include the log of a bank's age. \citet{Driscoll1998} standard errors in parentheses with bandwidth $h=2$; *, **, and *** indicate significance at the 10\%, 5\%, and 1\% level, respectively. 
        }
        \end{minipage}
 \end{table}%

\begin{table}[htpb]
\caption{\textbf{Regression Models Predicting Bank Failures: Richer Period-Specific Models for 1929-1934}}
\label{tab:predicting_failure_GD}
\begin{minipage}{1\textwidth}
        \begin{center}
             \scriptsize
        \begin{tabular}{lccccccc}
        \toprule
        Horizon $h$  & \multicolumn{5}{c}{Fail in next year } & 3 years & 5 years \\
        \cmidrule(lr){2-6} \cmidrule(lr){7-7} \cmidrule(lr){8-8}
        Withdrawals before failure   &  & & & & >7.5\% & & \\
        \cmidrule(lr){2-6} \cmidrule(lr){7-7} \cmidrule(lr){8-8}
        \input{./output/tables/05_tab_predicting_failure_GD} \\ 
        \bottomrule
        \end{tabular}
        \end{center}
        {\footnotesize Notes: This table presents OLS estimates of \eqref{eq:pred} with failure between $t$  and $t+h$ as the dependent variables for the 1929-1934 sample.  In addition to the reported  predictor variables, we also include the log of a bank's age. \citet{Driscoll1998} standard errors in parentheses with bandwidth $h=2$; *, **, and *** indicate significance at the 10\%, 5\%, and 1\% level, respectively. 
        }
        \end{minipage}
 \end{table}%

\begin{table}[htbp]

\caption{Predicting Bank Failures: True Positives, False Positives, True Negatives, and False Negatives from Linear Probability Model  \label{tab:true_false_positives}}
\begin{center}
\footnotesize
\begin{tabular}{ccccc}
\toprule
$c$ & TPR & FPR & \shortstack{TNR \\ (= 1 - FPR)} & \shortstack{ FNR \\ (= 1 - TPR)} \\
\midrule
\multicolumn{5}{c}{\textbf{Panel A: Historical Sample (1863-1934)}} \\ \midrule 
\input{./output/tables/99_TPR_FPR_TNR_FNR_historical_ols} \\
\midrule
\multicolumn{5}{c}{\textbf{Panel B: Modern Sample (1959-2024)}} \\ \midrule 
\input{./output/tables/99_TPR_FPR_TNR_FNR_modern_ols} \\
\bottomrule 
\end{tabular}
\end{center}
 {\footnotesize Notes: This table reports the true positive rate (TPR), false positive rate (FPR), true negative rate (TNR), and false negative rate (FNR) based on different cutoffs ($c$) for classifying failures. Specifically, we classify a bank as predicted to fail if the $\Pr[Failure_{b,t+1}]>c$. Predicted values are based on column (4) from \Cref{tab:predicting_failure_historical} in Panel A and (4) of \Cref{tab:predicting_failure_modern_era} in Panel B using a linear probability model.}
\end{table}

\begin{table}[htbp]

\caption{Predicting Bank Failures: True Positives, False Positives, True Negatives, and False Negatives from Logit Model  \label{tab:true_false_positives_logit}}
\begin{center}
\footnotesize
\begin{tabular}{ccccc}
\toprule
$c$ & TPR & FPR & \shortstack{TNR \\ (= 1 - FPR)} & \shortstack{ FNR \\ (= 1 - TPR)} \\
\midrule
\multicolumn{5}{c}{\textbf{Panel A: Historical Sample (1863-1934)}} \\ \midrule 
\input{./output/tables/99_TPR_FPR_TNR_FNR_historical_logit} \\
\midrule
\multicolumn{5}{c}{\textbf{Panel B: Modern Sample (1959-2024)}} \\ \midrule 
\input{./output/tables/99_TPR_FPR_TNR_FNR_modern_logit} \\
\bottomrule 
\end{tabular}
\end{center}
 {\footnotesize Notes: This table reports the true positive rate (TPR), false positive rate (FPR), true negative rate (TNR), and false negative rate (FNR) based on different cutoffs ($c$) for classifying failures. Specifically, we classify a bank as predicted to fail if the $\Pr[Failures_{b,t+1}]>c$. Predicted values are based on specification (4) from \Cref{tab:predicting_failure_historical} in Panel A and (4) of \Cref{tab:predicting_failure_modern_era} in Panel B using a logit model.}
\end{table}

\begin{table}[ht]
\caption{\textbf{Precision-Recall Curve for Predicting Bank Failures}  \label{tab:PR_AUC} }
\begin{minipage}{1\textwidth}
\begin{center}
\footnotesize
\begin{tabular}{lccccccc}
\toprule

\cmidrule(lr){2-8}
Prediction horizon $h$  & \multicolumn{5}{c}{1 year}  &\multicolumn{1}{c}{3 years}  &\multicolumn{1}{c}{5 years} \\             \cmidrule(lr){2-6}    \cmidrule(lr){7-7} \cmidrule(lr){8-8} \\

\midrule
& (1) & (2) & (3) & (4) & (5) & (6) & (7) \\ \midrule
\multicolumn{8}{c}{\textbf{Panel A: Historical Sample (1863-1934), Baseline Models in \Cref{tab:AUC}}} \\ \midrule  
\input{./output/tables/pr_auc_1863_1934.tex}  \\ \midrule

\multicolumn{8}{c}{\textbf{Panel B: Modern Sample (1959-2024), Baseline Models in \Cref{tab:AUC}}} \\ \midrule       
\input{./output/tables/pr_auc_1959_2024} \\ \midrule
\multicolumn{8}{c}{\textbf{Specification details}} \\ \midrule
Insolvency & \checkmark & & \checkmark  & \checkmark  & \checkmark  
& \checkmark  & \checkmark  \\
Noncore funding & & \checkmark & \checkmark  & \checkmark & \checkmark  & \checkmark  & \checkmark   \\
Insolvency $\times$ Noncore funding  & & & \checkmark  & \checkmark  & \checkmark  & \checkmark  & \checkmark    \\
Growth \& Aggregate conditions & & & & \checkmark   & \checkmark  & \checkmark    & \checkmark  \\
Deposit outflow before failure   &  & & & & >7.5\%  & & \\
Age controls & \checkmark  & \checkmark  & \checkmark  & \checkmark  & \checkmark  & \checkmark  & \checkmark   \\
\bottomrule
\end{tabular}
\end{center}
{\footnotesize Notes: This table reports the area under the precision-recall curve (PR-AUC), the mean failure rate, and precision at 10\% recall across different specifications, samples, and horizons using in-sample and pseudo-out-of-sample classification. Ratio refers to the ratio of the PR-AUC and the mean of the dependent variable, which would be the precision of a naive model that classified all observations as failures. The models used are equivalent to the models in the corresponding column of \Cref{tab:AUC}. The corresponding regression coefficients underlying the models are reported in \Cref{tab:predicting_failure_historical} (for Panel A) and \Cref{tab:predicting_failure_modern_era} (for Panel B).  The PR-AUC is based on predictions from linear probability models. We drop banks without information on deposit outflows before failure in column (5). This restricts the sample period to 1880-1934 in Panel (A) and 1993-2024 in Panel (B), as data on deposits at failure are not available before 1880 and 1993 respectively.}
\end{minipage}
 \end{table}%


\begin{table}[htbp]
\centering
\caption{Out-of-sample Predicted Probability of Failure Before Failure, 1875-1934  
\label{tab:oos_predicted} }
\begin{minipage}{1\textwidth}
\footnotesize
\begin{center}
\begin{tabular}{lccccccccc}
\toprule
   & \multicolumn{9}{c}{\bf Panel A1: OOS Predicted Probability of Failure ($h=1$)} \\ \midrule
  & & & \multicolumn{7}{c}{Share by Predicted Probability of Failure}  \\ \cmidrule(lr){4-10} Model & Avg. & Med. & [0,1\%) & [1\%,5\%)  & [5\%,10\%)  & [10\%,20\%)& [20\%,30\%) &  [30\%,40\%) & $\geq$40 \\
\midrule

\input{./output/tables/06_pred_prob_of_failure_baseline} \\

\input{./output/tables/06_pred_prob_of_failure_granular} \\
\midrule
 & \multicolumn{9}{c}{\bf Panel A2: OOS Predicted Probability of Failure ($h=3$)} \\ \midrule
\input{./output/tables/06_pred_prob_of_failure_baseline_3year} \\

\input{./output/tables/06_pred_prob_of_failure_granular_3year} \\
\midrule
& \multicolumn{9}{c}{\bf Panel B1: Required Excess Return (risk neutral)} \\ \midrule
   & & & \multicolumn{7}{c}{Share by Implied Required Excess Return }  \\ \cmidrule(lr){4-10}  & Avg. & Med. & <0.5\% & [0.5\%,1\%) & [1\%,2.5\%)  & [2.5\%,5\%)& [5\%,10\%) &  [10\%,15\%) & $\geq$15 \\
\midrule
\input{./output/tables/06_required_rate_risk_neutral} \\
\input{./output/tables/06_required_rate_risk_neutral_granular} \\

\midrule
& \multicolumn{9}{c}{\bf Panel B2: Required Excess Return (log-utility)} \\ \midrule
\input{./output/tables/06_required_rate_log} \\
\input{./output/tables/06_required_rate_log_granular} \\

\midrule
& \multicolumn{9}{c}{\bf Panel B3: Required Excess Return (CRRA, $\gamma=5$)} \\ 
\midrule
\input{./output/tables/06_required_rate_crra} \\
\input{./output/tables/06_required_rate_crra_granular} \\

\bottomrule
\end{tabular}
        \end{center}
 {\footnotesize Notes: Panels A1 and A2 report the distribution of the bank-level out-of-sample predicted probabilities of failure over the one-year and three-year horizon for failing banks as of the last call report before failure for all failures from 1873 through 1934. The predicted values in the baseline model reported in the first row result from estimating the model reported in Column (4) of \Cref{tab:predicting_failure_historical} using logit. The predicted probabilities of the model reported in the second row are from estimating the richer models that use a different set of ratios and different training sample across the National Banking Era, Early Federal Reserve, and Great Depression and are reported in column (4) of \Cref{tab:predicting_failure_NBEra}, \Cref{tab:predicting_failure_early_fed}, and \Cref{tab:predicting_failure_GD}, respectively using logit.  

  Panels B1 and B2 shows the distribution of the required excess return on deposits (trimmed at 100\%). We calculate the required excess return on deposits, $s_{b,t}$, as the solution to the following equation:
\[(1-\hat p_{b,t+1|t})u(1+r_t+s_{b,t}) + \hat p_{b,t+1|t} u(1-\ell_{t+1|t}) = u(1), \]
where $\ell_{t+1|t}$ is the loss rate on failures up to time $t$ and $r_t$ is the rate on treasury bills in year $t$. Note that this assumes that the certainty equivalent earns a return of zero. We assume $u(x)=x$ in Panel B1,  $u(c) = \ln(c)$ in Panel B3, and $u(c) = \frac{c^{1 - \gamma}}{1 - \gamma}$ with $\gamma=5$
 in Panel B3. 
        }
         \end{minipage}
\end{table}

\begin{table}[htpb]
\begin{center}
\begin{threeparttable}
\caption{\textbf{Predictability of Failures By Era and During Major Banking Crises} }\label{tab:auc_by_era}
\footnotesize 
\begin{tabular}{lcccccc}
\toprule
 
 \multicolumn{7}{c}{Panel A: 1863-1935} \\ \midrule
 & 1890 & 1893 & 1890-1896 & 1930-1933 & 1929-1931 & 1932-1933 \\ \midrule \input{./output/tables/04_pre_FDIC_auc_by_crisis} \\ \midrule 
\multicolumn{7}{c}{Panel B: 1959-2024} \\ \midrule
 & 1959-1981 & 1982-1994 & 1994-2006 & 2007-2024 & 1984-1990 & 2007-2013 \\ \midrule
\input{./output/tables/04_postwar_auc_by_era} \\ 
\bottomrule
\end{tabular}

\begin{tablenotes}
\footnotesize
\item  Notes:  This table reports the area under the receiver operating characteristic curve (AUC) by sample period. In Panel A, we calculate the AUC based on the predictions obtained from the model in column (4) of  \Cref{tab:predicting_failure_historical}. In Panel B, we calculate the AUC based on the predictions obtained from the model in column (4) of \Cref{tab:predicting_failure_modern_era}. 

\end{tablenotes}
\end{threeparttable}
	
\end{center}

\end{table}

\begin{table}[ht]
\centering
\begin{threeparttable}
\caption{\textbf{Fundamentals Predict Aggregate Rate of Bank Failures: Robustness Using Alternative Models  }  \label{tab:predicting_aggregate_robustness}}
\footnotesize 
    \begin{tabular}{lcccc}
\toprule
Dependent variable  & \multicolumn{4}{c}{Aggregate Failure Rate}  \\ \cmidrule(lr){2-5}  \\
    \hline 
    \multicolumn{5}{c}{\bf Panel A: LPM, Same Predictors as in Columns (1)-(4) of \Cref{tab:AUC} } \\ \hline
    \input{./output/tables/06_aggregate_predicted_actual_regs_rob.tex} \\ \hline

    \multicolumn{5}{c}{\bf Panel B: Logit, Same Predictors as in Columns (4) of \Cref{tab:AUC} } \\ \hline 
\input{./output/tables/06_aggregate_predicted_actual_regs_logit}
\\ \hline

    \multicolumn{5}{c}{\bf Panel C: LPM, Richer Period-Specific  Models} \\ \hline 
\input{./output/tables/06_aggregate_predicted_actual_regs_granular}
\\    
    
\bottomrule
\end{tabular}
\begin{tablenotes} 
\item 
\footnotesize
Notes: This table presents time series regressions of the annual aggregate failure rate in year $t$ on the average predicted failure rate $\overline p_{t|t-1}$.  The average predicted failure rate is constructed out-of-sample using an expanding sample that only incorporates information up to year $t-1$. The predicted failure rate in panel A is based on the models in columns (1) through (4) of \Cref{tab:AUC}. Panel B uses a logit model with the same predictors as in column (4) of \Cref{tab:AUC} on various time samples. Panel C uses a linear probability model with richer period-specific models from column (4) of \Cref{tab:predicting_failure_modern_era}, \Cref{tab:predicting_failure_NBEra}, \Cref{tab:predicting_failure_early_fed}, and \Cref{tab:predicting_failure_GD}.  Newey-West standard errors in parentheses with truncation parameter $S=1.3T^{1/2}$ following \cite{Lazarus2018HAR}. *, **, and *** indicate significance at the 10\%, 5\%, and 1\% level, respectively. 
\end{tablenotes}
\end{threeparttable}
\end{table}

\begin{table}
\centering
\begin{threeparttable}
\caption{\textbf{Asset Recovery Rates and Pre-Failure Market Structure 
 and Receivership Length}}
\label{tab:pred_recovery_rate_2}
\scriptsize
\begin{minipage}{0.7\textwidth}
 
        \begin{tabular}{lccccccc}
        \toprule
               Dependent variable & \multicolumn{7}{c}{Recovery Rate} \\
        \cmidrule(lr){2-8} 
        \input{./output/tables/07_predicting_recovery_rates_2} \\ 
        \bottomrule
        \end{tabular}
        \begin{tablenotes} 
\item 
\footnotesize
        Notes: This table reports results from a regression of the following form: 
        \[{\text{Recovery Rate}}_b = \tau_t+ \beta X_{b} +\epsilon_b,\] where the Recovery Rate$_b$ is calculated as the ratio of the total funds collected in receivership  over total assets held at suspension and additional assets received throughout receivership, $\tau_t$ is a set of year fixed effects, and  $X_b$ is a set of bank characteristics observed either before or in receivership. $X_b$ is either the number of banks active in the market before failure or the length of the receivership (in years) or balance sheet characteristics from before receivership (such as size measured in log of nominal assets or capitalization and loan ratio). *, **, and *** indicate significance at the 10\%, 5\%, and 1\% level, respectively. 

        \end{tablenotes}
        \end{minipage}
        \end{threeparttable}
 \end{table}%

 
\begin{table}[htpb]
\small
  \begin{center}
  \begin{threeparttable}
\caption{\textbf{Loss Rates for Uninsured Depositors in Bank Failures: Pre-FDIC versus Post-FDIC}}
\label{tab:depositor_losses}
\centering
\footnotesize
\begin{tabular}{lcccc}
\toprule
Era & Number of& Share of failures with  &  
Conditional & Unconditional   \\

&  failures &   losses to depositors &  
loss rate & loss rate  \\
 \\  \hline
\multicolumn{5}{c}{Panel A: Pre-FDIC} \\ \hline
\input{./output/tables/04_depositor_losses_preFDIC} 
\input{./output/tables/04_depositor_losses_preFDIC_all} \\ \midrule
\multicolumn{5}{c}{Panel B: Post-FDIC} \\ \hline
1992-2007 & 302 & 0.43 & 0.24 & 0.10 \\
2008-2022 & 536 & 0.06 & 0.43 & 0.03 \\
 All & 838 & 0.2 & 0.28 & 0.06  \\
		\bottomrule
    \end{tabular}
    		\tiny
		\begin{tablenotes} \item
\footnotesize
Notes: The loss rates reported in panel (A) are from the OCC's tables on national banks placed in receivership. The final loss rate for depositors does not account for interest payments or discounting. The data in panel (B) are as reported in \citet{FDIC2023}. The conditional loss rate is the loss rate for failures involving a loss for uninsured depositors.
\end{tablenotes}
\normalsize
	\end{threeparttable}
	\end{center}
\end{table}

\begin{table}[htbp]
\centering
\caption{Share of Fundamentally Insolvent Banks by $\rho$ and $v$ for Failures With and Without Runs} 
\label{tab:rho_v_by_run} 
\begin{minipage}{1\textwidth}
\begin{center}
\begin{tabular}{lccccccc}
\toprule
$\rho$ & $v = 0$ & $v = 2.5\%$ & $v = 5\%$ & $v = 7.5\%$ & $v = 10\%$ & $v = 15\%$ & $v = 20\%$ \\
\midrule
\multicolumn{8}{c}{Panel A: Failures with run (net deposit outflow larger than 7.5\%)}  \\ 
\midrule
\input{./output/tables/07_recovery_rho_v_run}  \\ 
\midrule
\multicolumn{8}{c}{Panel B: Failures without run (net deposit outflow less than 7.5\%)}  \\ 
\midrule
\input{./output/tables/07_recovery_rho_v_no_run} \\
\bottomrule
\end{tabular}
\end{center}
{\footnotesize Notes: This table is similar to \Cref{tab:rho_v}, but we report the share of banks for which the insolvency condition holds separately for banks with net deposit outflows above and below 7.5\%. The calculations are based on the sample of banks for which deposits at suspension are reported by the OCC.}
\end{minipage}
\end{table}

\begin{table}[htbp]
\centering
\caption{Sensitivity of the Share of Fundamentally Insolvent Banks to Recovery Rate Calculations}
\label{tab:recovery_rate_double_liability}
\begin{minipage}{1\textwidth}
        \begin{center}
\begin{tabular}{lccccccc}
\toprule
$\rho$ & $v = 0$ & $v = 2.5\%$ & $v = 5\%$ & $v = 7.5\%$ & $v = 10\%$ & $v = 15\%$ & $v = 20\%$ \\
\midrule
\multicolumn{8}{c}{Panel A: Double liability} \\
\midrule
\input{./output/tables/07_recovery_rho_v_double_liability} \\ \midrule
\multicolumn{8}{c}{Panel B: Deposits at suspension} \\
\midrule
\input{./output/tables/07_recovery_rho_v_alternative_deposits} \\
\bottomrule
\end{tabular}
        \end{center}
 {\footnotesize Notes:  This table reports the share of banks that are fundamentally insolvent defined as \[ \frac{1}{N}\sum_b   \mathbb{I}\left[\frac{1+v}{1-\rho} < \frac{\ell_b}{R_b} \right].  \] In both panels, the calculations are based on a sample of all national banks placed into the hands of a receiver from 1863 through 1934. In Panel A, we construct the recovery rate $R$ by dividing all funds recovered by the receiver---from the failed banks' assets as well as from shareholders due to their double liability---against the sum of assets held at suspension plus additional assets collected during the receivership. In Panel B, we calculate the leverage ratio $\ell$ by using the sum of deposits at suspension, offsets, and other liabilities (as opposed to proven claims, offsets, and other liabilities).
        }
         \end{minipage}
\end{table}

\begin{table}[ht!]
\small
\begin{center}
\begin{threeparttable}
\caption{\textbf{Duration of Receiverships, 1863-1941} \label{tab:receivership_length}
} 
\centering
\footnotesize
   \begin{tabular}{cccccccccc}
\toprule

\multicolumn{4}{c}{Length of receivership (in years)} &\multicolumn{6}{c}{Share of receiverships by length} \\
\cmidrule(lr){1-4} \cmidrule(lr){5-10} Average  &  Median & p25 & p75 & <1 year & 1-2 years & 2-4 years & 4-8 years &  8-16 years & >16 years  \\  \midrule
 \input{./output/tables/08_receivership_length.tex} \\
 \bottomrule

    \end{tabular}
 	
		\begin{tablenotes} \item
Notes: This table shows the distribution of the number of days elapsed between the date a receiver was appointed and the date on which the receivership was closed, for all receiverships started between 1863 and 1934 that were already closed by 1941. 

\end{tablenotes}
\normalsize
\end{threeparttable}
\end{center}
\end{table}

%% file: output/tables/03_tab_sumstats_prewar.tex
{
\def\sym#1{\ifmmode^{#1}\else\(^{#1}\)\fi}
\begin{tabular}{l*{1}{ccccccccc}}
\toprule
                    &           N&        Mean&   Std. dev.&         1st&        10th&        25th&        75th&        90th&        99th\\
\midrule
Failing bank        &     361,878&        0.18&        0.38&        0.00&        0.00&        0.00&        0.00&        1.00&        1.00\\
Surplus profit/equity&     361,576&        0.34&        0.19&        0.01&        0.11&        0.20&        0.49&        0.61&        0.79\\
Noncore funding/assets&     361,291&        0.01&        0.04&        0.00&        0.00&        0.00&        0.00&        0.05&        0.21\\
(Bills payable and rediscounts)/assets&     178,266&        0.01&        0.04&        0.00&        0.00&        0.00&        0.00&        0.04&        0.18\\
Equity/assets       &     361,213&        0.23&        0.13&        0.07&        0.10&        0.13&        0.31&        0.42&        0.59\\
Loans/assets        &     361,144&        0.52&        0.26&        0.13&        0.30&        0.42&        0.64&        0.71&        0.81\\
Deposits/assets     &     355,621&        0.63&        0.20&        0.12&        0.33&        0.50&        0.79&        0.86&        0.92\\
Liquid assets/assets&     355,794&        0.18&        0.11&        0.04&        0.07&        0.10&        0.23&        0.32&        0.53\\
OREO/loans          &      63,126&        0.02&        0.04&        0.00&        0.00&        0.00&        0.01&        0.05&        0.18\\
3-year asset growth (real)&     346,631&       -0.00&        0.69&       -1.94&       -0.77&       -0.31&        0.30&        0.77&        1.94\\
\bottomrule
\end{tabular}
}

%% file: output/tables/03_tab_sumstats_postwar.tex
{
\def\sym#1{\ifmmode^{#1}\else\(^{#1}\)\fi}
\begin{tabular}{l*{1}{ccccccccc}}
\toprule
                    &           N&        Mean&   Std. dev.&         1st&        10th&        25th&        75th&        90th&        99th\\
\midrule
Failing bank        &   2,528,198&        0.06&        0.24&        0.00&        0.00&        0.00&        0.00&        0.00&        1.00\\
Net income/assets   &     692,602&        0.01&        0.02&       -0.03&        0.00&        0.01&        0.01&        0.02&        0.03\\
Noncore funding/assets&   2,441,622&        0.36&        0.17&        0.00&        0.10&        0.25&        0.49&        0.56&        0.70\\
Other borrowed money/assets&   2,441,640&        0.01&        0.04&        0.00&        0.00&        0.00&        0.00&        0.05&        0.18\\
Deposits/assets     &   2,440,557&        0.86&        0.10&        0.45&        0.79&        0.85&        0.91&        0.92&        0.94\\
Time deposits/assets&   2,403,753&        0.35&        0.16&        0.00&        0.11&        0.24&        0.48&        0.55&        0.67\\
Equity/assets       &   2,439,490&        0.10&        0.07&        0.04&        0.06&        0.07&        0.11&        0.14&        0.35\\
Loans/assets        &   1,926,615&        0.57&        0.16&        0.08&        0.36&        0.47&        0.68&        0.77&        0.89\\
Liquid assets/assets&   2,441,047&        0.38&        0.16&        0.06&        0.17&        0.26&        0.49&        0.60&        0.80\\
Brokered deposits/assets&   1,484,932&        0.01&        0.08&        0.00&        0.00&        0.00&        0.00&        0.03&        0.23\\
NPL/loans           &   1,397,374&        0.02&        0.03&        0.00&        0.00&        0.00&        0.02&        0.04&        0.13\\
LLP/loans           &     471,045&        0.01&        0.43&       -0.00&        0.00&        0.00&        0.01&        0.01&        0.07\\
NIM                 &   1,941,223&        0.01&        0.01&       -0.02&       -0.00&        0.00&        0.02&        0.03&        0.05\\
3-year asset growth (real)&   2,089,168&        0.14&        0.31&       -0.38&       -0.11&       -0.01&        0.23&        0.42&        1.32\\
\bottomrule
\end{tabular}
}

%% file: output/tables/05_tab_auc_NBEra.tex
AUC (in-sample)     &       0.760&       0.780&       0.818&       0.847&       0.869&       0.796&       0.773

%% file: output/tables/05_tab_auc_oos_NBEra.tex
AUC (out-of-sample) &       0.736&       0.776&       0.806&       0.829&       0.833&       0.783&       0.766\\
\cmidrule(lr){1-8} N&       96044&       95871&       95871&       94466&       71591&       94466&       94466\\
No of Banks         &        5701&        5698&        5698&        5542&        5128&        5542&        5542\\
Mean of dep. var.   &         .37&         .37&         .37&         .37&         .24&           1&         1.6

%% file: output/tables/05_tab_auc_early_fed.tex
AUC (in-sample)     &       0.815&       0.827&       0.876&       0.892&       0.900&       0.845&       0.788

%% file: output/tables/05_tab_auc_oos_early_fed.tex
AUC (out-of-sample) &       0.816&       0.801&       0.880&       0.904&       0.905&       0.814&       0.759\\
\cmidrule(lr){1-8} N&      110443&      110491&      110443&      109529&      106636&      109529&      109529\\
No of Banks         &        9500&        9500&        9500&        9433&        9182&        9433&        9433\\
Mean of dep. var.   &         .61&         .61&         .61&         .61&         .34&         2.4&         5.3

%% file: output/tables/05_tab_auc_GD.tex
AUC (in-sample)     &       0.748&       0.773&       0.822&       0.833&       0.823&       0.806&       0.808

%% file: output/tables/05_tab_auc_oos_GD.tex
AUC (out-of-sample) &       0.638&       0.730&       0.731&       0.727&       0.674&       0.705&       0.736\\
\cmidrule(lr){1-8} N&       35102&       35104&       35102&       34996&       34720&       34996&       34996\\
No of Banks         &        7461&        7461&        7461&        7451&        7303&        7451&        7451\\
Mean of dep. var.   &         3.5&         3.5&         3.5&         3.5&         1.7&         9.8&          12

%% file: output/tables/05_tab_predicting_failure_historical.tex
                    &         (1)   &         (2)   &         (3)   &         (4)   &         (5)   &         (6)   &         (7)   \\
\cmidrule(lr){1-8} \textbf{\textit{Solvency:}}&               &               &               &               &               &               &               \\
\cmidrule(lr){1-8} - Surplus/Equity&       -3.38***&               &       -1.57***&       -1.19** &       -0.54** &       -2.87** &       -2.78** \\
                    &      (1.16)   &               &      (0.58)   &      (0.49)   &      (0.27)   &      (1.12)   &      (1.31)   \\
\cmidrule(lr){1-8} \textbf{\textit{Funding:}}&               &               &               &               &               &               &               \\
- Noncore Funding/Assets&               &       34.81***&       64.47***&       64.06***&       32.93***&      120.80***&      142.57***\\
                    &               &     (10.05)   &     (16.18)   &     (15.13)   &      (6.71)   &     (18.20)   &     (17.06)   \\
\cmidrule(lr){1-8} \textbf{\textit{Solvency $\times$ Funding:}}&               &               &               &               &               &               &               \\
- Surplus/Equity $\times$  Noncore Fund./Assets&               &               &      -91.66***&      -94.14***&      -47.13***&     -139.37***&     -142.95***\\
                    &               &               &     (19.35)   &     (19.02)   &      (8.58)   &     (17.54)   &     (22.17)   \\
\cmidrule(lr){1-8} \textbf{ \textit{Bank Growth:}}&               &               &               &               &               &               &               \\
- Q1 of Growth from t-3 to t&               &               &               &        0.87***&        0.56***&        1.83***&        2.12***\\
                    &               &               &               &      (0.30)   &      (0.21)   &      (0.66)   &      (0.79)   \\
- Q2 of Growth from t-3 to t&               &               &               &        0.12*  &        0.08*  &        0.40** &        0.41*  \\
                    &               &               &               &      (0.07)   &      (0.05)   &      (0.19)   &      (0.22)   \\
- Q4 of Growth from t-3 to t&               &               &               &       -0.10   &       -0.06   &       -0.30   &       -0.43*  \\
                    &               &               &               &      (0.08)   &      (0.05)   &      (0.19)   &      (0.23)   \\
- Q5 of Growth from t-3 to t&               &               &               &       -0.08   &       -0.06   &       -0.04   &        0.09   \\
                    &               &               &               &      (0.08)   &      (0.04)   &      (0.16)   &      (0.26)   \\
\cmidrule(lr){1-8} \textbf{\textit{Aggregate Conditions:}}&               &               &               &               &               &               &               \\
- GDP Growth from t-3 to t&               &               &               &       -4.07   &       -0.56   &       -6.68   &       -3.93   \\
                    &               &               &               &      (2.52)   &      (0.93)   &      (4.52)   &      (5.02)   \\
- CPI Inf. from t-3 to t&               &               &               &       -0.91   &       -0.65*  &       -3.45   &       -4.23   \\
                    &               &               &               &      (0.59)   &      (0.35)   &      (2.16)   &      (3.64)   \\
\cmidrule(lr){1-8} N&      294574   &      294253   &      294247   &      290088   &      262636   &      290088   &      290088   \\
No of Banks         &       12536   &       12535   &       12535   &       12428   &       11851   &       12428   &       12428   \\
Mean of dep. var.   &         .79   &         .79   &         .79   &          .8   &         .45   &         2.5   &         4.1   

%% file: output/tables/05_tab_predicting_failure_modern_era.tex
                    &         (1)   &         (2)   &         (3)   &         (4)   &         (5)   &         (6)   &         (7)   \\
\cmidrule(lr){1-8} \textbf{\textit{Solvency:}}&               &               &               &               &               &               &               \\
- Net Income/Assets &      -72.64***&               &       16.54***&       17.70***&        1.56** &       20.79***&       20.17***\\
                    &     (25.45)   &               &      (3.92)   &      (4.13)   &      (0.68)   &      (3.88)   &      (3.69)   \\
\cmidrule(lr){1-8} \textbf{\textit{Noncore funding:}}&               &               &               &               &               &               &               \\
- Noncore Funding/Assets&               &        2.41***&        5.20***&        5.46***&        0.52***&       11.07***&       13.66***\\
                    &               &      (0.82)   &      (0.77)   &      (0.82)   &      (0.12)   &      (1.90)   &      (2.35)   \\
\cmidrule(lr){1-8} \textbf{\textit{Solvency $\times$ Noncore funding:}}&               &               &               &               &               &               &               \\
- NI/Assets $\times$ Noncore Fund./Assets&               &               &     -426.31***&     -428.38***&      -42.93***&     -660.42***&     -684.30***\\
                    &               &               &     (67.48)   &     (67.63)   &     (11.23)   &     (94.50)   &     (95.61)   \\
\cmidrule(lr){1-8} \textbf{ \textit{Bank Growth:}}&               &               &               &               &               &               &               \\
- Q1 of Growth from t-3 to t&               &               &               &        0.13** &       -0.00   &        0.40***&        0.56***\\
                    &               &               &               &      (0.06)   &      (0.02)   &      (0.12)   &      (0.15)   \\
- Q2 of Growth from t-3 to t&               &               &               &       -0.06*  &        0.00   &       -0.11** &       -0.11** \\
                    &               &               &               &      (0.03)   &      (0.01)   &      (0.05)   &      (0.05)   \\
- Q4 of Growth from t-3 to t&               &               &               &        0.04   &        0.01   &        0.10*  &        0.22** \\
                    &               &               &               &      (0.03)   &      (0.01)   &      (0.05)   &      (0.09)   \\
- Q5 of Growth from t-3 to t&               &               &               &       -0.07   &       -0.00   &        0.24   &        0.80** \\
                    &               &               &               &      (0.05)   &      (0.01)   &      (0.16)   &      (0.34)   \\
\cmidrule(lr){1-8} \textbf{\textit{Aggregate Conditions:}}&               &               &               &               &               &               &               \\
- GDP Growth from t-3 to t&               &               &               &        0.67   &        0.25   &        3.78   &        4.57   \\
                    &               &               &               &      (0.93)   &      (0.36)   &      (3.92)   &      (6.03)   \\
- CPI Inf. from t-3 to t&               &               &               &        0.06   &        0.40** &       -0.42   &        1.05   \\
                    &               &               &               &      (0.24)   &      (0.19)   &      (0.91)   &      (1.85)   \\
\cmidrule(lr){1-8} N&      633407   &      633404   &      633404   &      590645   &      215320   &      590645   &      590645   \\
No of Banks         &       23107   &       23106   &       23106   &       22348   &       13900   &       22348   &       22348   \\
Mean of dep. var.   &         .32   &         .32   &         .32   &         .34   &        .032   &         .97   &         1.5  

%% file: output/tables/05_tab_auc_historical_glm.tex
 
AUC (in-sample)     &       0.685&       0.798&       0.803&       0.863&       0.857&       0.802&       0.736

%% file: output/tables/05_tab_auc_oos_historical_glm.tex
AUC (out-of-sample) &       0.768&       0.838&       0.836&       0.854&       0.843&       0.813&       0.775\\
\cmidrule(lr){1-8} N&      294574&      294253&      294247&      290088&      262636&      290088&      290088\\
No of Banks         &       12536&       12535&       12535&       12428&       11851&       12428&       12428\\
Mean of dep. var.   &         .79&         .79&         .79&          .8&         .45&         2.5&         4.1

%% file: output/tables/05_tab_auc_modern_era_glm.tex
AUC (in-sample)     &       0.946&       0.850&       0.949&       0.949&       0.900&       0.902&       0.856

%% file: output/tables/05_tab_auc_oos_modern_era_glm.tex
         
AUC (out-of-sample) &       0.916&       0.830&       0.938&       0.939&       0.893&       0.879&       0.825\\
\cmidrule(lr){1-8} N&      633407&      633404&      633404&      590645&      215320&      590645&      590645\\
No of Banks         &       23107&       23106&       23106&       22348&       13900&       22348&       22348\\
Mean of dep. var.   &         .32&         .32&         .32&         .34&        .032&         .97&         1.5

%% file: output/tables/05_tab_auc_historical_by_size.tex
AUC (in-sample)     &       0.846&       0.864&       0.864&       0.866&       0.869

%% file: output/tables/05_tab_auc_oos_historical_by_size.tex
AUC (out-of-sample) &       0.830&       0.855&       0.846&       0.856&       0.831\\
\cmidrule(lr){1-6} N&       56575&       58105&       58386&       58540&       58482\\
No of Banks         &        6548&        6831&        6106&        4784&        3127\\
Mean of dep. var.   &         1.3&         .93&         .69&         .57&         .49

%% file: output/tables/05_tab_auc_modern_era_by_size.tex
AUC (in-sample)     &       0.954&       0.951&       0.951&       0.943&       0.949

%% file: output/tables/05_tab_auc_oos_modern_era_by_size.tex
                    &            &            &            &            &            \\
AUC (out-of-sample) &       0.952&       0.944&       0.941&       0.940&       0.938\\
\cmidrule(lr){1-6} N&      122948&      122309&      121169&      117936&      106283\\
No of Banks         &        7251&       10155&       10912&        9614&        6695\\
Mean of dep. var.   &         .44&         .32&         .29&         .27&         .37\\

%% file: output/tables/05_tab_predicting_failure_NBEra.tex
                    &         (1)   &         (2)   &         (3)   &         (4)   &         (5)   &         (6)   &         (7)   \\
\cmidrule(lr){1-8} \textbf{\textit{Solvency:}}&               &               &               &               &               &               &               \\
- Surplus/Equity    &       -1.26***&               &       -0.60***&       -0.55***&       -0.15*  &       -1.88***&       -2.49***\\
                    &      (0.26)   &               &      (0.14)   &      (0.13)   &      (0.08)   &      (0.33)   &      (0.40)   \\
- Dividend Payout Restricted&        1.71***&               &        0.70***&        0.67***&        0.65***&        1.44***&        1.69***\\
                    &      (0.56)   &               &      (0.24)   &      (0.23)   &      (0.19)   &      (0.50)   &      (0.59)   \\
- Equity/Assets     &       -1.03***&               &       -0.93***&       -1.64***&       -0.92***&       -4.40***&       -6.73***\\
                    &      (0.29)   &               &      (0.23)   &      (0.26)   &      (0.25)   &      (0.66)   &      (0.91)   \\
- Loans/Assets      &        1.32***&               &        0.53** &        0.38   &       -0.14   &        1.66***&        3.07***\\
                    &      (0.34)   &               &      (0.26)   &      (0.24)   &      (0.19)   &      (0.41)   &      (0.54)   \\
\cmidrule(lr){1-8} \textbf{\textit{Noncore funding:}}&               &               &               &               &               &               &               \\
- (Bills Payable + Rediscounts)/Assets&               &       17.38***&       27.59***&       27.48***&       17.39***&       53.57***&       71.43***\\
                    &               &      (3.79)   &      (5.11)   &      (5.36)   &      (3.74)   &     (12.40)   &     (13.28)   \\
\cmidrule(lr){1-8} \textbf{\textit{Solvency $\times$ Noncore funding:}}&               &               &               &               &               &               &               \\
- Surplus/Equity $\times$ (Bills Pay. + Redis.)/Assets&               &               &      -57.92***&      -56.57***&      -35.02***&     -102.55***&     -130.79***\\
                    &               &               &     (12.34)   &     (13.31)   &      (9.70)   &     (26.93)   &     (26.14)   \\
- Div. Restricted $\times$ (Bills Pay. + Redis.)/Assets&               &               &       52.00***&       52.19***&       36.98** &       53.83***&       43.61** \\
                    &               &               &     (17.40)   &     (17.05)   &     (15.94)   &     (17.55)   &     (18.73)   \\
\cmidrule(lr){1-8} \textbf{ \textit{Bank Growth:}}&               &               &               &               &               &               &               \\
- Q1 of Growth from t-3 to t&               &               &               &        0.46***&        0.31***&        0.90***&        0.98***\\
                    &               &               &               &      (0.12)   &      (0.09)   &      (0.27)   &      (0.28)   \\
- Q2 of Growth from t-3 to t&               &               &               &        0.14***&        0.12** &        0.19** &        0.10   \\
                    &               &               &               &      (0.05)   &      (0.06)   &      (0.09)   &      (0.10)   \\
- Q4 of Growth from t-3 to t&               &               &               &        0.03   &        0.02   &       -0.02   &       -0.03   \\
                    &               &               &               &      (0.08)   &      (0.06)   &      (0.09)   &      (0.09)   \\
- Q5 of Growth from t-3 to t&               &               &               &       -0.07   &       -0.11** &        0.04   &        0.05   \\
                    &               &               &               &      (0.05)   &      (0.05)   &      (0.16)   &      (0.24)   \\
\cmidrule(lr){1-8} \textbf{\textit{Aggregate Conditions:}}&               &               &               &               &               &               &               \\
- GDP Growth from t-3 to t&               &               &               &       -1.69***&       -0.91** &       -3.94***&       -4.80** \\
                    &               &               &               &      (0.47)   &      (0.41)   &      (1.48)   &      (1.89)   \\
- CPI Inf. from t-3 to t&               &               &               &       -0.71   &       -0.87   &       -1.32   &       -0.93   \\
                    &               &               &               &      (0.62)   &      (0.70)   &      (1.68)   &      (2.36)   \\
\cmidrule(lr){1-8} N&       96044   &       95871   &       95871   &       94466   &       71591   &       94466   &       94466   \\
No of Banks         &        5701   &        5698   &        5698   &        5542   &        5128   &        5542   &        5542   \\
Mean of dep. var.   &         .37   &         .37   &         .37   &         .37   &         .24   &           1   &         1.6   

%% file: output/tables/05_tab_predicting_failure_early_fed.tex
                    &         (1)   &         (2)   &         (3)   &         (4)   &         (5)   &         (6)   &         (7)   \\
\cmidrule(lr){1-8} \textbf{\textit{Solvency:}}&               &               &               &               &               &               &               \\
\cmidrule(lr){1-8} - Surplus/Equity&       -4.01***&               &       -1.63***&       -1.18***&       -0.59***&       -4.80***&       -8.67***\\
                    &      (1.24)   &               &      (0.53)   &      (0.35)   &      (0.19)   &      (1.40)   &      (2.40)   \\
- Loans/Assets      &        0.56***&               &        0.22*  &        0.22*  &        0.12*  &        1.03*  &        2.32*  \\
                    &      (0.19)   &               &      (0.12)   &      (0.13)   &      (0.07)   &      (0.55)   &      (1.23)   \\
- Equity/Assets     &       -2.06** &               &       -1.99** &       -3.11** &       -1.93** &      -15.23***&      -33.95***\\
                    &      (0.88)   &               &      (0.97)   &      (1.28)   &      (0.80)   &      (4.28)   &      (8.61)   \\
\cmidrule(lr){1-8} \textbf{\textit{Noncore funding:}}&               &               &               &               &               &               &               \\
- Noncore funding/Assets&               &       25.31***&       52.45***&       52.32***&       31.88***&      114.89***&      138.08***\\
                    &               &      (6.50)   &     (12.57)   &     (11.83)   &      (8.86)   &     (13.46)   &     (13.72)   \\
\cmidrule(lr){1-8} \textbf{\textit{Solvency $\times$ Noncore funding:}}&               &               &               &               &               &               &               \\
- Surplus/Equity $\times$  Noncore funding/Assets&               &               &      -82.99***&      -83.86***&      -51.31***&     -162.95***&     -176.14***\\
                    &               &               &     (20.15)   &     (19.28)   &     (14.22)   &     (21.30)   &     (20.15)   \\
- Loans/Assets $\times$ Noncore funding/Assets&               &               &        0.67   &        0.64   &       -0.11   &        0.15   &       -1.79   \\
                    &               &               &      (0.87)   &      (0.81)   &      (0.34)   &      (2.73)   &      (3.97)   \\
\cmidrule(lr){1-8} \textbf{ \textit{Bank Growth:}}&               &               &               &               &               &               &               \\
- Q1 of Growth from t-3 to t&               &               &               &        1.02***&        0.61***&        2.91***&        3.91***\\
                    &               &               &               &      (0.36)   &      (0.21)   &      (0.98)   &      (1.21)   \\
- Q2 of Growth from t-3 to t&               &               &               &        0.07** &        0.06** &        0.60***&        0.91***\\
                    &               &               &               &      (0.03)   &      (0.03)   &      (0.18)   &      (0.27)   \\
- Q4 of Growth from t-3 to t&               &               &               &       -0.07   &       -0.03   &       -0.35** &       -0.97***\\
                    &               &               &               &      (0.06)   &      (0.04)   &      (0.16)   &      (0.33)   \\
- Q5 of Growth from t-3 to t&               &               &               &       -0.13   &       -0.08   &       -0.67** &       -1.30** \\
                    &               &               &               &      (0.11)   &      (0.07)   &      (0.28)   &      (0.53)   \\
\cmidrule(lr){1-8} \textbf{\textit{Aggregate Conditions:}}&               &               &               &               &               &               &               \\
- GDP Growth from t-3 to t&               &               &               &        2.04*  &        1.57** &        3.44   &        3.16   \\
                    &               &               &               &      (1.07)   &      (0.76)   &      (2.22)   &      (7.14)   \\
- CPI Inf. from t-3 to t&               &               &               &       -0.64** &       -0.38*  &       -3.60** &       -7.53*  \\
                    &               &               &               &      (0.31)   &      (0.21)   &      (1.67)   &      (4.09)   \\
\cmidrule(lr){1-8} N&      110443   &      110491   &      110443   &      109529   &      106636   &      109529   &      109529   \\
No of Banks         &        9500   &        9500   &        9500   &        9433   &        9182   &        9433   &        9433   \\
Mean of dep. var.   &         .61   &         .61   &         .61   &         .61   &         .34   &         2.4   &         5.3   

%% file: output/tables/05_tab_predicting_failure_GD.tex
                    &         (1)   &         (2)   &         (3)   &         (4)   &         (5)   &         (6)   &         (7)   \\
\cmidrule(lr){1-8} \textbf{\textit{Solvency:}}&               &               &               &               &               &               &               \\
- Equity/Assets     &       -9.53***&               &      -10.82***&      -15.26***&       -7.69** &      -43.56***&      -54.47***\\
                    &      (2.85)   &               &      (2.63)   &      (4.42)   &      (3.60)   &     (12.93)   &     (18.28)   \\
- Surplus/Equity    &      -10.99***&               &       -4.08** &       -2.38** &       -1.76   &      -10.21** &       -9.82*  \\
                    &      (3.90)   &               &      (1.99)   &      (0.94)   &      (1.10)   &      (4.68)   &      (5.52)   \\
- Dividend Payout Restricted&        3.34***&               &        1.79***&        1.51***&        0.27** &        1.60** &        1.04   \\
                    &      (0.46)   &               &      (0.34)   &      (0.31)   &      (0.13)   &      (0.72)   &      (1.20)   \\
- Loans/Assets      &       12.01***&               &        4.51***&        4.10***&        2.31** &       13.31***&       17.73***\\
                    &      (1.85)   &               &      (0.93)   &      (1.35)   &      (1.12)   &      (3.64)   &      (4.25)   \\
\cmidrule(lr){1-8} \textbf{\textit{Noncore funding:}}&               &               &               &               &               &               &               \\
- (Bills Payable + Rediscounts)/Assets&               &       99.29***&      169.76***&      166.42***&       57.92***&      243.52***&      245.73***\\
                    &               &      (8.51)   &     (13.08)   &     (13.71)   &     (19.59)   &     (19.78)   &     (19.71)   \\
\cmidrule(lr){1-8} \textbf{\textit{Solvency $\times$ Noncore funding:}}&               &               &               &               &               &               &               \\
- Surplus/Equity $\times$ (Bills Pay. + Redis.)/Assets&               &               &     -224.58***&     -218.83***&      -60.57***&     -217.17***&     -191.81***\\
                    &               &               &     (16.47)   &     (18.20)   &     (20.40)   &     (25.53)   &     (18.72)   \\
- Div. Payout Restr. $\times$ (Bills Pay. + Redis.)/Assets&               &               &        9.55   &       10.15   &       12.57   &       -7.65   &      -14.82   \\
                    &               &               &     (14.71)   &     (13.53)   &      (8.60)   &     (18.11)   &     (14.63)   \\
\cmidrule(lr){1-8} \textbf{ \textit{Bank Growth:}}&               &               &               &               &               &               &               \\
- Q1 of Growth from t-3 to t&               &               &               &        2.80***&        1.70** &        4.16** &        4.35** \\
                    &               &               &               &      (1.07)   &      (0.79)   &      (1.76)   &      (1.93)   \\
- Q2 of Growth from t-3 to t&               &               &               &        0.52*  &        0.26   &        1.57***&        1.72***\\
                    &               &               &               &      (0.27)   &      (0.22)   &      (0.60)   &      (0.66)   \\
- Q4 of Growth from t-3 to t&               &               &               &       -0.86***&       -0.46** &       -2.16***&       -2.35** \\
                    &               &               &               &      (0.28)   &      (0.20)   &      (0.81)   &      (0.98)   \\
- Q5 of Growth from t-3 to t&               &               &               &       -0.81** &       -0.36   &       -1.92** &       -2.49** \\
                    &               &               &               &      (0.35)   &      (0.24)   &      (0.83)   &      (1.25)   \\
\cmidrule(lr){1-8} \textbf{\textit{Aggregate Conditions:}}&               &               &               &               &               &               &               \\
- GDP Growth from t-3 to t&               &               &               &       -0.98   &        0.13   &      106.34   &      143.51   \\
                    &               &               &               &     (45.05)   &     (30.56)   &     (95.01)   &    (105.54)   \\
- CPI Inf. from t-3 to t&               &               &               &       -0.22   &        9.62   &     -132.53   &     -153.03   \\
                    &               &               &               &     (68.87)   &     (46.03)   &    (145.79)   &    (170.23)   \\
\cmidrule(lr){1-8} N&       35102   &       35104   &       35102   &       34996   &       34720   &       34996   &       34996   \\
No of Banks         &        7461   &        7461   &        7461   &        7451   &        7303   &        7451   &        7451   \\
Mean of dep. var.   &         3.5   &         3.5   &         3.5   &         3.5   &         1.7   &         9.8   &          12   

%% file: output/tables/99_TPR_FPR_TNR_FNR_historical_ols.tex
0.8\%&0.856&0.311&0.689&0.144\\
1\%&0.818&0.252&0.748&0.182\\
1.5\%&0.726&0.147&0.853&0.274\\
2\%&0.659&0.096&0.904&0.341\\
2.5\%&0.597&0.073&0.927&0.403\\
3\%&0.545&0.059&0.941&0.455\\
4\%&0.468&0.040&0.960&0.532\\
5\%&0.390&0.028&0.972&0.610\\
10\%&0.151&0.006&0.994&0.849

%% file: output/tables/99_TPR_FPR_TNR_FNR_modern_ols.tex
0.3\%&0.958&0.367&0.633&0.042\\
0.5\%&0.952&0.279&0.721&0.048\\
1\%&0.933&0.144&0.856&0.067\\
1.5\%&0.908&0.083&0.917&0.092\\
2\%&0.889&0.055&0.945&0.111\\
2.5\%&0.866&0.040&0.960&0.134\\
3\%&0.837&0.031&0.969&0.163\\
4\%&0.786&0.021&0.979&0.214\\
5\%&0.738&0.015&0.985&0.262\\
10\%&0.512&0.004&0.996&0.488

%% file: output/tables/99_TPR_FPR_TNR_FNR_historical_logit.tex
0.8\%&0.774&0.203&0.797&0.226\\
1\%&0.721&0.154&0.846&0.279\\
1.5\%&0.623&0.088&0.912&0.377\\
2\%&0.542&0.058&0.942&0.458\\
2.5\%&0.487&0.042&0.958&0.513\\
3\%&0.437&0.033&0.967&0.563\\
4\%&0.367&0.022&0.978&0.633\\
5\%&0.317&0.017&0.983&0.683\\
10\%&0.183&0.006&0.994&0.817

%% file: output/tables/99_TPR_FPR_TNR_FNR_modern_logit.tex
0.3\%&0.899&0.128&0.872&0.101\\
0.5\%&0.850&0.063&0.937&0.150\\
1\%&0.753&0.024&0.976&0.247\\
1.5\%&0.700&0.015&0.985&0.300\\
2\%&0.656&0.011&0.989&0.344\\
2.5\%&0.617&0.009&0.991&0.383\\
3\%&0.592&0.008&0.992&0.408\\
4\%&0.543&0.006&0.994&0.457\\
5\%&0.508&0.005&0.995&0.492\\
10\%&0.395&0.003&0.997&0.605

%% file: output/tables/pr_auc_1863_1934.tex
PR-AUC (in-sample)  &  0.026 &  0.059 &  0.076 &  0.087 &  0.042 &  0.145 &  0.143 \\
PR-AUC (out-of-sample) &  0.033 &  0.082 &  0.109 &  0.110 &  0.045 &  0.170 &  0.184 \\
Mean of dep. var. &  0.008 &  0.008 &  0.008 &  0.008 &  0.004 &  0.025 &  0.025 \\
Ratio (in-sample) &  3.250 &  7.382 &  9.544 & 10.909 &  9.441 &  5.764 &  5.673 \\
Ratio (out-of-sample) &  4.054 & 10.129 & 13.380 & 13.485 &  9.364 &  6.605 &  7.148 \\
Precision at 10\% recall (in-smp.) &  0.059 &  0.121 &  0.157 &  0.177 &  0.073 &  0.236 &  0.232 \\
Precision at 10\% recall (o.o.s.) &  0.054 &  0.147 &  0.181 &  0.190 &  0.071 &  0.296 &  0.308 

%% file: output/tables/pr_auc_1959_2024.tex
PR-AUC (in-sample)  &  0.212 &  0.034 &  0.297 &  0.298 &  0.050 &  0.282 &  0.283 \\
PR-AUC (out-of-sample) &  0.218 &  0.022 &  0.288 &  0.287 &  0.040 &  0.273 &  0.279 \\
Mean of dep. var. & 0.0032 & 0.0032 & 0.0032 & 0.0034 & 0.0003 & 0.0097 & 0.0097 \\
Ratio (in-sample) & 66.128 & 10.677 & 92.814 & 87.826 & 155.200 & 29.107 & 29.255 \\
Ratio (out-of-sample) & 56.932 &  5.742 & 75.395 & 75.115 & 102.984 & 25.029 & 25.637 \\
Precision at 10\% recall (in-smp.) &  0.330 &  0.050 &  0.453 &  0.453 &  0.071 &  0.509 &  0.508 \\
Precision at 10\% recall (o.o.s.) &  0.372 &  0.027 &  0.466 &  0.465 &  0.068 &  0.507 &  0.516 

%% file: output/tables/06_pred_prob_of_failure_baseline.tex
Baseline&0.07&0.02&0.40&0.32&0.10&0.08&0.04&0.03&0.04

%% file: output/tables/06_pred_prob_of_failure_granular.tex
Richer model&0.18&0.06&0.23&0.25&0.11&0.11&0.06&0.05&0.18

%% file: output/tables/06_pred_prob_of_failure_baseline_3year.tex
Baseline&0.14&0.05&0.13&0.38&0.15&0.12&0.07&0.05&0.12

%% file: output/tables/06_pred_prob_of_failure_granular_3year.tex
Richer Model&0.32&0.20&0.06&0.20&0.10&0.13&0.08&0.07&0.34

%% file: output/tables/06_required_rate_risk_neutral.tex
Baseline&0.04&0.01&0.46&0.13&0.16&0.09&0.07&0.03&0.06

%% file: output/tables/06_required_rate_risk_neutral_granular.tex
Richer model&0.10&0.02&0.28&0.10&0.16&0.11&0.10&0.06&0.20

%% file: output/tables/06_required_rate_log.tex
Baseline&0.05&0.01&0.41&0.13&0.17&0.10&0.07&0.04&0.08

%% file: output/tables/06_required_rate_log_granular.tex
Richer model&0.10&0.02&0.26&0.11&0.16&0.11&0.11&0.06&0.20

%% file: output/tables/06_required_rate_crra.tex
Baseline&0.13&0.05&0.28&0.04&0.08&0.10&0.12&0.09&0.30

%% file: output/tables/06_required_rate_crra_granular.tex
Richer model&0.17&0.10&0.16&0.03&0.08&0.10&0.12&0.13&0.39

%% file: output/tables/04_pre_FDIC_auc_by_crisis.tex
AUC                 &       0.851&       0.819&       0.796&       0.758&       0.776&       0.861

%% file: output/tables/04_postwar_auc_by_era.tex
AUC                 &       0.848&       0.951&       0.862&       0.966&       0.956&       0.967

%% file: output/tables/06_aggregate_predicted_actual_regs_rob.tex
                    &         (1)   &         (2)   &         (3)   &         (4)   \\
\cmidrule(lr){1-5} Predicted failure rate, \( \overline p_{t|t-1} \) &        1.86***&        2.64***&        1.57***&        1.25***\\
                    &      (0.34)   &      (0.77)   &      (0.59)   &      (0.24)   \\
Constant            &        0.14   &       -0.31   &        0.01   &        0.08   \\
                    &      (0.09)   &      (0.20)   &      (0.09)   &      (0.05)   \\
\cmidrule(lr){1-5} N&         117   &         117   &         117   &         117   \\
$R^2$               &        0.21   &        0.47   &        0.40   &        0.40   \\
\Cref{tab:AUC} model&      Col. 1   &      Col. 2   &      Col. 3   &      Col. 4   \\
Sample              &        Full   &        Full   &        Full   &        Full   

%% file: output/tables/06_aggregate_predicted_actual_regs_logit.tex
                    &         (1)   &         (2)   &         (3)   &         (4)   \\
\cmidrule(lr){1-5} Predicted failure rate, \( \overline p_{t|t-1} \) &        0.73***&        0.68***&        1.24***&        1.00***\\
                    &      (0.06)   &      (0.04)   &      (0.11)   &      (0.15)   \\
Constant            &        0.17   &        0.31   &       -0.14*  &        0.00   \\
                    &      (0.11)   &      (0.19)   &      (0.08)   &      (0.08)   \\
\cmidrule(lr){1-5} N&         116   &          60   &          55   &         110   \\
$R^2$               &        0.50   &        0.47   &        0.82   &        0.50   \\
Sample              &        Full   &   1874-1934   &   1970-2024   &Exclude 1929-1934   

%% file: output/tables/06_aggregate_predicted_actual_regs_granular.tex
                    &         (1)   &         (2)   &         (3)   &         (4)   \\
\cmidrule(lr){1-5} Predicted failure rate, \( \overline p_{t|t-1} \) &        0.83***&        0.83***&        0.97***&        0.78***\\
                    &      (0.08)   &      (0.10)   &      (0.15)   &      (0.15)   \\
Constant            &        0.11** &        0.16   &        0.06   &        0.10** \\
                    &      (0.05)   &      (0.11)   &      (0.04)   &      (0.04)   \\
\cmidrule(lr){1-5} N&          97   &          41   &          55   &          92   \\
$R^2$               &        0.59   &        0.56   &        0.81   &        0.58   \\
Sample              &        Full   &   1874-1934   &   1970-2024   &Exclude 1929-1934   

%% file: output/tables/07_predicting_recovery_rates_2.tex
                    &         (1)   &         (2)   &         (3)   &         (4)   &         (5)   &         (6)   &         (7)   \\
\cmidrule(lr){1-8} 2 banks in market before failure&       -1.90** &       -3.38***&       -1.24*  &               &               &               &               \\
                    &      (0.80)   &      (0.77)   &      (0.72)   &               &               &               &               \\
3 banks             &       -3.34***&       -5.61***&       -1.98*  &               &               &               &               \\
                    &      (1.28)   &      (1.24)   &      (1.18)   &               &               &               &               \\
4 or more banks     &       -3.27***&       -8.91***&       -3.70***&               &               &               &               \\
                    &      (1.20)   &      (1.24)   &      (1.24)   &               &               &               &               \\
Receivership length (in years)&               &               &               &       -0.51***&        0.11   &       -0.78***&       -0.15   \\
                    &               &               &               &      (0.12)   &      (0.11)   &      (0.12)   &      (0.11)   \\
Loans/assets        &               &      -27.74***&      -17.77***&               &               &      -27.96***&      -17.60***\\
                    &               &      (2.01)   &      (2.02)   &               &               &      (2.05)   &      (2.06)   \\
Surplus/equity      &               &        7.84***&        6.33***&               &               &       11.21***&        8.29***\\
                    &               &      (2.00)   &      (1.84)   &               &               &      (2.01)   &      (1.87)   \\
Size (log(assets))  &               &        3.01***&        1.76***&               &               &        2.52***&        1.73***\\
                    &               &      (0.35)   &      (0.34)   &               &               &      (0.37)   &      (0.34)   \\
\cmidrule(lr){1-8} N&        2717   &        2708   &        2704   &        2502   &        2498   &        2494   &        2490   \\
$R^2$               &       0.006   &       0.129   &       0.313   &       0.007   &       0.272   &       0.131   &       0.316   \\
Year FE             &               &               &  \checkmark   &               &  \checkmark   &               &  \checkmark   

%% file: output/tables/04_depositor_losses_preFDIC.tex
1863-1913 (NB Era)&533&0.68&0.39&0.27\\
1914-1918 (Early Fed)&732&0.91&0.52&0.47\\
1929-1933 (Depr., pre-Holiday)&1031&0.90&0.43&0.39\\
1933-1934 (Depr., post-Holiday)&605&0.65&0.28&0.18 \\

%% file: output/tables/04_depositor_losses_preFDIC_all.tex
All&2901&0.81&0.42&0.34

%% file: output/tables/07_recovery_rho_v_run.tex
0\%&0.88&0.86&0.84&0.82&0.79&0.73&0.66\\
5\%&0.84&0.82&0.79&0.75&0.72&0.65&0.58\\
10\%&0.78&0.74&0.71&0.67&0.63&0.57&0.51\\
20\%&0.59&0.56&0.53&0.50&0.48&0.43&0.37

%% file: output/tables/07_recovery_rho_v_no_run.tex
0\%&0.77&0.72&0.69&0.64&0.59&0.52&0.45\\
5\%&0.68&0.63&0.58&0.55&0.51&0.44&0.39\\
10\%&0.57&0.54&0.50&0.46&0.43&0.38&0.34\\
20\%&0.40&0.38&0.35&0.33&0.31&0.27&0.23

%% file: output/tables/07_recovery_rho_v_double_liability.tex
0\%&0.67&0.64&0.61&0.58&0.54&0.48&0.43\\
5\%&0.61&0.57&0.54&0.50&0.48&0.42&0.37\\
10\%&0.53&0.49&0.47&0.44&0.41&0.36&0.32\\
20\%&0.38&0.36&0.33&0.31&0.29&0.25&0.22

%% file: output/tables/07_recovery_rho_v_alternative_deposits.tex
0\%&0.86&0.84&0.82&0.80&0.77&0.72&0.67\\
5\%&0.82&0.79&0.77&0.74&0.71&0.66&0.59\\
10\%&0.76&0.73&0.70&0.67&0.64&0.58&0.52\\
20\%&0.61&0.57&0.54&0.51&0.48&0.42&0.36

%% file: output/tables/08_receivership_length.tex
5.8&6&4&7&.054&.041&.18&.51&.097&.02

%% file: 99_appendix_data.tex
\clearpage 

\section{Data Appendix}

\label{app:data}

\subsection{Appendix B1: Call Reports}

\label{sec:data_callreports}

\paragraph{OCC Annual Report to Congress: 1863 through 1941}

We use two main data sources on bank balance sheets. Data on national bank balance sheets from 1863 through 1941 are from the Office of the Comptroller of the Currency's (OCC) Annual Report to Congress. For most of the sample, the balance sheets were reported as of September or October of each year, but from 1928 onward, the reporting date shifted to the end of each year.

The data become first available in 1863. \Cref{fig:nb_balance_sheet_1900} shows an example of a balance sheet from the report from 1900. Until 1904, the report has up to three balance sheets on one page, each as its own table. These data have been digitized by \citet{Carlson2022} and made available online. 

Note that the format of the tables changes in 1905. Starting in 1905, balance sheets for multiple banks are reported in tables across two pages. \Cref{fig:nb_balance_sheet_1933} shows an example of the format after 1905 from the annual report to congress of 1933. We digitize these data also using the techniques discussed in \citet{Correia2022}.

To construct the panel, we compile a list of all significant bank events and their dates. Events include chartering, liquidations, and receiverships. The list of events is based on data manually collected by \citet{vanBelkum1968}, augmented by \citet{KeyBankData}, and which we validate by using information from the 1941 ``Alphabetical List of Banks'' \citep{AlphabeticalList}, as well as the corresponding OCC Annual Reports.

{
\setstretch{1.0}
  \begin{figure}[!ht]
    
    \begin{center}
    \caption{Example of a Balance Sheet Reported in the OCC's Annual Report to Congress from 1900. \label{fig:occ-bs1}}   \label{fig:nb_balance_sheet_1900}
    \includegraphics[width=0.9\textwidth]{"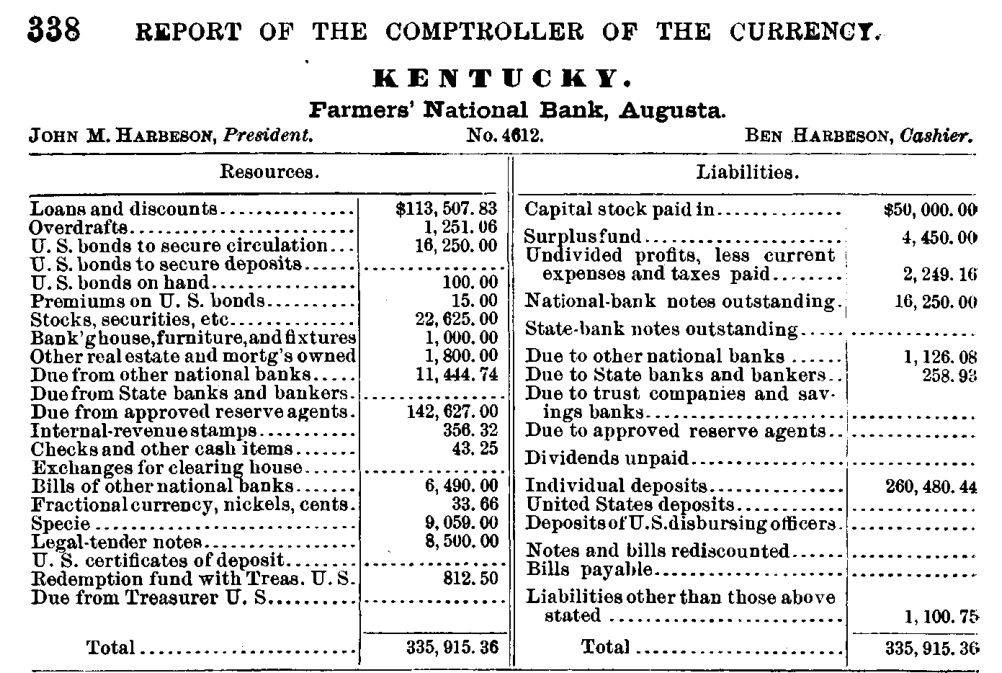"}
    \end{center}
    
  \end{figure}
}

{
\setstretch{1.0}
\begin{landscape}
  \begin{figure}[!ht]

    \begin{center}
    \caption{Example of a Balance Sheet Reported in the OCC's Annual Report to Congress from 1933. \label{fig:occ-bs2}}     \label{fig:nb_balance_sheet_1933}
    \includegraphics[width=1.4\textwidth]{"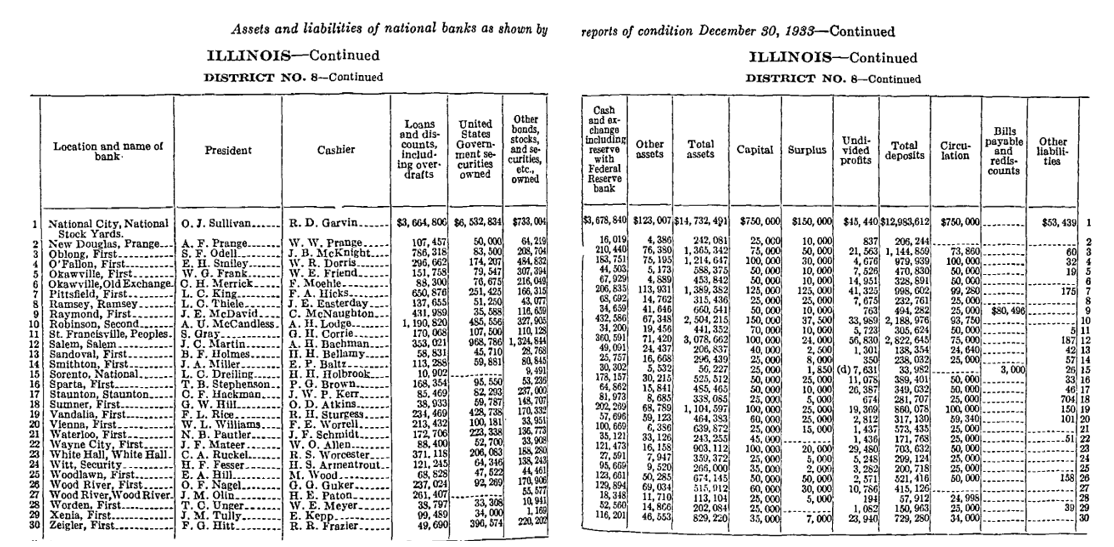"}
    \end{center}
    
  \end{figure}
\end{landscape}
}

\subsubsection{FFIEC Reports of Condition and Income: 1959 through 2024}
\label{sec:data_callreports_modern}
For the modern, contemporary banking system, we use the Federal Financial Institutions Examination Council (FFIEC) Consolidated Reports of Condition and Income (``Call Report''). These data provide bi-annual and since 1974 quarterly information on balance sheets and income statements  on a consolidated basis for all commercial banks operating in the United States and regulated by the FRS, the FDIC, and the OCC. 

We construct the data using the following FFIEC forms:

\begin{itemize}
    \item FFIEC 010: Domestic Report of Condition (< \$100 million in Total Assets). It was filed from June 10, 1959 to December 31, 1983. Prior to December 31, 1960, this was reported by FRS member banks and a handful of nonmember banks. Afterwards, it was reported by all insured commercial banks, savings banks, nondeposit trust companies, and industrial banks with domestic offices only. From December 31, 1978 to December 31, 1983, it was reported by all banks with domestic offices only with less than \$100 million in total assets. From 1964 to 1972, data is reported semiannually. Otherwise, data is reported quarterly.   \Cref{fig:ffiec010-snippet} shows an example of the historical balance sheet reporting form used in 1967. 
 \item FFIEC 011: Consolidated Report of Income (< \$100 million in Total Assets). It was filed from December 31, 1960 to December 31, 1983 by all insured commercial banks, savings banks, nondeposit trust companies, and industrial banks. From December 31, 1978 to December 31, 1983, it was reported by all banks with less than \$100 million in total assets. From 1960 to 1975, data is reported annually at the end of Q4. From 1976 to 1982, data is reported semiannually in Q2 and Q4 for savings banks and nondeposit trust companies. Otherwise, data is reported quarterly on a year-to-date basis. \Cref{fig:ffiec013-snippet} shows an example of the income statement reporting form from 1967.

\item FFIEC 012: Consolidated Report of Condition for a Bank and its Domestic Subsidiaries (Domestic Offices Only). It was filed from December 31, 1978 to December 31, 1983. It was reported by all insured commercial banks, savings banks, nondeposit trust companies, and industrial banks with domestic offices only. Data is reported quarterly. 

\item FFIEC 013: Consolidated Report of Income for a Bank and its Domestic and Foreign Subsidiaries. It was filed from December 31, 1978 to December 31, 1983 by all insured commercial banks, savings banks, nondeposit trust companies, and industrial banks with domestic offices and foreign branches, foreign subsidiaries, Edge Act and agreement subsidiaries, and/or branches in Puerto Rico or U.S. territories and possessions. From 1978 to 1982, data is reported semiannually in Q2 and Q4 for savings banks and nondeposit trust companies. Otherwise, data is reported quarterly on a year-to-date basis. 

\item FFIEC 014: Consolidated Report of Condition for a Bank and its Domestic and Foreign Subsidiaries. It was filed from June 30, 1969 to December 31, 1983. This form was filed by all insured commercial banks with domestic offices and foreign branches, foreign subsidiaries, Edge Act and agreement subsidiaries, and/or branches in Puerto Rico or U.S. territories and possessions. Prior to 1974, the respondent panel consisted of member banks of the Federal Reserve System with domestic and foreign offices. Prior to 1976, foreign holdings were reported semiannually during the June/December reports. Otherwise, data is reported quarterly.  

\item FFIEC 031: Consolidated Reports of Condition and Income for a Bank with Domestic and Foreign Offices. It has been filed since March 31, 1984 to present. It is filed by all national banks, state member banks, insured state nonmember banks, and savings associations with branches or consolidated subsidiaries in a foreign country, Puerto Rico, or a U.S. territory or possession, or have international banking facilities. As of June 30, 2018, it is also filed by banks with domestic offices only, if they have total consolidated assets of \$100 billion or more. Additionally, as of March 2020 banks with domestic offices only are required to file this form if they are an advanced approaches institution. Data is reported quarterly. 

\item FFIEC 032: Consolidated Reports of Condition and Income for a Bank with Domestic Offices Only and Total Assets of \$300 Million or More. It was filed from March 31, 1984 to December 31, 2000. It was filed by all national banks, state member banks, and insured state nonmember banks with domestic offices only and total assets of \$300 million or more. Data is reported quarterly. 

\item FFIEC 033: Consolidated Reports of Condition and Income for a Bank with Domestic Offices Only and Total Assets of \$100 Million or More But Less Than \$300 Million. It was filed from March 31, 1984 to December 31, 2000. It was filed by all national banks, state member banks, and insured state nonmember banks with domestic offices only and total assets of \$100 million or more, but less than \$300 million. Data is reported quarterly. 

\item FFIEC 034: Consolidated Reports of Condition and Income for a Bank with Domestic Offices Only and Total Assets Less Than \$100 Million. It was filed from March 31, 1984 to December 31, 2000. It was filed by all national banks, state member banks, and insured state nonmember banks with domestic offices only and total assets less than \$100 million. Data is reported quarterly. 

\item FFIEC 041: Consolidated Reports of Condition and Income for a Bank with Domestic Offices Only. It has been filed since March 31, 2001 to present. It replaced the FFIEC 032, FFIEC 033, and FFIEC 034 forms. It is filed by all national banks, state member banks, insured state nonmember banks, and savings associations with domestic offices only. Data is reported quarterly. 

\item FFIEC 051: Consolidated Reports of Condition and Income for a Bank with Domestic Offices Only and Total Assets Less than \$5 Billion. It has been filed since March 31, 2017 to present. From March 31, 2017 to June 30, 2019 it was filed by all national banks, state member banks, insured state nonmember banks, and savings associations with domestic offices only and total assets less than \$1 billion. From September 30, 2019 to present, it is filed by all national banks, state member banks, insured state nonmember banks, and savings associations with domestic offices only and total assets less than \$5 billion. Note that while some banks may fall under these criteria, they may elect to file the FFIEC 041 or be required to for regulatory capital purposes or as instructed by their primary federal regulator. Data is reported quarterly. 

\end{itemize}
We document the construction of the  variables from  line items in table \Cref{tab:line-items}.

\begin{table}[tbp] \centering
\caption{\textbf{Definitions of variales cased on call report line items.}} \label{tab:line-items}
\scriptsize
\newcolumntype{R}{>{\raggedleft\arraybackslash}X}
\newcolumntype{L}{>{\raggedright\arraybackslash}X}
\newcolumntype{C}{>{\centering\arraybackslash}X}

\begin{tabularx}{\linewidth}{@{}lCCC@{}}

\toprule
{Item}&{Series}&{Item Number}&{Valid Period} \tabularnewline
\midrule \addlinespace[\belowrulesep]
Assets&RCON&2170&1959--12--31 to present \tabularnewline
Equity&RCON&3210&1959--12--31 to present \tabularnewline
Deposits&RCON&2200&1959--12--31 to present \tabularnewline
Loans&RCON&1400&1959--12--31 to present \tabularnewline
&&2122&1976--03--31 to present \tabularnewline
Cash&RCON&0010&1959--12--31 to present \tabularnewline
Securities &RCON&0400 + 0600 + 0900 + 0950&1959--06--10 to 1976--03--31 \tabularnewline
&&0390&1976--03--31 to 1993--12--31 \tabularnewline
&&1754 + 1773&1994--03--31 to present \tabularnewline
C\&I loans&RCON&1600&1959--12--31 to 1984--03--31 \tabularnewline
&&1766&1984--03--31 to present  \tabularnewline
Real Estate Loans&RCON&1410&1959--12--31 to present \tabularnewline
Consumer Loans&RCON&1975&1959--12--31 to present \tabularnewline
Credit Card Loans&RCON&2008&1967--12--31 to 2000--12--31 \tabularnewline
&&B538&2001--03--31 to present \tabularnewline
Financial Loans&RCON&1495&1959--06--10 to 1983--12--31 \tabularnewline
&&1505 + 1510 + 1517 + 1756 +1757 &1976--03--31 to 2000--12--31 \tabularnewline
&&B531 + B534 + B535&2001--03--31 to present \tabularnewline
Time Deposits&RCON&2514&1961--04--12 to 1983--12--31 \tabularnewline
&RCON&2604 + 6648&1984--03--31 to 2009--12--31  \tabularnewline
&RCON&J473 + J474 + 6648&2010--03--31 to present  \tabularnewline
Demand Deposits&RCON&2210&1959--12--31 to present \tabularnewline
Brokered Deposits&RCON&2365&1983--09--30 to present \tabularnewline
Insured Deposits&RCON&2702&1983--06--30 to 2006--03--31 \tabularnewline
&RCON&F045 + F049&2006--06--30 to present \tabularnewline
Uninsured Deposits&RCON&2710 - (2722*100) &1983--06--30 to 1992--12--31 \tabularnewline
&RCON&5597&1993--03--31 to present \tabularnewline
Loan Loss Provisions&RIAD&4230&1969--12--31 to present \tabularnewline
Net Income &IADX&5106&1960--12--31 to 1968--12--31 \tabularnewline
&RIAD&4340&1969--12--31 to present \tabularnewline
Non-Performing Loans&RCON&1403 + 1407&1982--12--31 to present  \tabularnewline
Total Interest Income&RIAD&4107&1984--03--31 to present \tabularnewline
Total Interest Expenses&RIAD&4170 + 4180 + 4190 + 4200&1969--12--31 to 1978--09--30 \tabularnewline
&RIAD&4170 + 4180 + 4185 + 4200&1978--12--31 to 1983--12--31 \tabularnewline
&RIAD&4073&1984--03--31 to present  \tabularnewline
Salaries and Employee Benefits&RIAD&4135&1969--12--31 to present  \tabularnewline
Number of Full-Time Employees&RIAD&4150&1969--12--31 to present  \tabularnewline
\bottomrule 

\end{tabularx}
\end{table}

{
\setstretch{1.0}
  \begin{figure}[!ht]
    
    \begin{center}
    \caption{Example of FFIEC 010 Reporting Form from 1967. \label{fig:ffiec010-snippet}}
    \includegraphics[width=0.9\textwidth]{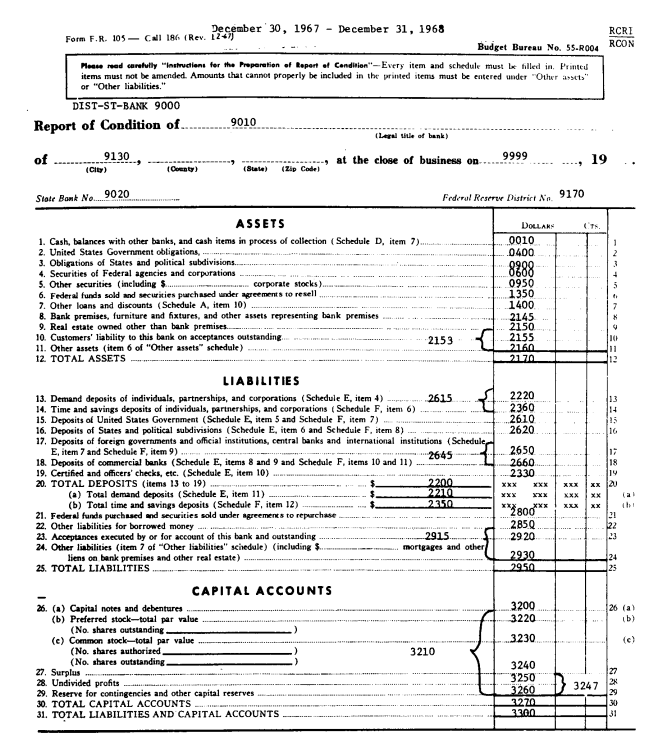}
    \end{center}
    
  \end{figure}
}

{
\setstretch{1.0}
  \begin{figure}[!ht]
    
    \begin{center}
    \caption{Example of FFIEC 013 Reporting Form from 1967. \label{fig:ffiec013-snippet}}
    \includegraphics[width=0.9\textwidth]{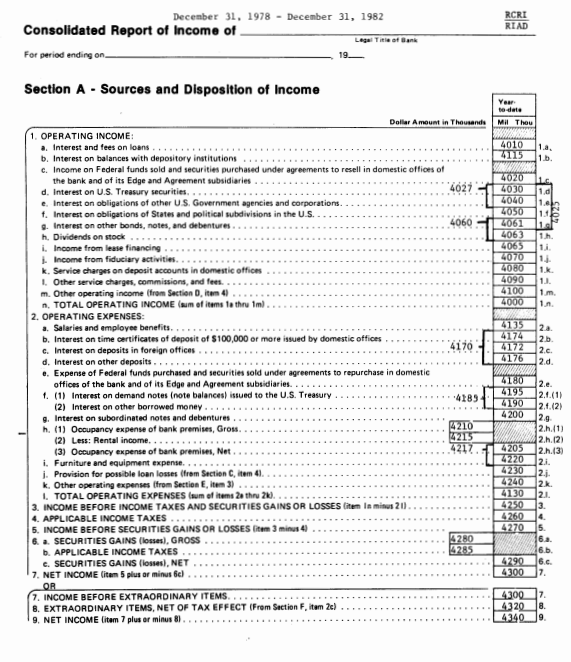}
    \end{center}
    
  \end{figure}
  }

{\color{red}
\paragraph{Details on Key Variables}

}

\clearpage 
\subsection{Data on National Bank Receiverships}
\label{sec:data_receiverships}

We digitize an extensive set of tables with information on national banks in charge of receivers from the OCC's Annual Report to Congress from each year between 1920 to 1941. The 1920 OCC Annual Report contains information on all receiverships from the first receivership in 1865 until receiverships initiated in August 1920. We further digitize the tables on receiverships for each year from 1921 to 1941 to record information on receiverships initiated after 1920, as well as receiverships initiated before 1920 but terminated after 1920.

These data contain a range of information, including the date the receiver was appointed; the date the bank was finally closed; deposits at suspension; assets at suspension; an estimate of the breakdown of assets at suspension by good, doubtful, or worthless assets; additional assets received after the date of suspension; total collections from assets; total collections from shareholder assessments; dividends paid to claim holders; amount of claims proved; secured and preferred liabilities paid; and legal expenses. The data on deposits outstanding at the time of suspension are available starting in 1880. These tables also contain the OCC's assessment of the cause of failure. We harmonize variable definitions over time, as some variable definitions are subject to change. For example, from 1927 onwards, secured and preferred liabilities paid includes offsets allowed and settled, but it excludes offsets before 1927.

We calculate three main statistics from the receivership data.
\begin{itemize}
    \item \textbf{Recovery Rate, $R$}:  We calculate the recovery rate $R$ as the amount ``collected from assets'' divided by the sum of ``nominal assets at date of suspension'' and ``additional assets received since date of suspension''.
    \item \textbf{Total Liabilities at Failure, $D$}, we calculate $D$ as the ``amount of claims proved'' (which refers to claims held by depositors, including unsecured interbank deposits) plus ``offsets allowed and settled'' and ``Loans paid and other disbursements'' (which in later years is reported as ``secured and preferred liabilities paid except through dividend, including offsets'' and includes secured claims such as other borrowed money or bills payable and rediscounts) paid out at failure. We proxy  secured and preferred liabilities at failure by the total effective payments made to these types of claims throughout the receivership. Hence, we implicitly assume full recovery rate for non-deposit liabilities. While we have no information on the actual recovery rates of these types of liabilities, observe that if the effective recovery rate were less than 1, we would be underestimating total liabilities at failure. Underestimating total liabilities, in turn, implies that the  share of fundamentally insolvent banks is even higher than estimated.
    \item \textbf{Deposit Recovery:} For the depositor recovery rate, we use the dividends paid (in \%), reported by the OCC. This reflects the dividends paid relative to the amount of claims proved. The depositor recovery rate does not include interest or account for the time value of money. \Cref{fig:depositor_recovery} shows that the depositor recovery in the first year amounts to only about half of the final recovery. 
\end{itemize}

To ensure robustness, we also study the sensitivity of our findings to calculating the solvency ratio in \Cref{eq:insolvent} in different ways. In our baseline approach, we use the sum of collected from assets and offsets allowed and settled, divided by the sum of the amount of claims proved, secured and preferred liabilities paid, and offsets allowed and settled. We also consider two alternative approaches. First, we calculate the denominator as deposits at suspension, secured and preferred liabilities paid, and offsets allowed and settled. Second, we calculate the numerator as sum of collected from assets, offsets allowed and settled, and collected from shareholders. \Cref{tab:recovery_rate_double_liability} shows the share of fundamentally insolvent banks when using these alternative ways of constructing the solvency ratio. 

\subsubsection{Causes of Failures as Classified By the OCC}
\label{sec:appendix_failure_reasons}

\begin{figure}[htpb]
\caption{\textbf{Classification of Causes of Failure by the OCC across Time}}
\label{fig:failure_reasons_classification_across_time}
\centering 
\includegraphics[width=1.0\textwidth]
{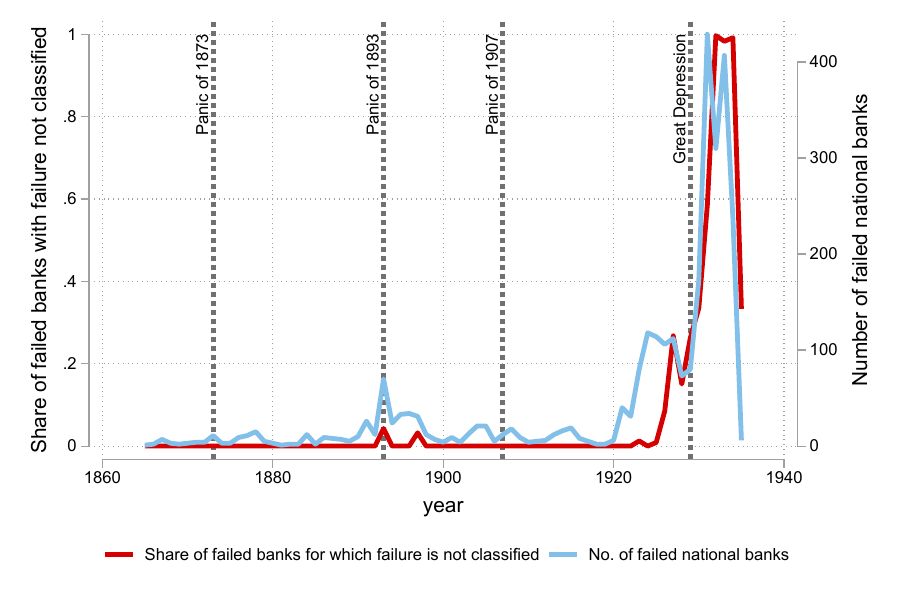}

   \begin{minipage}{\textwidth}
\footnotesize
Notes: This figure shows the share of failed national banks for which the OCC did not provide a cause of failure (left y-axis) and the number of failed national banks (right y-axis) from 1863 through 1935. 
\end{minipage}
\end{figure}

From 1863 to 1928, the OCC classified the ``apparent cause of failure'' for almost all bank failures. For 1929, 1930, and 1931, the OCC classified the cause of failure for 78\%, 75\%, and 48\% of failures, respectively. The OCC did not classify the cause of failure for failures occurring in 1932 and 1933. However, we were able to obtain the cause of failure for 12 failures from 1934-1937 from the OCC's 1937 Annual Report to Congress. See \Cref{fig:failure_reasons_classification_across_time} for the share of failures not classified by year.

We group the detailed cause of failure classifications from the OCC into one of the following broad categories: 
\begin{itemize}
   \item \textbf{Excessive lending:} Excessive lending refers to a bank lending more than 10\% of its paid-in capital to a single counterparty, which was not permitted by the national banking act.     
   \item \textbf{Economic conditions:} We classify failure as caused by external economic factors whenever the OCC cited the trigger of failure being related to things outside of a banks control such as crop losses, a deterioration of local economic conditions, robbery, or other shocks.
   \item \textbf{Fraud:} We classify a failure as due to fraud when the OCC cited misbehavior from bankers as the cause of failure. Fraud can be related to dishonesty of a bank employee or owner  and excessive loans to insiders. 
   \item \textbf{Governance:} We classify a failure being due to governance issues if bad management practices are cited as the cause of failure   
   \item \textbf{Losses:} We refer to the cause of failure being due to losses when the bank is subject to losses or unable to realize on assets,  injudicious banking practices, or depleted reserves.
    \item \textbf{Run:} We classify a run as being the cause of failure when the OCC reports the bank was closed by a run or anticipation of a run or heavy withdrawals. 
\end{itemize}
\Cref{tab:app_cause_of_failure} shows the detailed mapping to these categories.

\paragraph{Relation to the literature.}
\cite{Calomiris1991} analyzes the cause of failure from this same source, but they only use data from a subset of years in the pre-1914 sample in which they identified a banking panic. They find that asset losses and fraud were the predominant causes of failure during panic years. Even in banking panic years, the OCC only identified one failure due to a bank run. They concluded that ``the fact that the Comptroller only attributed one failure to a bank run per se shows that the \textit{direct} link between bank runs and bank failures during panics was not important'' \citep[][p. 154]{Calomiris1991}.

Using classifications from the Federal Reserve Board of Governors, \citet{Richardson2007} finds that, for the period 1929 through 1933, the main cause of failure of Federal Reserve member banks was asset losses, but illiquidity from heavy withdrawals also played a contributing role.  The evidence from the historical sample is also consistent with a detailed study conducted by the OCC of 171 bank failures between 1979 and 1987 \citep{OCC1988}. That study argued that the ``major cause of decline for problem banks continues to be poor asset quality that eventually erodes a bank's capital.'' \citet{OCC1988} write (also highlighted by \citet{Acharya2012}): ``Management-driven weaknesses played a significant role in the decline of 90 percent of the failed and problem banks the OCC evaluated. Many of the difficulties the banks experienced resulted from inadequate loan policies, problem loan identification systems, and systems to ensure compliance with internal policies and banking law. In other cases, directors' or managements' overly aggressive behavior resulted in imprudent lending practices and excessive loan growth that forced the banks to rely on volatile liabilities and to maintain inadequate liquid assets. Insider abuse and fraud were significant factors in the decline of more than one-third of the failed and problem banks the OCC evaluated... Economic decline contributed to the difficulties of many of the failed and problem banks... Rarely, however, were economic factors the sole cause of a bank's decline.''

Poor asset quality was most often caused by poor management decisions and practices, such as imprudent lending practices, excessive loan growth, and fraud. For instance, \cite{Bennett2015} find that fraud was a primary or contributing cause of failure in 24\% of failures based on a sample of failures between 1989 and 2007.


\clearpage
\begin{longtable}{p{13cm}|p{4cm}}

\caption{\normalsize \textbf{OCC Causes of Failure Classification}.\label{tab:app_cause_of_failure}}
\tiny
\\

\textit{OCC Cause of Failure} & \textit{Simplified Label}   
 \\  \hline
\input{./output/tables/appendix_cause_of_failure_list} \\ 
\end{longtable}


%% file: output/tables/appendix_cause_of_failure_list.tex
Crop loss and depreciation of securities&Economic conditions\\
Crop loss&Economic conditions\\
Deflation&Economic conditions\\
Local financial conditions&Economic conditions\\
Local financial depression from unforeseen agricultural or industrial disaster&Economic conditions\\
Excessive loans and failure of large debtors&Excessive lending\\
Excessive loans to officers and directors&Excessive lending\\
Excessive loans to others and depreciation of securities&Excessive lending\\
Excessive loans to others and investments in real estate and mortgages&Excessive lending\\
Excessive loans to others, injudicious banking, and depreciation of securities&Excessive lending\\
Excessive loans&Excessive lending\\
Failure of large debtors&Excessive lending\\
Defalcation by cashier&Fraud\\
Defalcation by former cashier&Fraud\\
Defalcation of officers and depreciation of securities&Fraud\\
Defalcation of officers and excessive loans to others&Fraud\\
Defalcation of officers and fraudulent management&Fraud\\
Defalcation of officers&Fraud\\
Dishonesty of an officier of employee and local financial depression from unforeseen agricultural or industrial disaster&Fraud\\
Dishonesty of an officier of employee&Fraud\\
Dishonesty&Fraud\\
Excessive loans to officers and directors and depreciation of securities&Fraud\\
Excessive loans to officers and directors and investments in real estate and mortgages&Fraud\\
Forgeries and embezzlement&Fraud\\
Fraudulent management&Fraud\\
Fraudulent management and closed by run&Fraud\\
Fraudulent management and depreciation of securities&Fraud\\
Fraudulent management and injudicious banking&Fraud\\
Fraudulent management and local financial conditions&Fraud\\
Fraudulent management, defalcation of officers, and depreciation of securities&Fraud\\
Fraudulent management, excessive loans to officers and directors, and depreciation of securities &Fraud\\
Fraudulent management, excessive loans to officers and directors, and excessive loans to others&Fraud\\
Fraudulent management, injudicious banking, investments in real estate and mortgages, and depreciation of securities&Fraud\\
Fraudulent management&Fraud\\
Irregularities of president and speculation in real estate&Fraud\\
Irregularities&Fraud\\
Wrecked by assistant cashier&Fraud\\
Wrecked by cashier and president and by excessive loans to themselves&Fraud\\
Wrecked by defalcation by bookkeeper&Fraud\\
Wrecked by president&Fraud\\
Wrecked by the cashier&Fraud\\
Bad management&Governance\\
Incompetent management and dishonesty of an officier of employee&Governance\\
Incompetent management and local financial depression from unforeseen agricultural or industrial disaster&Governance\\
Incompetent management&Governance\\
Bad paper taken over from old organization&Losses\\
Bad paper&Losses\\
Deficient reserve and unable to realize on loans&Losses\\
Depleted reserve&Losses\\
Depleted reserve and shrinkage of deposits&Losses\\
Depreciation of securities&Losses\\
Formerly in voluntary liquidation&Losses\\
General stringency of the money market, shrinkage in values, and imprudent methods of banking&Losses\\
Injudicious banking and adverse business conditions&Losses\\
Injudicious banking and depreciation of securities&Losses\\
Injudicious banking and excessive loans to officers and others&Losses\\
Injudicious banking and failure of large debtors&Losses\\
Injudicious banking&Losses\\
Insufficient credit&Losses\\
Investment in real estate mortgages and depreciation of securities&Losses\\
Investments in real estate and mortgages and depreciation of securities&Losses\\
Large losses and defalcation&Losses\\
Large losses and injudicious banking&Losses\\
Large losses in loans and discounts&Losses\\
Large losses, withdrawals, and insufficient credit&Losses\\
Large losses&Losses\\
Receiver appointed after sale of assets, and stockholders to vote to place bank in liquidation&Losses\\
Receiver appointed after voluntary liquidation&Losses\\
Receiver appointed to assess stockholders&Losses\\
Receiver appointed to levy and collect stock assessment covering deficiency in value of assets sold, or to complete unfinished liquidation&Losses\\
Receiver appointed to levy and collect stock assessment covering deficiency in value of assets sold&Losses\\
Unable to realize on assets&Losses\\
Unable to realize on loans and failure of stockholders to pay balance due on capital&Losses\\
Unable to realize on loans&Losses\\
Information not available&No information\\
Robbery and burning of bank&Other\\
Temporary suspension&Other\\
Temporary suspension to adjust settlement on adverse judgment&Other\\
Closed by directors in anticipation of run&Run\\
Closed by run&Run\\
Directors closed due to rumor of run&Run\\
Heavy withdrawals and lack of public confidence&Run\\
Heavy withdrawals&Run\\
Inability to meet demands&Run\\
Large demands and depleted cash&Run\\
Local financial conditions and closed by run&Run\\